\documentclass[12pt]{article}

\usepackage{latexsym}
\usepackage{amssymb,amsfonts,amsmath}
\usepackage{graphicx} 
\usepackage{indentfirst}
\usepackage{bbm}
\usepackage{amssymb}
\usepackage{verbatim}
\usepackage{amsmath, amsthm,amssymb}
\usepackage{mathrsfs}
\usepackage{hyperref}
\usepackage{amsfonts}
\usepackage{dsfont}
\usepackage{cite}
\usepackage{tikz}
\usepackage{enumitem}

\parskip=\medskipamount

\newcommand {\cA}{{\cal A}}

\newcommand {\cC}{{\cal C}}
\newcommand {\cD}{{\cal D}}
\newcommand {\cE}{{\cal E}}

\newcommand {\cH}{{\cal H}}

\newcommand {\cL}{{\cal L}}
\newcommand {\cM}{{\cal M}}
\newcommand {\cN}{{\cal N}}

\newcommand {\cQ}{{\cal Q}}

\newcommand {\cS}{{\cal S}}

\newcommand {\cU}{{\cal U}}
\newcommand {\cV}{{\cal V}}
\newcommand {\cW}{{\cal W}}

\def\a{\alpha}

\def\b{\beta}

\def\d{\delta}
\def\e{\epsilon}
\def\f{\phi}
\def\g{\gamma}
\def\G{\Gamma}

\def\k{\kappa}
\def\l{\lambda}
\def\m{\mu}

\def\o{\omega}

\def\q{\theta}
\def\r{\rho}
\def\s{\sigma}
\def\t{\tau}

\def\x{\xi}
\def\z{\zeta}

\def\F{\Phi}
\def\J{\Psi}
\def\L{\Lambda}
\def\O{\Omega}

\def\S{\Sigma}
\def\U{\Upsilon}

\def\tr{{\rm tr}}
\def\rd{{\rm d}}
\def\ri{{\rm i}}
\def\re{{\rm e}}

\newcommand{\ve}{\varepsilon}                            
\newcommand{\cDB}{{\bar\cD}}                            

\newcommand{\ab}{{\a\b}}

\newcommand{\pa}{\partial}                           
\newcommand{\hf}{\frac12}

\newcommand{\vf}{\varphi}

\newcommand{\be}{\begin{equation}}
\newcommand{\ee}{\end{equation}}
\newcommand{\bea}{\begin{eqnarray}}
\newcommand{\eea}{\end{eqnarray}}
\newcommand{\non}{\nonumber}
\newcommand{\1}{{\underline{1}}}
\newcommand{\2}{{\underline{2}}}

\newcommand{\bm}[1]{\mbox{\boldmath$#1$}}

\def\double #1{#1{\hbox{\kern-2pt $#1$}}}

\newcommand{\RomanNumeralCaps}[1]
{\MakeUppercase{\romannumeral #1}}

\newif\ifdtup

\def\de{{\nabla}}                                         
                                         
\newcommand{\bsubeq}{\begin{subequations}}
\newcommand{\esubeq}{\end{subequations}}

\numberwithin{equation}{section}

\newcommand{\sSL}{\mathsf{SL}}

\newcommand{\sSO}{\mathsf{SO}}

\newcommand{\sOSp}{\mathsf{OSp}}

\begin{document}

\begin{titlepage}
\begin{flushright}
November, 2020 \\
Revised version: June, 2021
\end{flushright}
\vspace{5mm}

\begin{center}
{\Large \bf 
Higher-spin gauge models with (1,1) supersymmetry in AdS${}_3$: 
Reduction  to (1,0)  superspace}
\\ 
\end{center}

\begin{center}

{\bf
Daniel Hutchings, Jessica Hutomo
and Sergei M. Kuzenko} \\
\vspace{5mm}

\footnotesize{
{\it Department of Physics M013, The University of Western Australia\\
35 Stirling Highway, Crawley W.A. 6009, Australia}}  
~\\

\vspace{2mm}
~\\
\texttt{Email: daniel.hutchings@research.uwa.edu.au,
jessica.hutomo@research.uwa.edu.au, sergei.kuzenko@uwa.edu.au}
\vspace{2mm}

\end{center}

\begin{abstract}
\baselineskip=14pt
In three dimensions, there are two types of $\cN=2$ anti-de Sitter (AdS) supersymmetry, which are denoted 
 (1,1) and (2,0). They are characterised by different supercurrents  and support different families of higher-spin gauge 
 models (massless and massive) which were constructed in Refs. \cite{HKO} and \cite{HK18}
for the (1,1) and  (2,0) cases, respectively, using superspace techniques. It turns out that the precise difference between the (1,1) and (2,0) higher-spin supermultiplets can be pinned down by reducing these gauge theories to (1,0) AdS superspace. The present paper is devoted to the $(1,1) \to (1,0)$ AdS superspace reduction. In conjunction with the outcomes of the $(2,0) \to (1,0)$ AdS superspace reduction carried out in Ref. \cite{HK19}, we demonstrate that every known higher-spin theory with (1,1) or (2,0) AdS supersymmetry decomposes into a sum of two off-shell (1,0) supermultiplets which belong to four series of inequivalent higher-spin gauge models.
The latter are reduced to components.
\end{abstract}
\vspace{5mm}

\vfill
\end{titlepage}

\newpage
\renewcommand{\thefootnote}{\arabic{footnote}}
\setcounter{footnote}{0}

\tableofcontents{}
\vspace{1cm}
\bigskip\hrule

\allowdisplaybreaks

\section{Introduction}

To study $\cN$-extended  supersymmetric theories in $d$ dimensions, 
it is advantageous to deal with a formulation that permits some amount of supersymmetry to be realised manifestly. 
In general there exist two superspace settings to achieve this. 
One of them makes use of the standard 
$\cN$-extended superspace (sometimes endowed with additional commuting variables
\cite{Rosly,GIKOS,KLR})
and provides a manifestly supersymmetric formulation. The other employs a smaller  $\widetilde \cN$-extended superspace, with $\widetilde \cN < \cN$, to keep manifest only $\widetilde \cN$ supersymmetries. 
Both approaches have found numerous applications in the literature.
For instance, it is known that the geometric properties of general $\cN=2$ supersymmetric 
nonlinear $\s$-models in Minkowski space ${\mathbb M}^4$ \cite{A-GF}
are remarkably transparent if these theories  are realised in $\cN=1$ superspace 
\cite{HKLR,BX}. One of the two supersymmetries is manifest and off-shell in  this setting, 
while the second supersymmetry is hidden and on-shell (the commutator of the first and the second supersymmetries closes {\it only} on-shell). 
On the other hand, in order to construct general off-shell $\cN=2$ supersymmetric nonlinear $\s$-models, manifestly supersymmetric techniques are indispensable, and 
there exist two powerful $\cN=2$ superspace approaches:
(i) the  harmonic superspace \cite{GIKOS,GIOS};
and (ii) the  projective superspace \cite{KLR,LR-projective1,LR-projective2}. 
One of the conceptual virtues of these manifestly supersymmetric formulations is the possibility to generate $\cN=2$ nonlinear $\s$-model actions (and thus hyperk\"ahler metrics) 
from Lagrangians of {\it arbitrary} functional form.\footnote{This possibility was realised for the first time in an important paper \cite{GIOS2}, where it was demonstrated that the most general interacting Lagrangian of the $q^+$ hypermultiplet superfields can be treated as the hyperk\"ahler potential parametrising the complete set of hyperk\"ahler metrics.}
Analogous results exist for general   $\cN=2$ supersymmetric nonlinear $\s$-models in four-dimensional anti-de Sitter space (AdS${}_4$), 
which were originally formulated as off-shell theories 
in $\cN=2$ AdS superspace  \cite{KT-M08}. This approach makes $\cN=2$ supersymmetry 
manifest, but the hyperk\"ahler geometry of the $\s$-model target space is hidden.
Some time later,  the most general $\cN=2$ supersymmetric $\s$-models in 
$\rm AdS_4$ were constructed 
using a formulation in terms of $\cN=1$ covariantly chiral superfields
\cite{BKsigma1,BKsigma2}. One of the two supersymmetries is hidden in this approach, 
but the geometric properties of 
the $\cN=2$ supersymmetric  $\s$-models in $\rm AdS_4$ become transparent.
Specifically, the target space must be a non-compact hyperk\"ahler manifold 
endowed with a Killing vector field which generates an
$\sSO(2)$ group of rotations on the two-sphere of complex structures. 
The two $\s$-model formulations developed in \cite{KT-M08} and  \cite{BKsigma1,BKsigma2}
are related via the $\cN=2 \to \cN=1$ AdS${}_4$ superspace reduction worked out  in \cite{BKLT-M}.

Extended supersymmetric theories in AdS${}_3$ possess exceptionally many superspace realisations. We recall that the (connected) AdS  group in three dimensions is not a simple group, 
\bea
\sSO_0(2,2) \cong \Big( \sSL(2, {\mathbb R}) 
\times \sSL( 2, {\mathbb R}) \Big)/{\mathbb Z}_2~,
\eea
and so are its supersymmetric extensions  
${\sOSp} (p|2; {\mathbb R} ) \times  {\sOSp} (q|2; {\mathbb R} )$.
This means that  there are several species of $\cN$-extended AdS supersymmetry
  \cite{AT}, 
  which are known as the  
  $(p,q)$ AdS supersymmetry types,
where the integers $p \geq q\geq 0$ are such that $\cN=p+q$.   
In principle, field theories possessing the  $(p,q)$ AdS supersymmetry may
be realised in the so-called $(p,q)$ AdS superspace  \cite{KLT-M12},
${\rm AdS}^{(3|p,q)} $,
which may be viewed as a maximally supersymmetric solution 
of the $(p,q)$ AdS supergravity  \cite{AT}, with anti-de Sitter space
$\rm AdS_3$ being the  bosonic body of the superspace.\footnote{In order 
to realise field theories with the  $(p,q)$ AdS supersymmetry in ${\rm AdS}^{(3|p,q)} $,
 for $p+q\leq 4$ one can employ
the off-shell supergravity methods developed in \cite{KLT-M11,KT-M11,KLRST-M}.}
More specifically, within the supergravity framework of \cite{KLT-M12,KLT-M11} 
the superspace ${\rm AdS}^{(3|p,q)} $
originates as  a maximally symmetric supergeometry with covariantly constant torsion and 
curvature 
generated by a symmetric real  torsion $S^{IJ}= S^{JI}$, with the structure-group indices $I, J$ 
taking values from 1 to $\cN$.  
It turns out that $S^{IJ}$ is nonsingular, and the parameters $p$ and $q= \cN-p$ 
determine its  signature. Since the isometry group of ${\rm AdS}^{(3|p,q)} $ is 
${\sOSp} (p|2; {\mathbb R} ) \times  {\sOSp} (q|2; {\mathbb R} )$ and since $S^{IJ}$ is 
invariant under its compact subgroup  $ \sSO(p) \times {\sSO}(q)$, the global realisation 
of the superspace is
\bea
{\rm AdS}^{(3|p,q)} = \frac{ {\sOSp} (p|2; {\mathbb R} ) \times  {\sOSp} (q|2; {\mathbb R} ) } 
{ {\sSL}( 2, {\mathbb R}) \times {\sSO}(p) \times {\sSO}(q)}~.
\label{1.2}
\eea
In fact, starting from the superspace geometry of $\cN$-extended conformal supergravity 
\cite{KLT-M11}
and restricting the torsion to be covariantly constant and Lorentz invariant, a general AdS 
superspace 
solution for $\cN\geq 4$ includes not only the torsion $S^{IJ}$,  but also 
a completely antisymmetric curvature  $X^{IJKL} = X^{[IJKL]} $. 
It turns out that the latter may be non-zero 
only if $S^{IJ} =S \d^{IJ}$,  which means $p=\cN$ and $q=0$. Such solutions correspond to 
exotic AdS superspaces,
${\rm AdS}^{3|2\cN}_{S,X}$, for which the isometry group is, in general,  a subgroup 
 of $\rm OSp (\cN|2; {\mathbb R} ) \times  SL (2, {\mathbb R} )$. The superspace
 ${\rm AdS}^{3|2\cN}_{S,X}$ is not conformally flat. All the superspaces \eqref{1.2} are conformally flat, see \cite{KLT-M12} for the technical  details.\footnote{It should be pointed out that the superconformal flatness of AdS  superspaces in diverse dimensions, including the $d=3$ case, was studied in Ref. \cite{BILS}
where the coset spaces \eqref{1.2} with $p+q=2$ were briefly discussed.}
 It should be mentioned that (1,0) AdS  superspace
   (or simply $\cN=1$ AdS superspace, ${\rm AdS}^{(3|1,0)} \equiv \rm AdS^{3|2} $) was introduced in \cite{GGRS}.

Consider a $(p,q)$ supersymmetric field theory formulated  in ${\rm AdS}^{(3|p,q)} $,
with $ p+q\geq 3$ and $p\geq q$.
As argued in \cite{BKT-M}, it can always be recast
as a supersymmetric theory realised in  $(2,0)$ AdS superspace, with $(p+q -2)$ 
supersymmetries being hidden. In the case $ p+q\geq 3$ and $p\geq q>0$, 
every supersymmetric field theory  in ${\rm AdS}^{(3|p,q)} $, can be reformulated as 
a theory in (1,1) AdS superspace. Such reformulations were developed in \cite{BKT-M} 
for general $(p,q)$ supersymmetric nonlinear $\s$-models in AdS${}_3$, with $p+q \leq4$. 
It is obvious that the list of possible superspace embeddings can be continued.
It is also clear that every extended supersymmetric theory in AdS${}_3$, with $p+q >1$, 
may be realised in $\cN=1$ AdS superspace.

This work is devoted to implementing the $(1,1) \to (1,0)$ AdS reduction of the $(1,1)$ supersymmetric higher-spin models in AdS${}_3$ (both massless and massive), which were constructed in \cite{HKO} as an extension of the earlier results in $\cN=2$ Minkowski superspace \cite{KO}.\footnote{The massless higher-spin theories proposed in \cite{HKO,KO} have natural counterparts in four dimensions \cite{KSP,KS93,KS94}, see also \cite{BK} for a review.}
There are two main motivations for this project. Firstly, the off-shell structure of the 
half-integer superspin
multiplets with (2,0) AdS supersymmetry \cite{HK18} drastically differs from that of their (1,1) counterparts \cite{HKO}.\footnote{It is pertinent to mention here that the (2,0) and (1,1) AdS supersymmetries support different supercurrent multiplets \cite{KT-M11}.}   
Direct comparison of these theories is difficult in a manifestly supersymmetric setting,  
 since they are formulated in different superspaces,  ${\rm AdS}^{(3|2,0)} $ and 
${\rm AdS}^{(3|1,1)} $, respectively. However, both families of higher-spin theories can be reformulated in the same (1,0) AdS superspace, and then the precise difference between the (2,0) and (1,1) higher-spin supermultiplets can be pinned down. 
The $(2,0) \to (1,0) $ reduction of the (2,0) supersymmetric higher-spin theories in AdS${}_3$  \cite{HK18} 
has already been carried out in \cite{HK19}. We will show that every massless higher-spin theory with (1,1) or (2,0) AdS supersymmetry decomposes into a sum of two off-shell (1,0) supermultiplets
which belong to the four series of inequivalent higher-spin gauge models constructed in \cite{HK19}.

The second motivation for this work is quantum mechanical. 
All known off-shell higher-spin $\cN=2$ supermultiplets in AdS${}_3$, with either (2,0) or (1,1) AdS supersymmetry \cite{HK18,HKO},
are reducible gauge theories (in the terminology of the Batalin-Vilkovisky 
quantisation \cite{BV}), similar to the massless higher-spin supermultiplets 
in AdS${}_4$ \cite{KS94}. The Lagrangian quantisation of such theories proves to be a nontrivial procedure. In four dimensions, the quantisation of the theories proposed in \cite{KS94} 
was achieved in \cite{BKS}. On the other hand, all off-shell higher-spin $\cN=1$ supermultiplets 
in AdS${}_3$ studied in this paper and \cite{HK19}, are irreducible gauge theories. They can be quantised using the Faddeev-Popov procedure \cite{FP}, as in the non-supersymmetric case, see e.g. \cite{GGS}. This opens the possibility to develop heat kernel techniques
for higher-spin theories in AdS${}^{3|2}$, by extending the four-dimensional 
results \cite{BK,McA2,BK86}.

This paper is organised as follows. Section \ref{Section2} provides a succinct review of ${\cN}=1$ supersymmetric field theory in AdS$_3$. In section \ref{Section3}, we review the structure of the four series of off-shell massless higher-spin ${\cN}=1$ supersymmetric models in AdS$_3$ constructed in \cite{HK19,KP}. A formalism to reduce every field theory with (1,1) AdS supersymmetry to ${\cN}=1$ AdS superspace is developed in section \ref{Section4}. This formalism is then applied throughout sections \ref{Section5}$-$\ref{Section8} to carry out the (1,1) $\to$ (1,0) AdS superspace reduction of massless higher-spin models with (1,1) AdS supersymmetry. These are detailed in sections \ref{Section5}$-$\ref{Section6} for the half-integer $(s+\hf)$ superspin case, and in sections \ref{Section7}$-$\ref{Section8} for the integer $(s)$ superspin case. Our results are summarised and discussed in section \ref{Section9}, where we also analyse the difference between the (2,0) and (1,1) higher-spin (massless and massive) supermultiplets. Our spinor notation and conventions are collected in appendix \ref{AppendixA}. In appendix \ref{AppB}, we study the component structure of the off-shell massless higher-spin $\mathcal{N}=1$ supersymmetric models in AdS$_3$ reviewed in section \ref{Section3}.


\section{${\cN}=1$ supersymmetric field theory in AdS$_3$} \label{Section2}

In this section we provide salient facts about the geometry of ${\cN}=1$ AdS superspace, AdS$^{3|2}$, and its isometries following \cite{KLT-M12}. We also recall certain duality transformations in AdS$^{3|2}$, following \cite{HK19}.

Let $z^\cM = (x^m, \q^\m)$ be local coordinates
parametrising  AdS$^{3|2}$. The geometry of AdS$^{3|2}$ is encoded in the set of covariant derivatives of the form
\be
\nabla_A = (\nabla_a, \nabla_\a) = E_A{}^M {\partial}_M + \hf \o_A{}^{bc}M_{bc}~, \label{N10}
\ee
where $E_A{}^M$ is the inverse vielbein and  $\o_A{}^{bc}$ the Lorentz connection. 
The explicit relation between Lorentz generators with two vector indices $(M_{ab})$, with
one vector index $(M_a)$ and with two spinor indices $(M_{\a \b})$ is described in appendix \ref{AppendixA}.
The covariant derivatives obey the following algebra \cite{GGRS}
\begin{subequations} \label{algN1}
\be
\{ \nabla_\a , \nabla_\b \} = 2\ri \nabla_{\a\b} - 4\ri|\m| M_{\a\b}~, \\
\ee
\be
\ [ \nabla_a, \nabla_\b ] = |\m| (\g_a)_\b{}^\g \nabla_\g~, \qquad \ [ \nabla_a, \nabla_b ] = -4 |\m|^2 M_{ab}~. \label{22b}
\ee
In spinor notation, eqs.\,\eqref{22b} take the form
\bea
\ [ \nabla_{\a \b}, \nabla_\g ] = -2|\m| \ve_{\g(\a}\nabla_{\b)}~, \qquad \ [ \nabla_{\a \b}, \nabla_{\g \d} ] = 4 |\m|^2 \Big(\ve_{\g(\a}M_{\b)\d} + \ve_{\d(\a} M_{\b)\g}\Big)~.
\eea
\end{subequations}
Here $|\m|>0 $ is a 
constant parameter, 
which determines the curvature of AdS$_{3}$.

The parameter $|\m|$ in \eqref{algN1} was denoted $\cS$ in \cite{KLT-M12}. However, in this paper we prefer to make use of 
the notation $|\m|$ which is more appropriate in the context of the $(1,1) \to (1,0)$ superspace reduction, which will be considered later. 
It should be remarked 
that the geometry of $\cN=1$ AdS superspace can also be described by 
the graded commutation relations which are obtained from \eqref{algN1} 
by replacing $|\m| \to -|\m|$. 
The two choices, $\cS=|\m|$ and $\cS = -|\m|$,  correspond to the so-called $(1,0)$ and $(0,1)$  
AdS superspaces \cite{KLT-M12}, which are different realisations of $\cN=1$ AdS superspace. The $(1,0)$ and $(0,1)$ AdS superspaces are naturally embedded 
in $(1,1)$ AdS superspace\footnote{ 
As follows from \eqref{1.3}, in $(1,1)$ AdS superspace
two subsets of covariant derivatives,  $( {\bm \de}_{a}, {\bm \de}_{\a}^{\1}) $ and 
$( {\bm \de}_{a}, {\bm \de}_{\a}^{\2}) $, form closed algebras of the type \eqref{algN1}, 
 with the curvature parameter being $|\m| $ and $ -|\m|$, respectively.}
 and are related to each other by a parity transformation.

One may derive several useful identities:
\begin{subequations}  \label{A8-mod}
	\bea 
	\nabla_\a \nabla_\b &=& \ri \nabla_{\a\b} - 2\ri|\m|M_{\a\b}+\frac{1}{2}\ve_{\a\b}\nabla^2~, \\
	\nabla^{\b} \nabla_\a \nabla_\b &=& 4\ri |\m|\nabla_\a~, \quad \{ \nabla^2, \nabla_\a  \} = 4\ri |\m|\nabla_\a~,\\
 \nabla^2 \nabla_\a &=& 2\ri |\m|\nabla_\a + 2\ri \nabla_{\a\b} \nabla^\b - 4\ri |\m| \nabla^\b M_{\a\b}~, \\
\qquad \ [ \nabla_{\a} \nabla_{\b}, \nabla^2 ]
&=& 0 \quad \Longrightarrow \quad \ [\nabla_{\a\b}, \nabla^2 ] = 0~,\\
\ [ \nabla_\a ,  \Box] &=& 2|\m| \nabla_{\a\b} \nabla^\b + 3|\m|^2 \nabla_\a~.
	\eea
\end{subequations}
where we have denoted $\nabla^2 = \nabla^\a \nabla_\a$ and $\Box = \de^a \de_a = -\hf \de^{\a \b} \de_{\a \b}$.
In particular, it follows from the algebra of the covariant derivatives that
\bea
-\frac{1}{4}\nabla^2\nabla^2 &=&  \Box - 2\ri|\m| \nabla^2+2|\m| \nabla^{\a\b}M_{\a\b} -2|\m|^2M^{\a\b}M_{\a\b}~.
\eea
This differential operator can be expressed in terms of the quadratic Casimir operator of the $\cN=1$ AdS supergroup \cite{KP}
\be \label{QuadraticCasimir}
\mathbb{Q} = -\frac{1}{4}\nabla^2 \nabla^2 + \ri |\m| \nabla^2, \qquad [\mathbb{Q}, \nabla_A] = 0~.
\ee

Given an arbitrary superfield $F$ and its complex conjugate $\bar F $, the following relation holds 
\bea
\overline{\nabla_{\a} F} = - (-1)^{\e(F)} \nabla_{\a} \bar F~,
\eea
where $\e(F)$ denotes the Grassmann parity of $F$. 

According to the general formalism of \cite{BK}, the isometry transformations of AdS$^{3|2}$ are generated by the Killing supervector fields,
\bea
\x = \x^B\nabla_B = \x^b\nabla_b + \x^\b\nabla_\b~, 
\eea
which by definition solve the master equation \cite{KLT-M12}
\be \label{Kcondition1-1}
[\x + \frac{1}{2}\z^{bc}M_{bc}, \nabla_A] = 0~,
\ee
for some Lorentz  parameter $\z^{bc} = - \z^{cb}$. 
It can be shown that the Killing equation \eqref{Kcondition1-1} is equivalent to the set of relations \cite{KLT-M12} (see also \cite{KP})
\begin{subequations}  \label{2.40d-1}
\bea
\nabla_{(\a}\x_{\b\g)} &=& 0~, \qquad \nabla_\b \x^{\b\a} =-6 \ri \x^\a~, \\
\nabla_{(\a}\z_{\b\g)} &=& 0~, \qquad  \nabla_\b \z^{\b\a} = -12 \ri |\m| \x^\a ~, \\
\nabla_\a \x_\b &=& \frac{1}{2} \z_{\a\b} +  |\m|   \x_{\a\b}~,
\eea
\end{subequations}
which imply 
\bsubeq \label{2.40e}
\bea
\de_a \xi_b + \de_b \xi_a &=& 0~,\\
\de_{\a \b}\xi^{\b}+3 |\m| \xi_{\a} &=& 0~, \\
\big( \ri \de^2 
+12  |\m| \big) \xi_{\a} &=&0 ~.
\eea
\esubeq
Thus, $\xi^a$ is a Killing vector. 
Given a tensor superfield $U(x, \q)$ (with suppressed indices) on AdS$^{3|2}$, its infinitesimal isometry transformation law is given by 
\bea
\d_\x U= \big (  \x^a \nabla_a + \x^\a \nabla_\a + \hf \z^{ab}M_{ab} \big ) U ~. \label{211}
\eea

To study the dynamics of $\cN = 1$ supersymmetric field theories
in AdS$_3$, a manifestly supersymmetric action principle is required. 
Such an action is associated with a real scalar Lagrangian $L$ and has the form
\bea
S = \int \rd^3x\,\rd^2\q E\, L~, \qquad E^{-1} = {\rm Ber}(E_A{}^{M})~.
\label{212}
\eea
In what follows, we make use of the notation $\rd^{3|2}z := \rd^3x \rd^2 \theta$. 

The component form of the action \eqref{212} is 
\bea
S = \frac{1}{4} \int \rd^3x \, e \, (\ri \nabla^2 + 8|\m|) L\,\big|_{\,\q = 0}~, \qquad 
e^{-1} = {\rm det}(e_a{}^{m})~.
\eea
Here the component inverse vielbein is defined as $e_{a}{}^{m}(x) = E_{a}{}^{m}|_{\theta=0}$. 
Making use of the $\cN=1$  AdS transformation law $\d_\xi L = \xi L$ and the identities \eqref{2.40d-1} and \eqref{2.40e}, it can be shown that the action
\eqref{212}  is invariant under the isometry group of AdS$^{3|2}$.

To conclude this section, we review the duality transformation described in  \cite{HK19}.
Consider a field theory in AdS${}^{3|2}$ which is formulated in terms of a real tensor 
superfield $V_{\a(n)}$ with the action
\bea
S^{\parallel} [ V_{\a(n)} ]= \int \rd^{3|2}z \, E\, \cL \big(\ri^{n+1}  \nabla_\b V_{\a(n)} \big)~.
\label{long5.1}
\eea 
We call $\nabla_\b V_{\a(n)} $ a longitudinal superfield, 
by analogy with a longitudinal vector field.
The theory \eqref{long5.1} has a dual formulation that is obtained by introducing 
a first-order action 
\bea
S_{\text{first-order}} = \int \rd^{3|2}z \, E\, \Big\{ \cL \big( \S_{\b; \, \a(n)} \big)
+ \ri^{n+1} \cW^{\b; \, \a(n) }  \S_{\b; \,\a(n)} \Big\}~.
\label{first-order}
\eea
Here $\S_{\b;\a(n)} $ is unconstrained and the Lagrange multiplier is 
\bea
 \cW_{\b; \,\a(n)} =
  \ri^{n+1} \Big( \nabla^{\g} \nabla_{\b} - 4\ri |\m| \delta^{\g}_{\,\b}\Big){\Psi}_{\g; \,\a(n)} ~,
  \qquad \nabla^\b \cW_{\b; \,\a(n)} =0~,
  \label{W5.3}
 \eea
for some unconstrained prepotential ${\Psi}_{\g; \,\a(n)} $. 
Varying \eqref{first-order} with respect to ${\Psi}_{\g; \,\a(n)} $ gives
\bea
\nabla^\b \nabla_\g \S_{\b; \,\a(n)} - 4\ri |\m| \S_{\g; \,\a(n)} =0 
\quad \implies \quad \S_{\b; \,\a(n)} = \ri^{n+1} \nabla_\b V_{\a(n)}~,
\eea
thus $S_{\text{first-order}} $ reduces to the original action \eqref{long5.1}.
On the other hand, integrating out 
$\S_{\b;\a(n)} $ from $S_{\text{first-order}}$ leads to a dual action of the form 
\bea
S^{\perp} [ {\Psi}_{\g; \,\a(n)} ]= \int \rd^{3|2}z \, E\,  
\cL_{\rm dual} \big(  \cW_{\b; \, \a(n) } \big)~,
\label{tran5.5}
\eea
which is invariant under gauge transformations
\bea
\d {\Psi}_{\g; \,\a(n)} = \ri^{n+1} \nabla_\g \eta_{\a(n)} ~.
\eea
It is natural to call the gauge-invariant field strength $\cW_{\b; \, \a(n) } $ a transverse superfield, due to constraint \eqref{W5.3}. The dual formulations \eqref{long5.1} and \eqref{tran5.5} are referred to as longitudinal and transverse, respectively. 
In the $n=1$ case, the duality transformation corresponds to the standard 
duality between the $\cN=1$ scalar and vector multiplets in three dimensions, see \cite{HitchinKLR}.

The above duality transformation naturally extends to Minkowski superspace ${\mathbb M}^{3|2}$ 
in the limit $|\m|\to 0$. Therefore, given two dually equivalent theories in $\rm AdS^{3|2}$, their flat superspace counterparts are also dually equivalent. The opposite is not always true. 
 For instance, the flat superspace counterparts of the massless half-integer superspin models 
\eqref{LongHalfIntAct} and \eqref{TransHIAct} are dual. However, these models are not dual in 
AdS${}_3$. The requirement of gauge invariance uniquely fixes the actions 
\eqref{LongHalfIntAct} and \eqref{TransHIAct} in  $\rm AdS^{3|2}$, including the presence of certain $|\m|$-dependent terms which are incompatible with duality invariance. 


\section{Massless higher-spin $\cN=1$ supermultiplets}  \label{Section3}

According to \cite{HK19}, for each superspin value $\hat{s}\geq 1$, where $\hat{s}$ is either integer $(\hat{s}=s)$ or half-integer $(\hat{s}=s+\frac{1}{2})$, there exist two off-shell formulations for a massless $\cN=1$ superspin-$\hat{s}$ multiplet in 
AdS$_3$. In the flat superspace limit, 
these models prove to related to each other via a superfield Legendre transformation. However, when uplifted to $\cN=1$ AdS superspace, it becomes apparent that these duality relations no longer hold. In this section, we review the explicit formulation of these respective theories, which were derived in \cite{HK19,KP}.

As is known, any massless multiplet of superspin $\hat s >1$ has no propagating degrees of freedom, and the notion of superspin is purely kinematical. It is used by analogy with massive supermultiplets. The concept of superspin is well defined in the massive case, and we follow the definition used in \cite{KT} in the super-Poincar\'e case. Specifically, 
if $n$ is  a positive integer,  an on-shell massive multiplet of superspin $n/2$
is described by a real symmetric rank-$n$ spinor superfield,
 $\cH_{\a_1 \cdots \a_n} = \bar \cH_{\a_1\dots \a_n} = \cH_{(\a_1 \cdots \a_n)}  $,
which obeys the differential conditions \cite{KNG}
\begin{subequations}
\bea
D^\b \cH_{\b \a_1 \cdots \a_{n-1}} &=& 0~ , 
\\
-\frac{\ri}{2} D^2  \cH_{\a_1 \dots \a_n} &=& m \s \cH_{\a_1 \dots \a_n}~, 
\qquad \s =\pm 1~.
\eea
\end{subequations}
Here, $m$ is a real constant of unit mass dimension, $D_\a$ is the spinor covariant derivative of $\cN=1$ Minkowski superspace, and $D^2 = D^\a D_\a$.
The superfield $\cH_{\a(n)}$ contains two ordinary 
on-shell massive fields, which are
\bea
\f_{\a_1 \dots \a_n} := \cH_{\a_1 \dots \a_n} |_{\q=0}~, \qquad 
\f_{\a_1 \dots \a_{n+1}} := \ri^{n+1} D_{(\a_1} \cH_{\a_2 \dots \a_{n+1})} |_{\q=0}~.
\eea
Their helicity values are $\frac{n}{2} \s $ and $\frac{n+1}{2}  \s$, respectively, see \cite{KT} for the technical details.

An off-shell massless superspin-$n/2$ gauge theory in AdS${}_3$ can be realised in terms of two superfields, of which one is universally a superconformal gauge prepotential $H_{\a(n)}$ and the other is a compensating multiplet. 
Here 
$H_{\a(n)}:=
H_{\a_1 .... \a_n}=H_{(\a_1...\a_n)}$ is a real symmetric rank-$n$ spinor  which is defined modulo gauge transformations of the form\footnote{In the flat superspace limit this gauge transformation 
reduces to that introduced in \cite{SK,KT}.}
\bea \label{SCSGI}
\d_\z H_{\a(n)} = \ri^n (-1)^{\lfloor n/2 \rfloor}\nabla_{(\a_1}\z_{\a_2 ... \a_n)}~,\
\eea
where the gauge parameter $\z_{\a(n-1)}$ is a real unconstrained superfield, and ${\lfloor x \rfloor}$ stands for  the floor function denoting the integer part of a real number $x \geq 0$.


\subsection{Massless half-integer superspin theories } 
There exist two off-shell formulations for the massless superspin-$(s+\frac{1}{2})$ multiplet, which are referred to as longitudinal and transverse. 


\subsubsection{Longitudinal formulation}

The longitudinal formulation is realised in terms of the real unconstrained dynamical variables
\be \label{MLHI}
\cV^\parallel_{(s+\frac{1}{2})} = \big \{ H_{\a(2s+1)} , L_{\a(2s-2)} \big \}~,
\ee
which are defined modulo gauge transformations of the form
\begin{subequations} \label{LFHIGT}
	\bea
	\d H_{\a(2s+1)} &=& \ri \nabla_{(\a_1} \z_{\a_2 ... \a_{2s+1})}~, \\
	\d L_{\a(2s-2)} &=& - \frac{s}{2(2s+1)}\nabla^{\b\g}\z_{\b\g\a(2s-2)}~,
	\eea
\end{subequations}
where the real gauge parameter $\z_{\a(2s)}$ is unconstrained. 
The unique gauge-invariant action 
formulated 
in terms of the superfields \eqref{MLHI} takes the following form
\bea \label{LongHalfIntAct}
&&{S^{\parallel}_{(s+\hf)}[H_{\a(2s+1)} ,L_{\a(2s-2)} ] 
	= \Big(-\hf \Big)^{s} 
	\int \rd^{3|2}z
	\, E\, \bigg\{-\frac{\ri}{2} H^{\a(2s+1)} {\mathbb{Q}} H_{\a(2s+1)} }
\non \\
&& -\frac{\ri}{8} \nabla_{\b} H^{\b \a(2s)} \nabla^2 \nabla^{\g}H_{\g \a(2s)}+\frac{\ri s}{4}{\nabla}_{\b \g}H^{\b \g \a(2s-1)} {\nabla}^{\rho \d}H_{\rho \d \a(2s-1)}
\non \\
&&  + (2s-1) L^{\a(2s-2)}\nabla^{\b \g} \nabla^{\d} H_{\b \g \d \a(2s-2)}
\non \\
&&  + 2 (2s-1)\Big( L^{\a(2s-2)} (\ri \nabla^2 - 4 |\m|) L_{\a(2s-2)}
- \frac{\ri}{s}(s-1) \nabla_{\b} L^{\b \a(2s-3)} \nabla^{\g}L_{\g \a(2s-3)}\Big)
\non \\
&&  + |\m| \Big(s \,\de_{\b}H^{\b \a(2s)} \de^{\g} H_{\g \a(2s)}+ \hf (2s+1)H^{\a(2s+1)}( \nabla^2-4 \ri |\m|)H_{\a(2s+1)} \Big)
\bigg\}~,
\eea
where $\mathbb{Q}$ is the quadratic Casimir operator 
$\eqref{QuadraticCasimir}$.
The action \eqref{LongHalfIntAct} was derived
in \cite{KP}. In the flat superspace limit, $|\m|\rightarrow 0$,  the action \eqref{LongHalfIntAct} coincides with the model derived in \cite{KT}.


\subsubsection{Transverse formulation}

The transverse formulation is constructed in terms of the real superfields
\be \label{THI}
\cV^\perp_{(s+\frac{1}{2})} = \big \{ H_{\a(2s+1)}, \U_{\b;\a(2s-2)} \big \}~, 
\ee
where $\U_{\b;\a(2s-2)}$ is a reducible superfield pertaining to the representation $\bm{ 2 \otimes (2s-1)}$ of $\text{SL}(2, \mathbb{R})$. The superfield $\U_{\b;\a(2s-2)}$ can be decomposed into irreducible components by the following rule
\be
\U_{\b;\a(2s-2)} = \U_{(\b;\a_1 ... \a_{2s-2})} + \frac{1}{2s-1} \sum_{k=1}^{2s-2}\varepsilon_{\b\a_k}\U^{\g;}{}_{\g\a_1 ... \hat{\a}_k ... \a_{2s-2}}~.
\ee
The dynamical variables \eqref{THI}  possess the following gauge freedom
\begin{subequations}  \label{GTTHI}
	\bea
	\d H_{\a(2s+1)} &=& \ri \nabla_{(\a_1}\z_{\a_2 ... \a_{2s+1})}~, \\
	\d \U_{\b ; \a(2s-2)} &=& \frac{\ri}{2s+1}\Big (\nabla^\g \z_{\g\b\a(2s-2)} + (2s+1)\nabla_\b \eta_{\a(2s-2)} \Big )~,
	\eea	
\end{subequations}	
where the gauge parameters $\z_{\a(2s)}$ and $\eta_{\a(2s-2)}$ are real unconstrained. The unique quadratic action which is invariant under gauge transformations \eqref{GTTHI} takes the form
\bea \label{TransHIAct}
&&S^{\perp}_{(s+\hf)}[{H}_{\a(2s+1)} ,\U_{\b; \,\a(2s-2)} ]
= \Big(-\hf \Big)^{s} 
\int \rd^{3|2}z \, E \,\bigg\{-\frac{\ri}{2} H^{\a(2s+1)} {\mathbb{Q}} H_{\a(2s+1)}
\non \\
&&   -\frac{\ri}{8} \nabla_{\b} H^{\b \a(2s)} \nabla^2 \nabla^{\g}H_{\g \a(2s)}+\frac{\ri}{8}{\nabla}_{\b \g}H^{\b \g \a(2s-1)} {\nabla}^{\rho \d}H_{\rho \d \a(2s-1)}
\non \\
&&   -\frac{\ri}{4}(2s-1) \O^{\b; \,\a(2s-2)} \nabla^{\g \d}H_{\g \d \b \a(2s-2)}  + \ri s (2s-1) |\m| \,\U^{\b ;\, \a(2s-2)} \O_{\b ; \, \a(2s-2)}
\non \\
&&   -\frac{\ri}{8}(2s-1)\Big(\O^{\b ;\, \a(2s-2)} \O_{\b ;\, \a(2s-2)}
-2(s-1)\O_{\b;}\,^{\b \a(2s-3)} \O^{\g ;}\,_{\g \a(2s-3)}  \Big) 
\non \\
&&   
+ |\m| \Big( H^{\a(2s+1)} \big( \nabla^2 - 4\ri |\m|\big) H_{\a(2s+1)} 
+ \hf \ \de_{\b}H^{\b \a(2s)} \de^{\g}H_{\g \a(2s)}
\Big) \bigg\}~,
\eea
where $\O_{\b ; \a(2s-2)}$ corresponds to the real $\cN=1$ field strength
\bea
\O_{\b;\, \a(2s-2)}
= -\ri \big( \nabla^{\g} \nabla_{\b} - 4\ri |\m| \delta_{\b}\,^{\g}\big){\U}_{\g; \,\a(2s-2)} ~,
\qquad \nabla^\b  \O_{\b;\, \a(2s-2)}=0~.
\eea
The above theory was introduced in \cite{HK19}.


\subsection{Massless integer superspin theories } 
Analogous to the half-integer case, there exist two off-shell formulations for the massless superspin-$s$ multiplet, which are referred to as longitudinal and transverse. 


\subsubsection{Longitudinal formulation}

The longitudinal formulation is realised in terms of the real unconstrained variables
\be
\cV^\parallel_{(s)} = \big \{ H_{\a(2s)}, V_{\a(2s-2)} \big  \}~,
\ee
which are defined modulo gauge transformations of the form
\begin{subequations} \label{LongGTI}
	\bea 
	\d H_{\a(2s)} &=& \nabla_{(\a_1}\z_{\a_2 ... \a_{2s})}~, \\
	\d V_{\a(2s-2)} &=& \frac{1}{2s}\nabla^\b \z_{\b\a(2s-2)}~,
	\eea
\end{subequations}
where the gauge parameter $\z_{\a(2s-1)}$ is real unconstrained. The unique action which is invariant under gauge transformations \eqref{LongGTI} assumes the form
\bea\label{action-t3}
\lefteqn{S^{\parallel}_{(s)}[H_{\a(2s)} ,V_{\a(2s-2)} ]
	= \Big(-\hf \Big)^{s} 
	\int 
	\rd^{3|2}z
	\, E \, \bigg\{
	\frac{1}{2} H^{\a(2s)} \big(\ri \de^2  +4 |\m|\big)H_{\a(2s)} }
\non \\
&&  
- \frac{\ri}{2}\de_{\b}H^{\b \a(2s-1)} \de^{\g}H_{\g \a(2s-1)}
-(2s-1) V^{\a(2s-2)} \nabla^{\b \g} H_{\b \g \a(2s-2)}
\\
&&  +(2s-1)\Big(\hf V^{\a(2s-2)} \big( \ri \nabla^2 +8s|\m|\big)V_{\a(2s-2)}
+ \ri (s-1) \nabla_{\b}V^{\b \a(2s-3)} \nabla^{\g}V_{\g \a(2s-3)} \Big) \bigg\}
~.
\non
\eea
Up to normalisation, action \eqref{action-t3} coincides with the off-shell $\cN=1$ supersymmetric action for the massless superspin-$s$ multiplet derived in \cite{KP}. Taking the flat-superspace limit, action \eqref{action-t3} reduces to the model derived in \cite{KT}.


\subsubsection{Transverse formulation}

The transverse formulation is constructed in terms of the real unconstrained superfields
\be
\cV^{\perp}_{(s)} = \big \{ H_{\a(2s)}, \J_{\b ; \a(2s-2)} \big \}~,
\ee
which have the following gauge freedom
\begin{subequations} \label{TIGT}
	\bea
	\d H_{\a(2s)} &=& \nabla_{(\a_1} \z_{\a_2 ... \a_{2s})}~, \\
	\label{psigt}
	\d \J_{\b ; \a(2s-2)} &=& - \z_{\b\a(2s-2)} + \ri \nabla_\b \eta_{\a(2s-2)}~,
	\eea
\end{subequations}
where the gauge parameters $\z_{\a(2s-1)}$ and $\eta_{\a(2s-2)}$ are real unconstrained. The gauge-invariant action is given by 
\bea \label{TIAction}
&&S^{\perp}_{(s)}[H_{\a(2s)} ,{\Psi}_{\b; \,\a(2s-2)} ]
= \Big(-\hf \Big)^{s} 
\int \rd^{3|2}z \,
E\, \bigg\{\frac{1}{2} H^{\a(2s)} (\ri \nabla^2 +8 s |\m|) H_{\a(2s)}
\non \\
&&  - \ri s \nabla_{\b} H^{\b \a(2s-1)} \nabla^{\g}H_{\g \a(2s-1)} 
-(2s-1) \cW^{\b ;\,\a(2s-2)} \nabla^{\g} H_{\g \b \a(2s-2)}
\non \\
&&   -\frac{\ri}{2} (2s-1)\Big(\cW^{\b ;\, \a(2s-2)} \cW_{\b ;\, \a(2s-2)}+\frac{s-1}{s} \cW_{\b;}\,^{\b \a(2s-3)} \cW^{\g ;}\,_{\g \a(2s-3)} \Big) 
\non\\
&&  
-2 \ri (2s-1) |\m| \Psi^{\b ;\, \a(2s-2)} \cW_{\b ; \, \a(2s-2)}
\bigg\}~,
\eea
where $\cW_{\b; \, \a(2s-2)}$ denotes the real  $\cN=1$ field strength
\bea
\cW_{\b;\, \a(2s-2)} =
-\ri \big( \nabla^{\g} \nabla_{\b} - 4\ri |\m| \delta_\b{}^\g \big){\Psi}_{\g; \,\a(2s-2)} ~,
\qquad \nabla^\b  \cW_{\b;\, \a(2s-2)}=0~.
\eea

For $s > 1$, the $\z$-gauge freedom \eqref{psigt}
can be used to impose the gauge condition
\bea
\J_{(\a_1;\, \a_2 \dots \a_{2s-1})} =0 \quad \Longleftrightarrow \quad 
\J_{\b ;\, \a(2s-2)} =  \sum_{k=1}^{2s-2}\ve_{\b \a_k} \vf_{\a_1 \dots \hat{\a}_k \dots \a_{2s-2}}~,
\label{3.355}
\eea
for some superfield $\varphi_{\a(2s-3)}$. 
The residual gauge freedom is characterised by 
\bea
 \z_{\a(2s-1)} = \ri \nabla_{(\a_1}  \eta_{\a_2 \dots \a_{2s-1})}~,
 \eea
 which means that $\eta_{\a(2s-2)}$ is the only independent gauge parameter. As a consequence, the model can be reformulated in terms of the following gauge superfields
\bea
\big\{H_{\a(2s)},~\vf_{\a(2s-3)}\big\}~,
\eea
which are defined modulo gauge transformations of the form
\bea
\d H_{\a(2s)} &=& -\nabla_{(\a_1 \a_2} \eta_{\a_3 \dots \a_{2s})}~,\\
\d \vf_{\a(2s-3)}&=& \ri \nabla^{\b} \eta_{\b \a(2s-3)}~.
\eea
The corresponding gauge-invariant action assumes the following form
\bea \label{GaugedTransIntAction}
&&S^{\perp}_{(s)}[H_{\a(2s)}, \varphi_{\a(2s-3)}] =  \Big ( - \frac{1}{2} \Big )^s   \int \rd^{3|2}z~E~ \bigg \{ \frac{1}{2}H^{\a(2s)}(\ri \nabla^2 +8s|\m|)H_{\a(2s)}~\non \\
&&-\ri s \nabla_\b H^{\b\a(2s-1)}\nabla^\g H_{\g\a(2s-1)} -2(s-1) \varphi^{\a(2s-3)}\nabla^\b\nabla^{\g\d}H_{\b\g\d\a(2s-3)} ~ \non \\
&&+ \frac{1}{s}(s-1) \Big ( 8 \ri (2s-1)|\m|^2\varphi^{\a(2s-3)}\varphi_{\a(2s-3)}+2(2s-3)|\m|\varphi^{\a(2s-3)}\nabla^2\varphi_{\a(2s-3)} \non \\
&&-2\ri \varphi^{\a(2s-3)}\mathbb{Q}\varphi_{\a(2s-3)} + \frac{\ri}{2(2s-1)}(2s-3)\nabla_\b \varphi^{\b\a(2s-4)}\nabla^2\nabla^\g \varphi_{\g\a(2s-4)} ~ \non \\
&&+ \frac{4}{2s-1}(2s-3)(3s-2)|\m|\nabla_\b \varphi^{\b \a(2s-4)}\nabla^\g \varphi_{\g \a(2s-4)}~ \non \\
&&-\frac{\ri}{2s-1}(2s-3)(s-2)\nabla_{\b\g}\varphi^{\b\g\a(2s-5)}\nabla^{\r \s}\varphi_{\r\s\a(2s-5)} \Big ) \bigg \}~.
\eea
The above theory was proposed in \cite{HK19}.
In the flat-superspace limit, the action \eqref{GaugedTransIntAction} coincides with the $\cN=1$ model (B.25) presented in \cite{HKO}.


\section{(1,1) $\rightarrow$ (1,0) AdS superspace reduction} \label{Section4}

In this section, we begin by reviewing the necessary aspects of $(1,1)$ AdS superspace which are then utilised to develop a consistent reduction procedure for field theories with (1,1) AdS supersymmetry to ${\cal N} = 1$ AdS superspace.


\subsection{(1,1) AdS superspace: Complex basis} \label{ss21}
We summarise the key results concerning (1,1) AdS superspace \cite{KT-M11, HKO} and superfield representations of the associated isometry group ${\rm OSp(1|2; {\mathbb{R}})} \times {\rm OSp(1|2; {\mathbb{R}})}$, following the presentation in \cite{HKO}. The geometry of (1,1) AdS
superspace, ${\rm AdS}^{(3|1,1)} $, can be described using either a real or complex basis for the spinor covariant derivatives. In this subsection we consider the formulation in the complex basis.

The covariant derivatives of ${\rm AdS}^{(3|1,1)} $ have the form
\bea
{\cD}_{\cA}=(\cD_{a}, \cD_{\a},\bar {\cD}^{\a})
=E_{\cA} + \O_{{\cA}}~,
\eea
where $z^{\cM} = (x^m, \theta^{\m}, \bar \theta_{\mu})$ are local complex superspace coordinates, while $E_{\cA}$ and $\O_{\cA}$ denote the inverse supervielbein and Lorentz connection,
\bea
E_{\cA} = E_{\cA}{}^{\cM}\frac{\pa}{\pa z^{\cM}}~, \qquad \O_{\cA} = \hf \O_{\cA}{}^{bc}M_{bc} = \hf \O_{\cA}{}^{\b \g}M_{\b \g}~.
\eea

The covariant derivatives satisfy the following algebra
\begin{subequations}  \label{1.1}
\bea
&& \qquad \{ \cD_\a , \bar \cD_\b \} = -2\rm i \cD_{\a \b} ~, \\
&& \qquad \{\cD_\a, \cD_\b \} = -4\bar \m\, M_{\a \b}~, \qquad
\{ {\bar \cD}_\a, {\bar \cD}_\b \} = 4\m\,M_{\a \b}~, \\
&& \qquad [ \cD_{ \a \b }, \cD_\g ] = -2 \rm i \bar \m\,\ve_{\g (\a} \bar \cD_{\b)}~,  \qquad
\,\,[\cD_{ \a \b }, { \bar \cD}_{\g} ] = 2 \rm i \m\,\ve_{\g (\a} \cD_{\b)}~,   \\
&&\quad \,\,\,\,[ \cD_{\a \b} , \cD_{ \g \d } ] = 4 \bar \m \m \Big(\ve_{\g (\a} M_{\b) \d}+ \ve_{\d (\a} M_{\b) \g}\Big)~,  
\eea
\end{subequations} 
with $\m\neq 0$ being a complex parameter determining the constant curvature of 
${\rm AdS}^{(3|1,1)} $. The phase of $\m = |\m | \re^{\ri \vf} $ can be given any fixed value by a re-definition $\cD_\a \to \re^{\ri \r} \cD_\a$ and $\bar \cD_\a \to \re^{-\ri \r} \bar \cD_\a$, with $\r$ constant.

Let ${G}_{\a(n)}$ be a symmetric rank-$n$ spinor superfield. It is said to be longitudinal linear if it obeys the following first-order constraint
\begin{subequations}
\bea
 \bar \cD_{(\a_1} {G}_{\a_2 \dots \a_{n+1} )} = 0  ~, \label{LL}
 \eea
 which implies 
 \bea
 \big(\bar \cD^2+2n\m \big){G}_{\a(n)} &=& 0~. \label{LL-sq}
 \eea
 \end{subequations}
 In the scalar case, $n=0$, the constraint \eqref{LL} becomes the condition of covariant chirality, $\bar \cD_\a {G}=0$.

A symmetric rank-$n$ spinor superfield 
${\G}_{\a(n)}$ is called transverse linear if it obeys the  first-order constraint
\begin{subequations}
 \bea
 \bar \cD^\b {\G}_{ \b \a_1 \dots \a_{n - 1} } = 0 ~,  \qquad n \neq 0~,   
 \label{TL}
 \eea
 which implies 
 \bea
 \big(\bar \cD^2-2(n+2)\m\big){\G}_{\a(n)} = 0~. \label{TL-sq}
 \eea
 \end{subequations}
In the $n=0$ case, the constraint \eqref{TL} is not defined.
However its corollary \eqref{TL-sq} is perfectly consistent, 
\bea
 \big(\bar \cD^2- 4\m\big){\G} = 0~, 
 \eea
 and defines a covariantly  linear scalar superfield ${\G}$.

The constraints \eqref{LL} and \eqref{TL} may be solved in terms of 
prepotentials ${\J}_{\a(n-1)}$ and ${\F}_{\a(n+1)}$ as follows:
\begin{subequations}
\bea
 {G}_{\a(n)}&=& \bar \cD_{(\a_1}
{\J}_{ \a_2 \dots \a_{n}) } ~, 
\label{2.19a}\\
{\G}_{\a(n)}&=& \bar \cD^\b
{\F}_{(\b \a_1 \dots \a_n )} ~.
\label{2.19b}
\eea
\end{subequations}
Provided the constraints \eqref{LL} and \eqref{TL} are the only conditions
imposed on ${G}_{\a(n)}$ and ${\G}_{\a(n)}$ respectively, 
the prepotentials ${\J}_{\a(n-1)}$ and ${\F}_{\a(n+1)}$ can be chosen 
to be  unconstrained complex. They are defined modulo
gauge transformations of the form
\begin{subequations} 
\bea
\d_\z {\J}_{\a(n-1)} 
&=&  \bar \cD_{(\a_1 }
{\z}_{\a_2 \dots \a_{n-1})} ~, \label{2.20aa}\\
\d_\x {\F}_{\a(n+1)} &=&  \bar \cD^\g
{\x}_{(\g \a_1 \dots \a_{n+1})} ~,
\eea
\end{subequations}
with the gauge parameters ${\z}_{\a(n-2)}$ and ${\x}_{\a(n+2)}$
being complex unconstrained. 

One can define projectors $P^{\perp}_{n}$ and $P^{||}_{n}$
on the spaces of  transverse linear and longitudinal linear superfields, respectively.
The projectors are 
\begin{subequations}
\bea
P^{\perp}_{n}&=& \frac{1}{4 (n+1)\m} (\bar \cD^2+2n\m) ~,\\
P^{||}_{n}&=&- \frac{1}{4 (n+1)\m} (\bar \cD^2-2(n+2)\m ) ~,
\eea
\end{subequations} 
with the properties 
\bea
\big(P^{\perp}_{n}\big)^2 =P^{\perp}_{n} ~, \quad 
\big(P^{||}_{n}\big)^2=P^{||}_{n}~,
\quad P^{\perp}_{n} P^{||}_{n}=P^{||}_{n}P^{\perp}_{n}=0~.
\eea
These projectors are three-dimensional (3D) cousins of those introduced by Ivanov and Sorin \cite{IS} 
in the case of $\cN=1$ AdS supersymmetry in four dimensions. 

Let  ${V}_{\a(n)} $ ($n \neq 0$) be an arbitrary complex tensor superfield. It can be represented
as a sum of transverse linear and longitudinal linear multiplets \cite{HKO}
\bea
{V}_{ \a(n)} = &-& 
\frac{1}{2 \mu (n+2)} \cDB^\g \cDB_{(\g} {V}_{ \a_1 \dots  \a_n)} 
- \frac{1}{2 \mu (n+1)} \cDB_{(\a_1} \cDB^{|\g|} {V}_{ \a_2 \dots \a_{n} ) \g} 
~ ,~~~
\eea
where indices placed between vertical bars (for example $|\g|$) are not subject to symmetrisation.
If we choose ${V}_{ \a(n)} $ to be  
either longitudinal linear (${G}_{ \a(n)} $)
or transverse linear (${\G}_{ \a(n)} $), then the above identity produces the relations \eqref{2.19a} and \eqref{2.19b}
for some prepotentials ${\J}_{\a(n-1)}$ and ${\F}_{\a(n+1)} $, respectively.

The isometry transformations of ${\rm AdS}^{(3|1,1)}$ are generated by real supervector fields ${\l}^{\cA} E_{\cA} $ which solve 
the Killing equation 
\bea
\Big{[}{\L}+\hf l^{ab}M_{ab},\cD_{\cC}\Big{]}=0~,
\eea
where 
\bea \label{KillingSVC}
\L= \l^{\cA} \cD_{\cA} =\l^a\cD_a+\l^\a\cD_\a+\bar \l_\a\bar \cD^\a~,
\qquad\bar{\l}^a=\l^a~,
\eea 
and $l^{ab}$ is some local Lorentz parameter.
As shown in \cite{KT-M11},
this equation implies that the parameters $\l^\a$ and $l^{ab}$ 
can be uniquely determined in terms of the vector  $\l^a$,
\bea
\l_\a =\frac{\ri}{6} \bar \cD^\b \l_{\a\b}~,\qquad
l_{\a\b} =2\cD_{(\a}\l_{\b)}~,
\label{Killing-1}
\eea
and the vector parameter obeys the equation
\bea
\cD_{(\a}\l_{\b\g)}=0 \quad \Longleftrightarrow \quad 
\bar\cD_{(\a}\l_{\b\g)}=0~.
\eea
In comparison with the 3D $\cN=2$ Minkowski superspace,
the specific feature of ${\rm AdS}^{(3|1,1)} $ is that any two of the three parameters 
 $\{ \l_{\ab}, \l_\a, l_{\a\b}\}$ can be expressed in terms of the third parameter. In particular,
\bea
\l_{\a\b} =\frac{\ri}{\m} \bar \cD_{(\a}\l_{\b)}~,\qquad 
\l_\a =
\frac{1}{12 \bar \m} \cD^\b   l_{\a\b}
~.
\label{2.7}
\eea
From \eqref{Killing-1} and \eqref{2.7} we deduce
\bea
\bar \cD^\a\l_\a=
\cD_\a \l^\a=0~.
\label{1,1-SK_1}
\eea
The solution to these equations is given in \cite{KT-M11}.


\subsection{(1,1) AdS superspace: Real basis}

It proves beneficial to realise the (1,1) AdS covariant derivatives in a real basis when performing the reduction procedure to $\cN=1$ AdS superspace. In accordance with \cite{KLT-M12}, the algebra of covariant derivative \eqref{1.1} can be converted to the real basis by (i) making a convenient choice  $\m =-\ri |\m|$;  and (ii) 
replacing the complex operators $\cD_\a , \bar \cD_\a$ with ${\bm \de}^I_{\a} = ({\bm \de}^{\1}_{\a} , {\bm \de}^{\2}_{\a} )$ defined by
\bea \label{realrep}
&& \qquad \cD_\a = \frac{1}{\sqrt{2}}({\bm \de}^{\1}_\a - \ri{\bm \de}^{\2}_\alpha) ~, 
\qquad \bar{\cD}_\a = - \frac{1}{\sqrt{2}}({\bm \de}^{\1}_\a + \ri{\bm \de}^{\2}_\alpha) ~.
\eea
Furthermore, we introduce the real coordinates $z^{\cM}= (x^m, \theta^{\mu}_I)$ which are used to parametrise $(1,1)$ AdS superspace. Choosing to define ${\bm \de_a} = \cD_a$, it can be shown that the algebra of $(1,1)$ AdS covariant derivatives assumes the following form in the real basis \eqref{realrep}
\begin{subequations}  \label{1.3}
\begin{alignat}{3}
\qquad \{{\bm \de}^{\1}_\a	, {\bm \de}^{\2}_\b \} &= 0 ~,& \\
\qquad \{ {\bm \de}^{\1}_\a , {\bm \de}^{\1}_\b \} &= 2 \ri {\bm \de}_{\a \b} - 4 \ri |\m| M_{\a \b} ~,&
 \qquad \{ {\bm \de}^{\2}_\a , {\bm \de}^{\2}_\b \} &= 2 \ri {\bm \de}_{\a \b} + 4 \ri |\m| M_{\a \b} ~,\\
\qquad [ {\bm \de}_{a} , {\bm \de}^{\1}_\a ] &= |\m| (\g_a)_\a{}^\b{\bm \de}^{\1}_\b ~,&
 \qquad [ {\bm \de}_{a} , {\bm \de}^{\2}_\a] &= - |\m| (\g_a)_\a{}^\b {\bm \de}^{\2}_\b ~, \\
\qquad [{\bm \de}_{a}, {\bm \de}_{b}] &= -4|\m|^2 M_{ab}~.&
\end{alignat}
\end{subequations} 
It is apparent from \eqref{1.3} that 
the operators ${\bm \de}_{a}$ and ${\bm \de}_{\a}^{\1}$ possess the following properties:
\begin{enumerate}
\item These operators form a closed algebra given by 
\begin{subequations} \label{1.4}
\be
\{ {\bm \de}_{\a}^{\1} , {\bm \de}^{\1}_\b \} = 2 \ri {\bm \de}_{\a\b} - 4 \ri |\m| M_{\a \b} ~,
\ee
\be
\quad [ {\bm \de}_{a} , {\bm \de}_{\b}^{\1} ] = |\m| (\g_a)_\b{}^\g {\bm \de}^{\1}_\g ~,
\quad [{\bm \de}_a, {\bm \de}_b] = -4|\m|^2 M_{ab}~.
\ee
\end{subequations}
\item Relations \eqref{1.4} coincide with the algebra of covariant derivatives of AdS$^{3|2}$, eq.\,\eqref{algN1}.
\end{enumerate}
These properties imply that AdS$^{3|2}$ is naturally realised as a surface embedded in $(1,1)$ AdS superspace. One can make an appropriate choice in the real Grassmann variables $\q^\m_I = (\q^\m_\1,\q^\m_\2)$ such that AdS$^{3|2}$ can be identified as the surface defined by $\q^\m_\2=0$ in ${\rm AdS}^{(3|1,1)} $. These properties enable the consistent reduction of any field theory with (1,1) AdS supersymmetry to ${\cal N} = 1$ AdS superspace.

We now wish to recast the fundamental properties of the Killing supervector fields of (1,1) AdS superspace \eqref{KillingSVC} in the real representation \eqref{realrep}. The isometries of (1,1) AdS superspace are generated by the (1,1) AdS Killing supervector fields,
\be \label{11KillingVector}
\L : = \l^\cA\bm \nabla_\cA = \l^a \bm \nabla_a + \l^\a_I \bm \nabla^I_\a~, \qquad I= \1, \2~,
\ee
which are defined to satisfy the Killing equation
\be \label{Kconstraint}
[\L + \frac{1}{2}l^{ab}M_{ab}, \bm \nabla_\cA] = 0~,
\ee
for some real Lorentz parameter $l^{ab} = - l^{ba}$. It can be shown that equation \eqref{Kconstraint} is equivalent to the set of equations
\begin{subequations} \label{KillingEqnSD}
\bea 
\bm \nabla^\1_\a \l^\1_\b &=& \frac{1}{2}l_{\a\b}+|\m|\l_{\a\b}~, \qquad \bm \nabla^\1_\a \l^\2_\b = 0 ~,\\
\bm \nabla^\2_\a \l^\2_\b &=& \frac{1}{2}l_{\a\b}-|\m|\l_{\a\b} ~,  \qquad \bm \nabla^\2_\a \l^\1_\b = 0 ~,\\
\bm \nabla^\1_\a   l_{\b\g} &=& 8 \ri |\m| \varepsilon_{\a(\b}\l^\1{}_{\g)}~, \qquad \bm \nabla^\2_\a   l_{\b\g} = -8 \ri |\m| \varepsilon_{\a(\b}\l^\2{}_{\g)}~, \\
\bm \nabla^I_\a \l_a &=& 2\ri (\g_a)_{\a\b} \l^{I\b}~,
\eea
\end{subequations}
and
\begin{subequations} \label{KillingEqnVD}
\bea \label{2.26a}
\bm \nabla_a \l_b &=&l_{ab}~, \\
\bm \nabla_a \l^\1_\a &=& |\m|(\g_a)_\a{}^\b \l ^\1_\b~, \qquad \bm \nabla_a \l^\2_\a = - |\m|(\g_a)_\a{}^\b \l ^\2_\b~, \label{SKeqn}\\
\bm \nabla_a l_{bc} &=& 8 |\m|^2 \eta_{a[b} \l_{c]}~.
\eea
\end{subequations}
Equations \eqref{KillingEqnSD} and \eqref{KillingEqnVD} can be recast in the equivalent form
\begin{subequations} \label{K-equiv}
\bea
\bm \nabla^I_{(\a} \l_{\b\g)} &=& 0~, \qquad \bm \nabla^I_{(\a} l_{\b\g)} = 0~, \\
\bm \nabla^\1_{(\a} \l^\1_{\b)} &=& \frac{1}{2}l_{\a\b} + |\m|\l_{\a\b}~, \qquad \bm \nabla^\2_{(\a} \l^\2_{\b)} = \frac{1}{2}l_{\a\b} - |\m|\l_{\a\b}      ~,\\
\bm \nabla^\1_{(\a} \l^\2_{\b)} &=& \bm \nabla^\2_{(\a} \l^\1_{\b)} = 0, \qquad \bm \nabla^{\a(I} \l^{J)}_\a = 0~, \\
\l^{\1 \a} &=& \frac{\ri}{6} \bm \nabla^\1_\b \l^{\a\b} = \frac{\ri}{12|\m|} \bm \nabla^\1_\b l^{\a\b}~, \label{2.27d} \\
\l^{\2 \a} &=& \frac{\ri}{6} \bm \nabla^\2_\b \l^{\a\b} = - \frac{\ri}{12|\m|} \bm \nabla^\2_\b l^{\a\b}~. \label{2.27e}
\eea
\end{subequations}
It follows from \eqref{2.26a} that the parameter $\l_a$ is a Killing vector field
\be
\bm \nabla_a \l_b + \bm \nabla_b \l_a = 0~,
\ee
and relations \eqref{SKeqn} are Killing spinor equations.

\subsection{Reduction from (1,1) to $\cN=1$ AdS superspace}
Given a tensor superfield $U(x,\q_I)$ on (1,1) AdS superspace, where indices have been suppressed, we define its bar-projection to $\cN=1$ AdS superspace by the rule
\be
U|:= U(x,\q_I)|_{\q_\2=0}
\ee
in a \textit{special coordinate system} which will be described below. Given the $(1,1)$ covariant derivative in the the real representation \eqref{realrep}
\be
\bm \nabla_\cA = (\bm \nabla_a, \bm \nabla^I_\a) = E_{\cA}{}^{\cM}\frac{\pa}{\pa z^{\cM}} + \hf \O_{\cA}{}^{bc}M_{bc} ~,
\ee
we define its $\cN=1$ projection by the rule
\be
\bm \nabla_\cA| = E_{\cA}{}^{\cM}|\frac{\pa}{\pa z^{\cM}} + \hf \O_{\cA}{}^{bc}|M_{bc} ~.
\ee
We use the freedom to perform general coordinate and local Lorentz transformations to impose the gauge
\be \label{reductiongauge}
\bm \nabla_a | = \nabla_a~, \qquad \bm \nabla^\1_\a| = \nabla_\a ~,
\ee
where $\nabla_A = (\nabla_a, \nabla_\a)$ is the set of covariant derivatives for AdS$^{3|2}$, see eq.\,\eqref{N10}. 
In the chosen coordinate system, the operator $\bm \nabla^\1_\a$ does not involve any partial derivatives with respect to $\q_\2$. Thus for any positive integer $k$, it follows that $(\bm \nabla^\1_{\a_1}...\bm \nabla^\1_{\a_k} U)| = \nabla_{\a_1}... \nabla_{\a_k} U|$. 

We now consider the $\cN=1$  projection of the $(1,1)$ AdS Killing supervector \eqref{11KillingVector}
\be
\L | = \x^a \nabla_a + \x^\a \nabla_\a + \e^\a \bm \nabla^\2_\a |~,
\ee
where we have introduced the $\cN=1$ superfields
\be
\x^a := \l^a |~, \qquad \x^\a := \l^\a_\1|~, \qquad \e^\a : = \l^\a_\2|~.
\ee
Additionally, we define the $\cN=1$ projection of the Lorentz parameter $l^{ab}$ to be
\be
\z^{ab} := l^{ab}|~.
\ee
It is important to note that the superfields $(\x^a,\x^\a, \z^{ab})$ parametrise the infinitesimal isometries of AdS$^{3|2}$. Such transformations are generated by
the Killing supervector fields, $\x = \x^a\nabla_a + \x^\a \nabla_\a$, satisfying the ${\cN}=1$ Killing equation, eq.~\eqref{Kcondition1-1}. Indeed, the relations \eqref{2.40d-1} and \eqref{2.40e} automatically follow from the $(1,1)$ AdS Killing equations \eqref{K-equiv}, upon projection.
The parameter $\e_\a$, which generates the second supersymmetry transformation, has the property
\bea
\nabla_a\e_\a &=& -|\m|(\g_a)_\a{}^\b \e_\b~. \label{2.40d}
\eea

Given the transformation law of a tensor superfield $U(x,\q_I)$ on $(1,1)$ AdS superspace
\be
\d_\L U= \big ( \L + \frac{1}{2}l^{ab}M_{ab} \big ) U~,
\ee
we find its projection to $\cN=1$ AdS superspace to be
\be \label{PTL}
\d_\L U| = \d_\x U| + \d_\e U| ~,
\ee
where 
\begin{subequations} \label{PTL2}
\bea \label{PTL2A}
 \d_\x U| &=& \big (  \x^a \nabla_a + \x^\a \nabla_\a + \hf \z^{ab}M_{ab} \big ) U| ~, \\
 \d_\e U| &=&  \e^\a \big (\bm \nabla^\2_\a U \big ) |~. \label{PTL2B}
\eea
\end{subequations}
The first transformation \eqref{PTL2A} coincides with the infinitesimal transformation generated by a Killing supervector in AdS$^{3|2}$, eq.~\eqref{211}. Thus, $U|$ can be identified as a tensor superfield on $\cN=1$ AdS superspace. The other transformation \eqref{PTL2B} corresponds to the second supersymmetry transformation, which is generated by $\e^\a$.


\subsection{The (1,1) AdS supersymmetric actions in AdS$^{3|2}$} \label{Subsect44}

Every supersymmetric field theory in $(1,1)$ AdS superspace can be reduced to $\cN=1$ AdS superspace. In the following subsection, we explore the necessary mathematical framework which will be employed to develop such a reduction procedure. 
As presented in \cite{KLT-M11,KT-M11,KLRST-M,BKT-M}, manifestly supersymmetric actions in (1,1) AdS superspace can be constructed by either
\begin{enumerate}
	\item Integrating a real scalar Lagrangian $\cL$ over the full $(1,1)$ AdS superspace,
	\bea \label{real11}
&&\int \rd^3 x \, \rd^2 \q\, \rd^2 \bar \q \, \bm E\,\cL = \frac{1}{16}\int \rd^3 x\,e\, (\cD^2- 16 \bar \m)(\bar \cD^2 - 4 \mu) \cL \Big |_{\q=0} \non\\
&&= \int \rd^3 x\,e\,\Big ( \frac{1}{16}\cD^\a (\bar{\cD}^2 - 6 \m)\cD_\a - \frac{\m}{4}\cD^2 - \frac{\bar{ \m}}{4}\bar{ \cD}^2+4\m \bar{\m} \Big )  \cL \Big |_{\q=0}~,
	\eea
with $\bm E^{-1} ={\rm Ber}(E_{\cA}{}^{\cM})$.  
\item Integrating a covariantly chiral Lagrangian $\cL_c$ over the chiral subspace,
\be \label{CI}
\int \rd^3 x\,\rd^2 \q \,\cE\,\cL_c  = - \frac{1}{4}\int \rd^3 x\,e\,(\cD^2 - 16\bar{ \m})\cL_c\Big |_{\q=0}~, \qquad \bar{ \cD}_{\a} \cL_c = 0~,
\ee	
where $\cE$ is the chiral density.
The two supersymmetric invariants are related by the rule
\bea
\int \rd^3 x \, \rd^2 \q\, \rd^2 \bar \q \, \bm E\,\cL = \int \rd^3 x\,\rd^2 \q \,\cE\,\cL_c~, \qquad \cL_c := -\frac{1}{4} (\bar \cD^2 - 4 \mu) \cL~. 
\eea
\end{enumerate}
In $(1,1)$ AdS superspace, every chiral action can always be recast as an integral over the full superspace 
\be \label{CFRelation}
\int \rd^3 x\,\rd^2 \q \,\cE\,\cL_c = \frac{1}{\mu} \int \rd^3 x \, \rd^2 \q\, \rd^2 \bar \q \, \bm E \,\cL_c~.
\ee
We will use the notation $\rd^{3|4}z :=\rd^3x\, \rd^2 \q \rd^2 \bar \q$ for the full superspace measure.

Instead of reducing the above supersymmetric actions to components, we wish to obtain a prescription which allows for their reduction to $\cN=1$ AdS superspace. The supersymmetric action in $\rm AdS^{3|2}$ is described by a real scalar Lagrangian $L$
\be \label{1.8}
S = \int \rd^{3|2}z \, E \, L = \frac{1}{4} \int \rd^3x \, e \, (\ri \nabla^2 + 8|\m|) L\,\big|_{\,\q = 0}~.
\ee
The action \eqref{real11} reduces to $\rm AdS^{3|2}$ as follows
\be \label{N1rule}
{\mathbb S} = \int \rd^{3|4}z \, \bm E \, \cL = -\frac{\ri}{4}\int \rd^{3|2} z \, E \, \Big \{ (\bm \nabla^\2)^2+8\ri|\m| \Big\} \cL \big|~,
\ee
By making use of the Killing equations \eqref{K-equiv} and \eqref{2.40d}, it can be shown that the action \eqref{N1rule} is invariant under $(1,1)$ AdS isometry transformations given by equation \eqref{PTL}.

In the remainder of this paper, we will carry out the $(1,1)\rightarrow (1,0)$ AdS superspace reduction of the massless higher-spin supermultiplets, and show that the reduced actions coincide with those presented in section \ref{Section3}.

\section{Massless half-integer superspin: Transverse formulation} \label{Section5}
In $(1,1)$ AdS superspace, there exist two off-shell formulations for the massless multiplet of half-integer superspin-$(s+\frac{1}{2})$, with $s \geq 2$ \cite{HKO}. These two theories, which are called transverse and longitudinal, prove to be dual to each other. In the following section, we develop the $(1,1) \rightarrow (1,0)$ reduction procedure for the transverse formulation.

\subsection{Transverse formulation}
According to \cite{HKO}, the transverse formulation for the massless superspin-$(s+\frac{1}{2})$ multiplet is described by the dynamical variables
\be
\cV^\perp_{(s+\hf )} = \big\{ \mathfrak{H}_{\a(2s)}, \G_{\a(2s-2)}, \bar{\G}_{\a(2s-2)} \big\} ~, \label{2.1}
\ee
where $\mathfrak{H}_{\a(2s)} = \mathfrak{H}_{(\a_1   ... \a_{2s})}$ is an unconstrained real superfield, and the complex superfield $\G_{\a(2s-2)}=\G_{(\a_1 ... \a_{2s-2})}$ is transverse linear \eqref{TL}. 
The superfields $\mathfrak{H}_{\a(2s)}$ and $ \G_{\a(2s-2)}$ are defined modulo gauge transformations of the form 
\begin{subequations} \label{tr-gauge-half}
\bea 
\d_\l \mathfrak{H}_{\a(2s)}&=& 
{\bar \cD}_{(\a_1} \l_{\a_2 \dots \a_{2s})}-{\cD}_{(\a_1}\bar {\l}_{\a_2 \dots \a_{2s})}
\equiv
 g_{\a(2s)}+ \bar { g}_{\a(2s)} ~, \label{H-gauge} \\ 
\d_\l \G_{\a(2s-2)}&=&
-\frac{1}{4}\bar{\cD}^{\b} 
\big( {\cD}^2  +2(2s-1) \bar \m \big)\bar{\l}_{\b\a(2s-2)}
\equiv \frac{s}{2s+1}\bar{\cD}^{\b}\cD^{\g}\bar{g}_{(\b \g \a{(2s-2)})}
~,
\label{gamma-gauge}
\eea
\end{subequations}
where the gauge parameter $\l_{\a(2s-1)}$ is complex unconstrained. By construction, the complex gauge parameter $g_{\a(2s)} : = {\bar \cD}_{(\a_1} \l_{\a_2 \dots \a_{2s})}$ is longitudinal linear \eqref{LL}. Note that gauge transformation \eqref{H-gauge} indicates that the superfield $\mathfrak{H}_{\a(2s)}$ corresponds to the $\cN=1$ superconformal gauge prepotential \cite{KO}. 
Up to normalisation, there exists a unique quadratic action which is invariant under the gauge transformations \eqref{tr-gauge-half}. This action takes the form
\bea \label{2.5}
&&\mathbb{S}^{\perp}_{(s+\hf)} [\mathfrak{H}_{\a(2s)},\G_{\a(2s-2)}, \bar{\G}_{\a(2s-2)}]
= \Big(-\hf \Big)^s \int \rd^{3|4}z \, \bm E \,
\bigg\{ 2s(2s-1)\bar{\m}\m \mathfrak{H}^{\a(2s)} \mathfrak{H}_{\a(2s)} ~ \non \\
&&+\frac{1}{8} \mathfrak{H}^{\a(2s)}  \cD^\b ({\bar \cD}^2- 6\mu)\cD_\b \mathfrak{H}_{\a(2s)} + \mathfrak{H}^{\a(2s)}\Big(\cD_{\a_1} {\bar \cD}_{\a_2} {\G}_{\a_3 \dots \a_{2s}}~ - {\bar \cD}_ {\a_1} \cD_{\a_2} \bar {\G}_{\a_3 \dots \a_{2s}} \Big)
~ \non \\
&&+ \frac{2s-1}{s} \bar {\G}^{\a(2s-2)} {\G}_{\a(2s-2)} + \frac{2s+1}{2s} 
\Big({\G}^{\a(2s-2)}{\G}_{\a(2s-2)} + \bar {\G}^{\a(2s-2)} \bar {\G}_{\a(2s-2)}\Big) \bigg\}~.
\eea

The above construction does not take into consideration the case where $s=1$, since the transverse linear constraint \eqref{TL} is ill-defined for $n=0$. However, corollary \eqref{TL-sq} is consistent for $n=0$ and defines a covariantly transverse linear scalar superfield \eqref{TL}. So $\G$ and $\G$ can be interpreted as compensators, which have the corresponding gauge transformations
\begin{subequations}
\bea
\d_\l \mathfrak{H}_{\a\b} &=& \bar{\cD}_{(\a}\l_{\b)} - \cD_{(\a} \bar{\l}_{\b)}~, \\
\d_\l \G &=& - \frac{1}{4} \bar{\cD}^\b(\cD^2+2\bar{\m})\bar{\l}_\b~, \label{variationgamma}
\eea
\end{subequations}
as a result of \eqref{tr-gauge-half}. It is simple to show that constraint \eqref{TL-sq} is consistent with the variation \eqref{variationgamma}. Choosing $s=1$ in \eqref{2.5} yields a gauge-invariant action which can be identified as the linearised action for non-minimal $(1,1)$ AdS supergravity \cite{KT-M11}.

\subsection{Reduction of gauge prepotentials to $\text{AdS}^{3|2}$}
We wish to reduce the gauge prepotentials \eqref{2.1} to $\cN=1$ AdS superspace. We start by reducing the superconformal gauge multiplet $\mathfrak{H}_{\a(2s)}$. Converting the longitudinal linear constraint  of $g_{\a(2s)}$ \eqref{LL} to the real representation \eqref{realrep} yields  
\be \label{2.6}
\bm \nabla^\2{}_{(\a_1} g_{\a_2...\a_{2s+1})} = \ri \bm \nabla^{\1}{}_{(\a_1} g_{\a_2...\a_{2s+1})}~.
\ee
Performing a Taylor expansion of $g_{\a(2s)}(\q^I)$ about $\q^\2$, and using \eqref{2.6}, we find the independent $\q^\2$~-components of $g_{\a(2s)}$ to be
\be
g_{\a(2s)} |~, \qquad \bm \nabla^{\underline{2}\beta}g_{\b\a(2s-1)}|~.
\ee
The gauge transformation \eqref{H-gauge} allows us to impose the gauge conditions
\be
\mathfrak{H}_{\a(2s)}|=0~, \qquad \bm \nabla^{\2\beta}\mathfrak{H}_{\b\a(2s-1)}|=0~. \label{2.8}
\ee
In this Wess-Zumino (WZ) gauge, we stay with the unconstrained real $\cN=1$ superfields\begin{subequations} 
\label{2.9}
\bea 
H_{\a(2s+1)} :&=&\ri \bm \nabla^{\2}{}_{(\a_1}\mathfrak{H}_{\a_2...\a_{2s+1})}|~,  \\
H_{\a(2s)} :&=&\frac{\ri}{4}(\bm \nabla^\2)^2\mathfrak{H}_{\a(2s)}|~.
\eea
\end{subequations}
The residual gauge freedom which preserves gauge conditions \eqref{2.8} are described by the unconstrained real $\cN = 1$ superfields
\begin{subequations}\label{2.10}
\bea 
g_{\a(2s)}|&=& - \frac{\ri}{2}\z_{\a(2s)}~, \qquad \bar \z_{\a(2s)} = \z_{\a(2s)}~, \\
\bm \nabla^{\2\b}g_{\b\a(2s-1)}|&=& \frac{2s+1}{2s}\z_{\a(2s-1)}~, \qquad \bar \z_{\a(2s-1)} =  \z_{\a(2s-1)}~.
\eea
\end{subequations}
From \eqref{2.10}, we can readily determine the gauge transformations of \eqref{2.9}
\begin{subequations} \label{2.11}
\bea 
\d H_{\a(2s+1)} &=& \ri \nabla_{(\a_1} \z_{\a_2 ... \a_{2s+1})}~, \\
\d H_{\a(2s)} &=& \nabla_{(\a_1}\z_{\a_2 ... \a_{2s})}~.
\eea
\end{subequations}

Next, we wish to reduce $\G_{\a(2s-2)}$ to $\cN =1$ AdS superspace. The superfield $\G_{\a(2s-2)}$ obeys the transverse linear constraint \eqref{TL}, which takes the following form in the real representation \eqref{realrep}
\be \label{2.12}
\bm \nabla^{\2\b}\G_{\b\a(2s-3)} = \ri \bm \nabla^{\1\b} \G_{\b\a(2s-3)}~.
\ee
It follows that $\G_{\a(2s-2)}(\q^I)$ has two independent $\q^\2$ -components
\be \label{2.13}
\G_{\a(2s-2)}| \ , \qquad  \bm \nabla^\2{}_{(\a_1}\G_{\a_2 ... \a_{2s-1})}|~.
\ee
Utilising the gauge transformation of $\G_{\a(2s-2)}$ \eqref{gamma-gauge} and the real representation \eqref{realrep}, we find
\begin{subequations} \label{2.14}
\bea 
\d \G_{\a(2s-2)} &=& \frac{\ri s}{2s+1}\Big (\bm \nabla^{\1\b} \bm \nabla^{\2\g} - \bm \nabla^{\b\g}\Big ) \bar {g}_{\b\g\a(2s-2)} ~, \\ 
\bm \nabla^\2{}_{(\a_1}\d \G_{\a_2...\a_{2s-1})} &=& \frac{s}{2s+1}\Big ( \bm \nabla^{\1 \b} \bm \nabla_{(\a_1}{}^\g \bar {g}_{\a_2 ... \a_{2s-1})\b\g}-(4s+1)|\m| \bm \nabla^{\1 \b} \bar {g}_{\b\a(2s-1)}~\\ \non
&+&  \bm {\nabla}_{\b}{}^\g \bm {\nabla}^{\1\b}\bar {g}_{\g\a(2s-1)} - \ri \bm {\nabla}_\b{}^\g \bm \nabla^{\2\b} \bar{g}_{\g\a(2s-1)}-\frac{1}{2}( \bm \nabla^{\1})^2 \bm \nabla^{\2\b} \bar {g}_{\b\a(2s-1)}~ \non \\
&-& (2s+1)\ri|\m|\bm \nabla^{\2\b}\bar {g}_{\b\a(2s-1)} -\ri \bm \nabla^{\b\g} \bm \nabla^{\2}{}_{(\a_1} \bar {g}_{\a_2 ... \a_{2s-1})\b\g}\Big )~. \non
\eea
\end{subequations}
From \eqref{2.14}, we can immediately read off the gauge transformations of the complex $\cN =1$ superfields \eqref{2.13}
\begin{subequations} \label{GTHITrans}
\bea \label{2.15a}
\d \G_{\a(2s-2)}| &=& - \frac{\ri}{2}\nabla^\b\z_{\b\a(2s-2)} + \frac{s}{2(2s+1)}\nabla^{\b\g}\z_{\b\g\a(2s-2)} ~, \\ 
\bm \nabla^\2{}_{(\a_1}\d \G_{\a_2...\a_{2s-1})}| &=& \frac{s}{2(2s+1)}\bigg ( \ri \nabla^\b\nabla_{(\a_1}{}^\g \z_{\a_2...\a_{2s-1})\b\g} + \ri \nabla_\b{}^\g \nabla^\b \z_{\g\a(2s-1)}~\\ \non
&-& (4s+1)\ri |\m|\nabla^\b\z_{\b\a(2s-1)} + \frac{2s+1}{s}\Big ((2s+1)\ri|\m|\z_{\a(2s-1)}  ~\\
&+&\ri \nabla_{(\a_1}{}^\b \z_{\a_2...\a_{2s-1})\b} + \frac{1}{2}\nabla^2\zeta_{\a(2s-1)} \Big) \bigg ) ~. \label{2.15b} \non
\eea
\end{subequations}
Let us express the ${\cN}=1$ superfields \eqref{2.13} in terms of their real and imaginary parts,
\begin{subequations}
\bea
\G_{\a(2s-2)}| &=&-L_{\a(2s-2)} -  \ri s V_{\a(2s-2)} ~,\\
\bm \nabla^\2{}_{(\a_1} \G_{\a_2...\a_{2s-1})}| &=&\frac{1}{2} \Big (\F_{\a(2s-1)} + \ri \O_{\a(2s-1)} \Big ).
\eea
\end{subequations}

It then follows from gauge transformations \eqref{2.11} and \eqref{GTHITrans} that we are in fact dealing with two different gauge theories. The first model is formulated in terms of the real unconstrained gauge superfields
\be \label{2.17}
\cV^\parallel_{(s+\frac{1}{2})}= \{ H_{\a(2s+1)}, L_{\a(2s-2)}, \F_{\a(2s-1)} \}~,
\ee
which are defined modulo gauge transformations of the form
\begin{subequations}\label{2.18}
\bea \label{2.18a}
\d H_{\a(2s+1)} &=& \ri \nabla_{(\a_1}\z_{\a_2 ... \a_{2s+1})}~, \\ \label{2.18b}
\d L_{\a(2s-2)} &=&- \frac{s}{2(2s+1)}\nabla^{\b\g}\z_{\b\g\a(2s-2)}~,\\
\d \F_{\a(2s-1)} &=& \frac{\ri s}{2s+1}\Big (\nabla^\b \nabla_{(\a_1}{}^\g \z_{\a_2...\a_{2s-1})\b\g} - (4s+1)|\m|\nabla^\b\z_{\b\a(2s-1)} ~ \\
&-& \nabla^{\b\g}\nabla_{\b} \z_{\g\a(2s-1)}\Big )~, \non
\eea
\end{subequations}
where the gauge parameter $\z_{\a(2s)}$ is real unconstrained. The other gauge theory is constructed in terms of the superfields
\be \label{2.19}
\cV^\parallel_{(s)} = \{ H_{\a(2s)}, V_{\a(2s-2)}, \O_{\a(2s-1)} \}~,
\ee
which possesses the gauge freedom
\begin{subequations} \label{2.20}
\bea \label{2.20a}
\d H_{\a(2s)} &=&  \nabla_{(\a_1}\z_{\a_2 ... \a_{2s})} ~, \\ \label{2.20b}
\d V_{\a(2s-2)} &=&  \frac{1}{2s}\nabla^\b \z_{\b\a(2s-2)}~, \\
\d \O_{\a(2s-1)} &=& (2s+1)|\m|\z_{\a(2s-1)} + \nabla_{(\a_1}{}^\b \z_{\a_2 ... \a_{2s-1})\b} - \frac{\ri}{2}\nabla^2\z_{\a(2s-1)}~,
\eea
\end{subequations}
where the parameter $\z_{\a(2s-1)}$ is real unconstrained. Applying the superspace reduction procedure to the action \eqref{2.5} yields two decoupled $\cN =1$ supersymmetric theories, each formulated in terms of the gauge fields \eqref{2.17} and \eqref{2.19} respectively
\bea \label{ReductionTransHI}
\mathbb{S}^{\perp}_{(s+\hf)} [\mathfrak{H}_{\a(2s)},\G_{\a(2s-2)}, \bar{\G}_{\a(2s-2)}] &=& S^\parallel_{(s+\frac{1}{2})}[H_{\a(2s+1)},L_{\a(2s-2)},  \F_{\a(2s-1)}] \non ~\\ 
&+& S^\parallel_{(s)}[H_{\a(2s)},V_{\a(2s-2)}, \O_{\a(2s-1)}]~.
\eea
Explicit expressions for the decoupled supersymmetric actions are given in the following subsection.


\subsection{Massless higher-spin $\cN =1$ supermultiplets}
The first of the decoupled $\cN=1$ supersymmetric actions, which is realised by the dynamical variables \eqref{2.17}, takes the following form
\bea \label{2.22}
&&S^\parallel_{(s+\frac{1}{2})}[H_{\a(2s+1)},L_{\a(2s-2)},\F_{\a(2s-1)}]= \Big (-\frac{1}{2} \Big )^s \Big ( - \frac{\ri}{8}\Big  )\int \rd^{3|2}z~ E ~\bigg \{ 2H^{\a(2s+1)}\mathbb{Q} H_{\a(2s+1)} \non ~\\
&&- 8(s-2)(2s+1)|\m|^2H^{\a(2s+1)}H_{\a(2s+1)} +  (2s+1)\ri|\m|H^{\a(2s+1)}\nabla^2H_{\a(2s+1)}~ \non \\
&&-4(2s+1)|\m|H^{\b\a(2s)}\nabla_\b{}^\g H_{\g\a(2s)}-\ri H^{\b\a(2s)}\nabla^2\nabla_\b{}^\g H_{\g\a(2s)}  - 4 H^{\b\g\a(2s-1)}\nabla_{\b\g}\F_{\a(2s-1)}~ \non \\
&&- 8\ri H^{\b\g\d\a(2s-2)}\nabla_{\b\g} \nabla_{\d}L_{\a(2s-2)} - 16\ri L^{\a(2s-2)}\nabla^\b \F_{\b\a(2s-2)} + \frac{2}{s}\F^{\a(2s-1)}\F_{\a(2s-1)} ~ \non \\
&&+ \frac{8\ri}{s(2s-1)} \Big ( 2(s-1)(4s-1)L^{\b\a(2s-3)}\nabla_\b{}^\g L_{\g\a(2s-3)}  ~ \non \\ &&-4s(4s^2-3s+1)|\m|L^{\a(2s-2)}L_{\a(2s-2)} 
- (3s-1)\ri L^{\a(2s-2)}\nabla^2L_{\a(2s-2)} \Big ) \bigg \}~.
\eea
Upon inspection, it is apparent that the superfield $\F_{\a(2s-1)}$ is auxiliary. So by making use of its equation of motion, 
\be
\F_{\a(2s-1)} = -s \nabla^{\b\g}H_{\b\g\a(2s-1)}-4\ri s \nabla_{(\a_1}L_{\a_2 ... \a_{2s-1})}~,
\ee
we can eliminate the auxiliary field $\F_{\a(2s-1)}$ from the action \eqref{2.22}. It can be shown that the resulting action coincides with the off-shell $\cN=1$ supersymmetric action for massless half-integer superspin in AdS$_3$ \eqref{LongHalfIntAct}.

The other $\cN =1$ theory is formulated in terms of the gauge superfields \eqref{2.19} 
\bea \label{2.23}
&&S^\parallel_{(s)}[H_{\a(2s)},V_{\a(2s-2)}, \O_{\a(2s-1)}] = \Big (-\frac{1}{2} \Big )^s \int \rd^{3|2}z~ E ~\bigg\{2 |\m| H^{\a(2s)}H_{\a(2s)} \non \\
&&+ \frac{\ri}{2} H^{\a(2s)}\nabla^2H_{\a(2s)} + 2s H^{\b\g\a(2s-2)}\nabla_{\b\g}V_{\a(2s-2)} - H^{\b\a(2s-1)}\nabla_\b \O_{\a(2s-1)} \non ~ \\
&&+ \frac{\ri}{2}\O^{\a(2s-1)}\O_{\a(2s-1)} + \frac{1}{2s-1} \Big (4s(s-1)^2V^{\b\a(2s-3)}\nabla_\b{}^\g V_{\g\a(2s-3)}~ \non \\
&&+4s(2s^3-2s+1)|\m|V^{\a(2s-2)}V_{\a(2s-2)} +\ri s (2s^2-2s+1)V^{\a(2s-2)}\nabla^2 V_{\a(2s-2)} \non \\
&&+(2s-1)V^{\a(2s-2)}\nabla^\b\O_{\b\a(2s-2)} \Big ) \bigg \}~.
\eea
The superfield $\O_{\a(2s-1)}$ is auxiliary, so upon elimination via its equation of motion
\be
\O_{\a(2s-1)} = - \ri \nabla^\b H_{\b\a(2s-1)} - \ri \nabla_{(\a_1}V_{\a_2 ... \a_{2s-1})}~,
\ee
we find that the resulting action coincides with the off-shell $\cN=1$ action massless superspin-$s$ multiplet in AdS$_{3}$ \eqref{action-t3}.


\subsection{Second supersymmetry transformations} \label{Subsect5.4}
As discussed in subsection \ref{Subsect44}, by construction, the $\cN=1$ reduced actions \eqref{2.22} and \eqref{2.23} are invariant under the second supersymmetry transformations \eqref{PTL2B}. For convenience, we recall the form of these transformations
\be \label{SST}
\d_\e  U| = \e^\a (\bm \nabla^\2_\a U)|~,
\ee
where $U(x,\q^I)$ is a $\cN=(1,1)$ superfield, with indices being suppressed. The second supersymmetry tranformations act on the $\cN=1$ fields \eqref{2.17} and \eqref{2.19} in the following fashion
\begin{subequations}
	\bea
	\d_\e \mathfrak{H}_{\a(2s)} | &=& -\ri \e^\b H_{\b\a(2s)} ~, \label{SST1} \\
	\d_\e \bm \nabla^{\2\b} \mathfrak{H}_{\b\a(2s-1)}| &=& 2 \ri \e^\b H_{\b\a(2s-1)}~, \label{SST2} \\
	\d_{\e} H_{\a(2s+1)} &=& -2\e_{(\a_1}H_{\a_2...\a_{2s+1})} ~, \\
	\d_{\e} H_{\a(2s)} &=& - \frac{\ri}{2} \Big ( (2s+1)|\m|\e^\b H_{\b\a(2s)} - \e^\b \nabla_\b{}^\g H_{\g\a(2s)} \Big ) ~,  \\
	\d_{\e}L_{\a(2s-2)}&=& - \frac{\ri}{2}\e^\b\O_{\b\a(2s-2)} + \frac{2s}{2s-1}(s-1) \e_{(\a_1} \nabla^\b V_{\a_2 ... \a_{2s-2})\b}~,  ~ \\
	\d_\e V_{\a(2s-2)}&=& \frac{\ri}{2s}\e^\b \F_{\b\a(2s-2)} - \frac{2}{s(2s-1)}(s-1)\e_{(\a_1} \nabla^\b L_{\a_2 ... \a_{2s-2})\b}~, \\
	\d_\e \F_{\a(2s-1)} &=& \frac{1}{2s-1} \Big ( 2s(2s-1) \e^\b \nabla_{\b(\a_1}V_{\a_2 ... \a_{2s-1})} + \ri s \e_{(\a_1}\nabla^2 V_{\a_2 ... \a_{2s-1})} ~ \\
	&+&4s(4s^2-3s+1)|\m|\e_{(\a_1}V_{\a_2 ... \a_{2s-1})} + (2s-1)\e_{(\a_1}\nabla^\b \O_{\a_2 ... \a_{2s-1})\b} ~ \non \\
	&-&4s(s-1)\e_{(\a_1}\nabla_{\a_2}{}^\b V_{\a_3 ... \a_{2s-1})\b} \Big ) ~, ~ \non\\
	\d_\e \O_{\a(2s-1)} &=& - \frac{1}{2s-1} \Big (2(2s-1) \e^\b \nabla_{\b(\a_1}L_{\a_2 ... \a_{2s-1})} + \ri  \e_{(\a_1}\nabla^2 L_{\a_2 ... \a_{2s-1})} ~ \\
	&+&4(4s^2-3s+1)|\m|\e_{(\a_1}L_{\a_2 ... \a_{2s-1})} + (2s-1)\e_{(\a_1}\nabla^\b \F_{\a_2 ... \a_{2s-1})\b} \non ~ \\
	&-&4(s-1)\e_{(\a_1}\nabla_{\a_2}{}^\b L_{\a_3 ... \a_{2s-1})\b} \Big ) ~.~ \non
	\eea
\end{subequations}

It is apparent from \eqref{SST1} and \eqref{SST2}  that the second supersymmetry transformation \eqref{SST} breaks the WZ gauge conditions \eqref{2.8}, which we recall for convenience
\be \label{GC}
\mathfrak{H}_{\a(2s)}|=0~, \qquad \bm \nabla^{\2\beta}\mathfrak{H}_{\b\a(2s-1)}|=0~.
\ee
 In order to resolve this, it is necessary to supplement the variation \eqref{SST} with the $\e$-dependent gauge transformations:
\bsubeq \label{edepgt}
\bea 
\d_{g(\e)} \mathfrak{H}_{\a(2s)} &=& g_{\a(2s)}(\e) + \bar{g}_{\a(2s)}(\e)~, \\
\d_{g(\e)} \G_{\a(2s-2)} &=& \frac{s}{2s+1} \bar \cD^{\b} \cD^{\g} \bar g_{\b \g \a(2s-2)} (\e)~.
\eea
\esubeq
The modified second supersymmetry transformations now take the following form
\bsubeq \label{MSST}
\bea 
\hat{\d}_\e \mathfrak{H}_{\a(2s)} &=&  \d_\e  \mathfrak{H}_{\a(2s)}  + \d_{g(\e)} \mathfrak{H}_{\a(2s)} ~, \\
\hat{\d}_\e \G_{\a(2s-2)} &=&  \d_\e  \G_{\a(2s-2)}  + \d_{g(\e)} \G_{\a(2s-2)} ~.
\eea
\esubeq
The sole purpose of introducing the $\e$-dependent gauge transformations \eqref{edepgt} is to restore the original Wess-Zumino gauge \eqref{GC}. Fixing the form of the ${\cN}=1$ components of $g_{\a(2s)} (\e)$ as 
\begin{subequations} \label{FRGP}
	\bea 
	g_{\a(2s)}(\e)| &=& \frac{\ri}{2}\e^\b H_{\b\a(2s)}~,\\
	\bm \nabla^{\2\b}g_{\b\a(2s-1)}(\e)| &=& - \ri \e^\b H_{\b\a(2s-1)}~,
	\eea
\end{subequations}
ensures that \eqref{MSST} takes us back to the original gauge \eqref{GC}.
The transformations \eqref{MSST} act on the ${\cN}=1$ superfields by the rule
\begin{subequations} \label{MSSTU}
	\bea
	\hat{\d}_{\e} H_{\a(2s+1)} &=&  -2\e_{(\a_1}H_{\a_2...\a_{2s+1})} ~, \\
	\hat{\d}_{\e} H_{\a(2s)} &=&- \frac{\ri}{4(2s+1)} \Big ( \ri \e^\b \nabla^2 H_{\b\a(2s)} 
	+4s\e_{(\a_1}\nabla^{\b\g}H_{\a_2 ... \a_{2s})\b\g} \\
	&+&2(2s+1)(4s+1)|\m|\e^\b H_{\b\a(2s)} - 2\e^\b \nabla_\b{}^\g H_{\g\a(2s)} \Big )~, \non \\
	\hat{\d}_{\e}L_{\a(2s-2)}&=&  - \frac{\ri}{2}\e^\b\O_{\b\a(2s-2)} + \frac{2s}{2s-1}(s-1) \e_{(\a_1} \nabla^\b V_{\a_2 ... \a_{2s-2})\b} \\
	&+& \frac{s}{2s+1}\e^\b \nabla^\g H_{\b\g\a(2s-2)}~, \non \\
	\hat{\d}_\e V_{\a(2s-2)}&=& \frac{\ri}{2s}\e^\b \F_{\b\a(2s-2)} - \frac{2}{s(2s-1)}(s-1)\e_{(\a_1} \nabla^\b L_{\a_2 ... \a_{2s-2})\b} ~ \\
	&+& \frac{\ri}{2(2s+1)}\e^\b \nabla^{\g\d}H_{\b\g\d\a(2s-2)}~, \non \\
	\hat{\d}_\e \F_{\a(2s-1)} &=& \frac{1}{2s-1} \Big ( 2s(2s-1) \e^\b \nabla_{\b(\a_1}V_{\a_2 ... \a_{2s-1})} + \ri s \e_{(\a_1}\nabla^2 V_{\a_2 ... \a_{2s-1})} ~ \\
	&+&4s(4s^2-3s+1)|\m|\e_{(\a_1}V_{\a_2 ... \a_{2s-1})} + (2s-1)\e_{(\a_1}\nabla^\b \O_{\a_2 ... \a_{2s-1})\b} ~ \non \\
	&-&4s(s-1)\e_{(\a_1}\nabla_{\a_2}{}^\b V_{\a_3 ... \a_{2s-1})\b} \Big ) - \frac{s}{2s+1} \e^\b \Big ( 4(s+1)|\m|  H_{\b\a(2s-1)} ~ \non\\
	&+& 2\nabla_{(\a_1}{}^\g H_{\a_2 ... \a_{2s-1})\b\g} - \ri \nabla^2 H_{\b\a(2s-1)} \Big )~, \non \\
	\hat{\d}_\e \O_{\a(2s-1)} &=& - \frac{1}{2s-1} \Big (2(2s-1) \e^\b \nabla_{\b(\a_1}L_{\a_2 ... \a_{2s-1})} + \ri  \e_{(\a_1}\nabla^2 L_{\a_2 ... \a_{2s-1})} ~ \\
	&+&4(4s^2-3s+1)|\m|\e_{(\a_1}L_{\a_2 ... \a_{2s-1})} + (2s-1)\e_{(\a_1}\nabla^\b \F_{\a_2 ... \a_{2s-1})\b} \non ~ \\
	&-&4(s-1)\e_{(\a_1}\nabla_{\a_2}{}^\b L_{\a_3 ... \a_{2s-1})\b} \Big )  + \frac{s}{2s+1}\e^\b\Big ( 2(2s-1)|\m| \nabla^\g H_{\b\g\a(2s-1)} \non ~ \\
	&-&  \nabla_{(\a_1}{}^\g \nabla^\d H_{\a_2 ... \a_{2s-1})\b\g\d} + \nabla^{\g \d} \nabla_{\g} H_{\b\d\a(2s-1)} \Big )~. \non
	\eea
\end{subequations}	

\section{Massless half-integer superspin: Longitudinal formulation} \label{Section6}
In this section, we develop the superspace reduction procedure for the longitudinal formulation, following the prescription advocated in section \ref{Section5}.

\subsection{Longitudinal formulation}
For $s \geq 2$, the longitudinal formulation 
for the massless superspin-$(s+\hf)$ multiplet 
is described in terms of the variables
\bea \label{3.1}
\cV^{\|}_{(s+\hf)} = \big\{ \mathfrak{H}_{\a(2s)}, G_{\a(2s-2)}, \bar{G}_{\a(2s-2)} \big\} ~.
\eea
Here, the real superfield $\mathfrak{H}_{\a(2s)}$ is identical to that of \eqref{2.1},
and the complex superfield ${G}_{\a(2s-2)}$ is longitudinal linear \eqref{LL}. Modulo an overall normalisation factor, the longitudinal formulation is uniquely described by the action
\bea \label{3.3}
&&\mathbb{S}^{\|}_{(s+\hf)}[\mathfrak{H}_{\a(2s)}, G_{\a(2s-2)}, \bar{G}_{\a(2s-2)}]= \Big(-\hf \Big)^{s}\int 
\rd^{3|4}z\, \bm E~ \bigg\{\frac{1}{8}\mathfrak{H}^{\a(2s)}
\cD^{\b}(\bar{\cD}^{2}-6\mu)\cD_{\b}
\mathfrak{H}_{\a(2s)} \non \\
&&+ 2s(s-1)\mu\bar{\mu} \mathfrak{H}^{\a(2s)}\mathfrak{H}_{\a(2s)}
-\frac{1}{16}([\cD_{\b},\bar{\cD}_{\g}]\mathfrak{H}^{\b \g \a(2s-2)})
[\cD^{\d},\bar{\cD}^{\r}]\mathfrak{H}_{\d \r \a(2s-2)}
 \non \\
&&+ \frac{s}{2}(\cD_{\b \g}\mathfrak{H}^{\b \g \a(2s-2)})
\cD^{\d \r}\mathfrak{H}_{\d \r \a(2s-2)}
\non \\
&&+ \frac{2s-1}{2s} \Big ( \ri
\cD_{\b \g}\mathfrak{H}^{\b \g \a(2s-2)} 
\left( G_{\a(2s-2)}-\bar{G}_{\a(2s-2)} \right)
+\frac{1}{s}
\bar{G}^{\a(2s-2)} G_{\a(2s-2)} \Big )
\non \\
&&-\frac{2s+1}{4s^{2}}\left(G^{\a(2s-2)}G_{\a(2s-2)}+\bar{G}^{\a(2s-2)} \bar {G}_{\a(2s-2)}\right)
\bigg\}~,
\label{long-action-half}
\eea
which is invariant under the gauge transformations 
\begin{subequations} \label{3.4}
\bea \label{3.4a}
\d_\l \mathfrak{H}_{\a(2s)}&=& 
{\bar \cD}_{(\a_1} \l_{\a_2 \dots \a_{2s})}-{\cD}_{(\a_1}\bar {\l}_{\a_2 \dots \a_{2s})} \equiv g_{\a(2s)} + \bar {g}_{\a(2s)} ~,\\ \label{3.4b}
\d_\l {G}_{\a(2s-2)}&=& 
-\frac{1}{4}\big( \bar{\cD}^{2} -4s\m\big)\cD^{\b}\l_{\a(2s-2)\b}
+\ri (s-1)\bar{\cD}_{(\a_{1}} \cD^{|\b\g|} \l_{\a_2 \dots \a_{2s-2})\b\g} \non ~\\
&\equiv& \Big(\frac{s}{2s+1}\cD^\b \bar \cD^\g + \ri s \cD^{\b\g}\Big )g_{\b\g\a(2s-2)}~,
\eea
\end{subequations}
where $\l_{\a(2s-1)}$ is complex unconstrained and $g_{\a(2s)}$ is  longitudinal linear, as in \eqref{H-gauge}. 

In the case where $s=1$, the compensator $G$ is covariantly chiral. If we introduce the field redefinition $G=3\s$ in action \eqref{3.3}, along with choosing $s=1$, then the model can be shown to coincide with the linearised action for minimal $(1,1)$ AdS supergravity \cite{KT-M11}. In the case $s=1$,  the gauge transformations \eqref{3.4} yield
\bea
\d_\l \mathfrak{H}_{\a\b} &=& \bar{\cD}_{(\a}\l_{\b)} - \cD_{(\a} \bar{\l}_{\b)}~, \\
\d_\l G &=& - \frac{1}{4}(\bar{\cD}^2-4\bar{\m})\cD^\b \l_\b~. \label{Gvariation}
\eea
It is an easy exercise to show that $\d_\l G$ is covariantly chiral.


\subsection{Reduction of  gauge prepotentials to AdS$^{3|2}$}

The reduction of the superconformal gauge prepotential $\mathfrak{H}_{\a(2s)}$ was addressed in section \ref{Section5}. We need only reduce the compensator $G_{\a(2s-2)}$ to $\cN =1$ AdS superspace. We start by converting the longitudinal constraint of $G_{\a(2s-2)}$ \eqref{LL}  to the real basis \eqref{realrep}
\be \label{3.5}
\bm \nabla^\2{}_{(\a_1}G_{\a_2...\a_{2s-1})}=\ri \bm \nabla^{\1}{}_{(\a_1}G_{\a_2...\a_{2s-1})}~.
\ee
Performing a Taylor expansion of $G_{\a(2s-2)}(\q^I)$ about $\q^\2$, and using \eqref{3.5}, we find the independent $\q^\2$  -components of $G_{\a(2s-2)}$ to be
\be \label{3.6}
G_{\a(2s-2)}|~, \qquad   \bm \nabla^{\2\b}G_{\b\a(2s-3)}|~.
\ee
Utilising the gauge transformations \eqref{3.4b} and the real representation \eqref{realrep}, we find
\begin{subequations}
\bea
\d G_{\a(2s-2)} = \frac{\ri s}{2s+1}\Big (2s\bm \nabla^{\b\g} - \bm \nabla^{\1 \b}\bm\nabla^{\2\g}\Big )g_{\b\g\a(2s-2)}~,\\
\bm \nabla^{\2\b}\d G_{\b\a(2s-3)} =\frac{s}{2s+1}\Big (2\ri s \bm \nabla^{\b\g}\bm \nabla^{\2\d}- \bm \nabla^{\b\g} \bm \nabla^{\1 \d} \Big ) g_{\b\g\d\a(2s-3)}~.
\eea
\end{subequations}
We can use the residual gauge freedom \eqref{2.10} to compute the gauge transformations of the complex $\cN=1$ superfields \eqref{3.6}
\begin{subequations} \label{3.8}
\bea
\d G_{\a(2s-2)}| = \frac{s^2}{2s+1} \nabla^{\b\g}\z_{\b\g\a(2s-2)} - \frac{\ri}{2}\nabla^\b\z_{\b\a(2s-2)}~ \label{3.8a} ,\\
\bm \nabla^{\2\b}\d G_{\b\a(2s-3)}| = \ri s \nabla^{\b\g}\z_{\b\g\a(2s-3)}  + \frac{\ri s}{2(2s+1)}\nabla^{\b\g} \nabla^\d \z_{\b\g \d\a(2s-3)}~. \label{3.8b}
\eea
\end{subequations}
It is useful to separate the complex superfields \eqref{3.6} into real and imaginary components\begin{subequations}
\bea
 G_{\a(2s-2)}| &=&-2sL_{\a(2s-2)} - \ri s V_{\a(2s-2)} ~,\\
\bm \nabla^{\2\b} G_{\b\a(2s-3)}|  &=&\frac{1}{2}\Big ( \F_{\a(2s-3)} + \ri \O_{\a(2s-3)} \Big )~.
\eea
\end{subequations}

It then becomes apparent from gauge transformations \eqref{2.11} and \eqref{3.8} that we are in fact dealing with two different gauge theories. The first gauge theory is formulated in terms of the dynamical variables
\be \label{3.10}
\cV^\parallel_{(s+\frac{1}{2})} = \big \{ H_{\a(2s+1)}, L_{\a(2s-2)}, \F_{\a(2s-3)} \big \}~,
\ee
which are defined modulo gauge transfomations of the form
\begin{subequations} \label{3.11}
\bea
\d H_{\a(2s+1)} &=& \ri \nabla_{(\a_1}\z_{\a_2 ...\a_{2s+1)}} ~, \label{3.11a} \\
\d L_{\a(2s-2)} &=& -\frac{s}{2(2s+1)} \nabla^{\b\g}\z_{\b\g\a(2s-2)}~,  \label{3.11b}\\
\d \F_{\a(2s-3)} &=& \frac{\ri s}{2s+1}\nabla^{\b\g}\nabla^\d\z_{\b\g\d\a(2s-3)}~.
\eea
\end{subequations}
The other gauge model is described in terms of the superfields
\be \label{3.12}
\cV^\parallel_{(s)} = \big \{H_{\a(2s)}, V_{\a(2s-2)}, \O_{\a(2s-3)} \big \}~,
\ee
which possess the following gauge freedom
\begin{subequations} \label{3.13}
\bea
\d H_{\a(2s)} &=& \nabla_{(\a_1}\z_{\a_2 ...\a_{2s})} ~, \label{3.13a}\\
\d V_{\a(2s-2)} &=& \frac{1}{2s}\nabla^\b\z_{\b\a(2s-2)}~, \label{3.13b}\\
\d \O_{\a(2s-3)} &=& 2s\nabla^{\b\g}\z_{\b\g\a(2s-3)}~.
\eea
\end{subequations}
After carrying out the reduction to $\cN=1$ AdS superspace, the action \eqref{3.3} decouples into two $\cN =1$ theories, which are formulated in terms of superfields \eqref{3.10} and \eqref{3.12} respectively
\bea \label{RATHI}
\mathbb{S}^{\parallel}_{(s+\hf)} [\mathfrak{H}_{\a(2s)},G_{\a(2s-2)}, \bar{G}_{\a(2s-2)}] =&& S^\parallel_{(s+\frac{1}{2})} [H_{\a(2s+1)},L_{\a(2s-2)},  \F_{\a(2s-3)}] \non ~\\ 
&&+ S^\parallel_{(s)}[H_{\a(2s)},V_{\a(2s-2)}, \O_{\a(2s-3)}]~.
\eea
In the following subsection, we provide the explicit forms of the decoupled $\cN=1$ actions.


\subsection{Massless higher-spin $\cN=1$ supermultiplets}
The first $\cN = 1$ gauge theory, described in terms of the dynamical variables \eqref{3.10}, takes the form 
\bea \label{3.15}
&&S^\parallel_{(s+\frac{1}{2})}[H_{\a(2s+1)}, L_{\a(2s-2)}, \F_{\a(2s-3)}] = \Big(-\hf \Big)^{s} \Big ( - \frac{\ri}{8} \Big ) \int 
\rd^{3|2}z \, E ~\bigg \{2H^{\a(2s+1)} \mathbb{Q} H_{\a(2s+1)}~ \non \\
&&+(2s+1)\ri|\m|H^{\a(2s+1)}\nabla^2 H_{\a(2s+1)}   -4(2s+1)|\m|H^{\b\a(2s)}\nabla_\b{}^\g H_{\g\a(2s)}-\ri H^{\b\a(2s)}\nabla^2\nabla_\b{}^\g H_{\g\a(2s)}~ \non \\
&&-8(s-2)(2s+1)|\m|^2H^{\a(2s+1)}H_{\a(2s+1)} - 2s\nabla_{\b\g}H^{\b\g\a(2s-1)}\nabla^{\d\r}H_{\d\r\a(2s-1)}  ~ \non \\
&&-\frac{4}{2s-1} \Big ( 16\ri |\m| (2s^3-2s+1)L^{\a(2s-2)}L_{\a(2s-2)} + 4(2s^2-1)L^{\a(2s-2)}\nabla^2L_{\a(2s-2)}~ \non \\
&&+16\ri  (s-1)^2 L^{\b\a(2s-3)}\nabla_\b{}^\g L_{\g\a(2s-3)}  + 2\ri (2s-1)^2\nabla_{\b\g}H^{\b\g\d\a(2s-2)}\nabla_\d L_{\a(2s-2)} ~ \non \\
&&+\frac{4\ri}{s} (s-1)L^{\b\a(2s-3)}\nabla_\b \F_{\a(2s-3)} + \frac{1}{s}(s-1)\F^{\a(2s-3)}\F_{\a(2s-3)} \Big ) \bigg \}~.
\eea
The action \eqref{3.15} is invariant under the gauge transformations \eqref{3.11}. We can eliminate the auxiliary field $\F_{\a(2s-3)} $ from \eqref{3.15} by using the equation of motion
\be
\F_{\a(2s-3)} = - 2 \ri \nabla^\b L_{\b\a(2s-3)}~.
\ee
The resulting action, up to an overall factor, then coincides with model \eqref{LongHalfIntAct}.

The other $\cN=1$ model which is constructed in terms of the dynamical variables \eqref{3.12} assumes the form
\bea \label{3.16}
&&S^\parallel_{(s)}[H_{\a(2s)},V_{\a(2s-2)}, \O_{\a(2s-3)}] = - \Big (-\frac{1}{2} \Big )^s \int \rd^{3|2}z~ E ~\bigg\{ (s-1)|\m|H^{\a(2s)}H_{\a(2s)} ~\non \\
&&-\frac{\ri}{4} H^{\a(2s)}\nabla^2H_{\a(2s)} + \frac{1}{2}H^{\b\a(2s-1)}\nabla_\b{}^\g H_{\g\a(2s-1)}-(2s-1)H^{\a(2s)}\nabla_{(\a_1 \a_2}V_{\a_3 ... \a_{2s})} \non ~ \\
&&-\frac{1}{4(2s-1)}\Big (2\ri s V^{\a(2s-2)}\nabla^2 V_{\a(2s-2)} -4 (s-1)(4s-1)V^{\b\a(2s-3)}\nabla_\b{}^\g V_{\g\a(2s-3)} ~ \non \\
&&+8s(4s^2-3s+1)|\m|V^{\a(2s-2)}V_{\a(2s-2)} +8(s-1)V^{\b\a(2s-3)}\nabla_\b \O_{\a(2s-3)}~ \non \\
&&-\frac{\ri}{s^2}(s-1)\O^{\a(2s-3)}\O_{\a(2s-3)} \Big ) \bigg \}~,
\eea
which is invariant under the gauge transformations \eqref{3.13}. It is evident that the superfield $\O_{\a(2s-3)}$ is auxiliary, so upon elimination via its equation of motion
\be
\O_{\a(2s-3)} = - 4 \ri s^2 \nabla^\b V_{\b\a(2s-3)}~,
\ee
one obtains the action \eqref{action-t3}.


\subsection{Second supersymmetry transformations}
Let us note that the reduced actions \eqref{RATHI}
\bea
\mathbb{S}^{\parallel}_{(s+\hf)} [\mathfrak{H}_{\a(2s)},G_{\a(2s-2)}, \bar{G}_{\a(2s-2)}] =&& S^\parallel_{(s+\frac{1}{2})} [H_{\a(2s+1)},L_{\a(2s-2)},  \F_{\a(2s-3)}] \non ~\\ 
&&+ S^\parallel_{(s)}[H_{\a(2s)},V_{\a(2s-2)}, \O_{\a(2s-3)}]~,
\eea
are invariant under second supersymmetry. More explicitly, the second supersymmetry transformation acts on the dynamical superfields \eqref{3.10} and \eqref{3.12} by the following rule
\begin{subequations}
	\bea	
	\d_\e \mathfrak{H}_{\a(2s)} | &=& -\ri \e^\b H_{\b\a(2s)} ~, \label{SSTHIL1}\\
	\d_\e \bm \nabla^{\2\b} \mathfrak{H}_{\b\a(2s-1)}| &=& 2 \ri \e^\b H_{\b\a(2s-1)}~,  \label{SSTHIL2}\\
	\d_{\e} H_{\a(2s+1)} &=& -2\e_{(\a_1}H_{\a_2...\a_{2s+1})} ~, \\
	\d_{\e} H_{\a(2s)} &=& - \frac{\ri}{2} \Big ( (2s+1)|\m|\e^\b H_{\b\a(2s)} - \e^\b \nabla_\b{}^\g H_{\g\a(2s)} \Big ) ~,  \\
	\d_\e L_{\a(2s-2)} &=& -\hf \e^\b\nabla_{(\b}V_{\a(2s-2))} + \frac{\ri (s-1)}{2s(2s-1)}\e_{(\a_1}\O_{\a_2...\a_{2s-2})}~, \\
	\d_\e V_{\a(2s-2)} &=& 2 \e^\b \nabla_{(\b}L_{\a(2s-2))} - \frac{\ri (s-1)}{s(2s-1)}\e_{(\a_1}\F_{\a_2...\a_{2s-2})} ~, \\
	\d_\e \F_{\a(2s-3)} &=& \frac{1}{2s-1}\e^\b \Big ( 2s(2s-3) \nabla_{(\a_1}{}^\g V_{\a_2 ... \a_{2s-3})\b\g} - \ri s \nabla^2 V_{\b\a(2s-3)} \non ~ \\
	&+&4s^2 \nabla_\b{}^\g V_{\g\a(2s-3)} - 4s(4s^2-5s+2)|\m| V_{\b\a(2s-3)}    ~ \non \\
	&-& 2(s-1)\nabla_{(\b}\O_{\a(2s-3))} \Big )~,  \\
	\d_\e \O_{\a(2s-3)} &=& \frac{1}{2s-1} \e^\b \Big ( 2\ri s \nabla^2 L_{\b\a(2s-3)} - 4s(2s-3) \nabla_{(\a_1}{}^\g L_{\a_2...\a_{2s-3})\b\g}  \non ~ \\
	&-&8s^2  \nabla_{\b}{}^\g L_{\g\a(2s-3)} + 8s(4s^2-5s+2)|\m| L_{\b\a(2s-3)}~  \non \\
	&+&2(s-1)\nabla_{(\b}\F_{\a(2s-3))} \Big )~.
	\eea
\end{subequations}
It is evident from the variations \eqref{SSTHIL1} and \eqref{SSTHIL2} that second supersymmetry is not compatible with the imposed WZ gauge conditions \eqref{GC}. Recall that the longitudinal and transverse formulations are both constructed in terms of the real superfield $\mathfrak{H}_{\a(2s)}$, which is defined modulo gauge transformations \eqref{H-gauge}. We used this gauge freedom to impose \eqref{GC} in both theories.

Using a similar approach as in subsection \ref{Subsect5.4}, the WZ gauge conditions \eqref{GC} in the longitudinal formulation can be restored provided we accompany the second supersymmetry transformations with the following $\e$-dependent gauge transformations:
\bsubeq \label{edepgt-HIL}
\bea 
\d_{g(\e)} \mathfrak{H}_{\a(2s)} &=& g_{\a(2s)}(\e) + \bar{g}_{\a(2s)}(\e)~, \\
\d_{g(\e)} G_{\a(2s-2)} &=&\bigg( \frac{s}{2s+1} \cD^{\b} \bar \cD^{\g} + \ri s \cD^{\b \g} \bigg) g_{\b \g \a(2s-2)} (\e)~, 
\eea
\esubeq
where
\begin{subequations} 
	\bea 
	g_{\a(2s)}(\e)| &=& \frac{\ri}{2}\e^\b H_{\b\a(2s)}~,\\
	\bm \nabla^{\2\b}g_{\b\a(2s-1)}(\e)| &=& - \ri \e^\b H_{\b\a(2s-1)}~.
	\eea
\end{subequations}
The modified second supersymmetry transformations now read
\bsubeq \label{MSST-HIL}
\bea 
\hat{\d}_\e \mathfrak{H}_{\a(2s)} &=&  \d_\e  \mathfrak{H}_{\a(2s)}  + \d_{g(\e)} \mathfrak{H}_{\a(2s)} ~, \\
\hat{\d}_\e G_{\a(2s-2)} &=&  \d_\e  G_{\a(2s-2)}  + \d_{g(\e)} G_{\a(2s-2)} ~.
\eea
\esubeq
The transformations \eqref{MSST-HIL} act on the $\cN=1$ superfields \eqref{3.10} and \eqref{3.12} as follows
\begin{subequations}
	\bea	
	\hat{\d}_{\e} H_{\a(2s+1)} &=&  -2\e_{(\a_1}H_{\a_2...\a_{2s+1})} ~, \\
	\hat{\d}_{\e} H_{\a(2s)} &=&- \frac{\ri}{4(2s+1)} \Big ( \ri \e^\b \nabla^2 H_{\b\a(2s)} 
	+4s\e_{(\a_1}\nabla^{\b\g}H_{\a_2 ... \a_{2s})\b\g} \\
	&+&2(2s+1)(4s+1)|\m|\e^\b H_{\b\a(2s)} - 2\e^\b \nabla_\b{}^\g H_{\g\a(2s)} \Big )~, \non \\
	\hat{\d}_\e L_{\a(2s-2)} &=& -\hf \e^\b\nabla_{(\b}V_{\a(2s-2))} + \frac{\ri (s-1)}{2s(2s-1)}\e_{(\a_1}\O_{\a_2...\a_{2s-2})}\\ 
	&-& \frac{1}{2(2s+1)}\e^\b\nabla^\g H_{\b\g\a(2s-2)}~, \non \\
	\hat{\d}_\e V_{\a(2s-2)} &=& 2 \e^\b \nabla_{(\b}L_{\a(2s-2))} - \frac{\ri (s-1)}{s(2s-1)}\e_{(\a_1}\F_{\a_2...\a_{2s-2})} ~ \\
	&-& \frac{\ri s}{2s+1} \e^\b \nabla^{\g\d} H_{\b\g\d \a(2s-2)}~, \non \\
	\hat{\d}_\e \F_{\a(2s-3)} &=& \frac{1}{2s-1}\e^\b \Big ( 2s(2s-3) \nabla_{(\a_1}{}^\g V_{\a_2 ... \a_{2s-3})\b\g} - \ri s \nabla^2 V_{\b\a(2s-3)}  ~ \\
	&+&4s^2 \nabla_\b{}^\g V_{\g\a(2s-3)} - 4s(4s^2-5s+2)|\m| V_{\b\a(2s-3)}  ~ \non \\
	&-&2(s-1)\nabla_{(\b}\O_{\a(2s-3))} \Big ) + \frac{4s^2}{2s+1} \e^\b \nabla^{\g\d} H_{\b\g\d\a(2s-3)}~, \non \\
	\hat{\d}_\e \O_{\a(2s-3)} &=& \frac{1}{2s-1} \e^\b \Big ( 2\ri s \nabla^2 L_{\b\a(2s-3)} - 4s(2s-3) \nabla_{(\a_1}{}^\g L_{\a_2...\a_{2s-3})\b\g}   ~ \\
	&-&8s^2  \nabla_{\b}{}^\g L_{\g\a(2s-3)} + 8s(4s^2-5s+2)|\m| L_{\b\a(2s-3)}~  \non \\
	&+&2(s-1)\nabla_{(\b}\F_{\a(2s-3))} \Big ) + \frac{s}{2s+1}\e^\b \nabla^{\g\d} \nabla^\l H_{\b\g\d\l\a(2s-3)}~. \non
	\eea
\end{subequations}


\section{Massless integer superspin: Longitudinal formulation} \label{Section7}
In \cite{HKO}, two dually equivalent off-shell formulations, called transverse and longitudinal, for the massless multiplets of integer superspin were developed in $(1,1)$ AdS superspace. In the following section, we reduce the longitudinal model to $\cN=1$ AdS superspace.

\subsection{Longitudinal formulation}
Given an integer $s\geq1$, the longitudinal formulation 
for the massless superspin-$s$ multiplet 
is realised in terms of the following dynamical variables
\bea
\cV^{\|}_{(s)} = \{U_{\a(2s-2)}, G_{\a(2s)}, \bar{G}_{\a(2s)} \} ~.
\label{4.1}
\eea
Here $U_{\a(2s-2)}$ is an unconstrained real superfield, and the complex superfield $G_{\a(2s)}$ is longitudinal linear. The dynamical superfields $U_{\a(2s-2)}$ and $G_{\a(2s)}$ are defined modulo gauge transformations of the form 
\begin{subequations}\label{4.3}
	\bea \label{4.3a}
	\d_L U_{\a(2s-2)}
	&=&\cD^{\b}L_{\b \a(2s-2)}-\bar{\cD}^{\b}\bar{L}_{\b \a(2s-2)} 
	\equiv \bar {\g}_{\a(2s-2)}+{\g}_{\a(2s-2)}
	~,  \\
	\d_L G_{\a(2s)} 
	&=&-\frac{1}{2}\bar{\cD}_{(\a_{1}}\big(\cD^{2}
	-2(2s+1) \bar \m \big)
	L_{\a_2 \dots \a_{2s})}
	\equiv \bar{\cD}_{(\a_{1}} \cD_{\a_{2}}\bar{\g}_{\a_3 \dots \a_{2s})}
	~.
	\label{4.3b}
	\eea
\end{subequations} 
Here the gauge parameter $L_{\a(2s-1)}$ is an unconstrained complex superfield, 
and ${\g}^{\a(2s-2)}:= {\bar \cD}_{\b}\bar{L}^{\b \a(2s-2)}$ is transverse linear.

Modulo an overall  normalisation factor, the longitudinal formulation for the massless superspin-$s$ multiplet is described by the action
\bea \label{4.4}
&&\mathbb{S}_{(s)}^{\|}[U_{\a(2s-2)},G_{\a(2s)}, \bar{G}_{\a(2s)}]
= \Big(-\hf\Big)^{s}\int 
\rd^{3|4}z\, \bm E \,
\bigg\{\frac{1}{8}U^{\a(2s-2)}\cD^{\g}({\bar \cD}^{2}-6\mu)\cD_{\g}U_{\a(2s-2)}
\non \\
&&+\frac{s}{2s+1}U^{\a(2s-2)}\Big(\cD^{\b} {\bar \cD}^{\g}
G_{\b \g \a(2s-2) }
-{\bar \cD}^{\b}{\cD}^{\g}\bar{G}_{\b \g \a(2s-2)} \Big) +\frac{s}{2s-1}  \bar{G}^{\a(2s)} G_{\a(2s)} \non \\
&&+ \frac{s}{2(2s+1)}\Big(G^{\a(2s)}G_{\a(2s)}+\bar{G}^{\a(2s)}\bar{G}_{\a(2s)}\Big) +2s(s+1)\mu\bar{\mu} U^{\a(2s-2)}U_{\a(2s-2)}
\bigg\}~,
\label{long-action-int}
\eea
which is invariant under the gauge transformations \eqref{4.3}.

\subsection{Reduction of gauge prepotentials to AdS$^{3|2}$}
We wish to reduce the gauge superfields \eqref{4.1} to $\cN=1$ AdS superspace. We start by reducing the superfield $U_{\a(2s-2)}$. Converting the transverse linear constraint of  $\g_{\a(2s-2)}$ \eqref{TL} to the real basis \eqref{realrep} gives
\be \label{TLReal}
\bm \nabla^{\2\b}\g_{\b\a(2s-3)}=\ri \bm \nabla^{\1\b} \g_{\b\a(2s-3)}~.
\ee
Taking a Taylor expansion of $\g_{\a(2s-2)}(\q^I)$ about $\q^\2$, then using \eqref{TLReal}, we find the independent $\q^\2$~-components of $\g_{\a(2s-2)}$ to be
\be
\g_{\a(2s-2)}|, \qquad \qquad \bm \nabla^\2{}_{(\a_1}\g_{\a_2...\a_{2s-1})}|~.
\ee

The gauge transformation \eqref{4.3a} allows us to choose the gauge conditions 
\be \label{4.8}
U_{\a(2s-2)}| = 0~.
\ee
It must be noted that the gauge condition \eqref{4.8} is less restrictive than those proposed in the analogous reduction procedure performed in Minkowski superspace,  as given in appendix B of \cite{HKO}. This was done in order to ensure that one of the decoupled $\cN=1$ actions coincides with \eqref{TIAction}. It can be shown that if one chooses to impose the same gauge conditions as detailed in \cite{HKO}, i.e
\be
U_{\a(2s-2)}| = 0~,  \qquad \bm \nabla^\2{}_{(\a_1}U_{\a_2 ... \a_{2s-1})}| = 0~,
\ee
 then one will obtain a $\cN=1$ action which would instead coincide with \eqref{GaugedTransIntAction}. 

The residual gauge symmetry which preserves the gauge condition \eqref{4.8} is described by the real unconstrained $\cN =1$ superfield
	\be
	\g_{\a(2s-2)}| = - \frac{\ri}{2}\z_{\a(2s-2)}~, \qquad \bar \z_{\a(2s-2)} = \z_{\a(2s-2)}~. \label{4.10}
	\ee 

We now wish to reduce the superfield $\bar{G}_{\a(2s)}$ to $\cN =1$ AdS superspace. Converting the longitudinal linear constraint of $\bar{G}_{\a(2s)}$ \eqref{LL} to the real basis yields
\be \label{4.12}
\bm \nabla^\2{}_{(\a_1}\bar{G}_{\a_2...\a_{2s+1})} = - \ri \bm \nabla^\1{}_{(\a_1}\bar{G}_{\a_1...\a_{2s+1})}~.
\ee
Upon reduction to $\cN=1$ AdS superspace, it follows from constraint \eqref{4.12} that $\bar{G}_{\a(2s)}$ has two independent $\q^\2$ -components:
\be \label{4.13}
\bar{G}_{\a(2s)}|~, \qquad \bm \nabla^{\2\b}\bar{G}_{\b\a(2s-1)}|~.
\ee
Making use of the gauge transformations \eqref{4.3b} and the real representation \eqref{realrep}, we find
\begin{subequations} \label{LIGvar}
	\bea
	\d \bar{G}_{\a(2s)} &=& \ri \Big (\bm \nabla^\1{}_{(\a_1}\bm \nabla^\2{}_{\a_2} + \bm \nabla_{(\a_1\a_2}\Big )\g_{\a_3...{\a_{2s})}}~,\\
	\bm \nabla^{\2\b}\d \bar{G}_{\b\a(2s-1)} &=& \frac{\ri}{2s} \Big ( 2s\bm \nabla_{(\a_1}{}^\b \bm \nabla^\2{}_{\a_2}\g_{\a_3...\a_{2s-1}\b)} + (4s^2-1)|\m| \bm \nabla^\2{}_{(\a_1}\g_{\a_2 ... \a_{2s-1})}~  \\ 
	&+& 2\ri(s+1) \bm \nabla_{\b(\a_1} \bm \nabla^{\1\b} \g_{\a_2 ... \a_{2s-1})} + 6\ri s(2s+1)|\m| \bm \nabla^\1{}_{(\a_1} \g_{\a_2 ... \a_{2s-1})} \non ~ \\
	&+&\frac{\ri}{2}(2s+1)(\bm \nabla^\1)^2\bm \nabla^\2{}_{(\a_1}\g_{\a_2 ... \a_{2s-1})} - 2\ri (s-1)\bm \nabla_{(\a_1}{}^\b\bm \nabla^\1{}_{\a_2}\g_{\a_3 ... \a_{2s-1})\b} \Big ) ~. \non
	\eea
\end{subequations}
Using the residual gauge freedom \eqref{4.10}, we can directly compute the gauge transformations of the $\cN =1$ superfields \eqref{4.13} from \eqref{LIGvar}
\begin{subequations} \label{4.15}
	\bea
	 \d \bar {G}_{\a(2s)}| &=& \frac{1}{2} \Big (\nabla_{(\a_1\a_2}\z_{\a_3 ... \a_{2s})} + 2 \ri \nabla_{(\a_1}\big (\t_{\a_2...\a_{2s})} + \ri \tilde{\t}_{\a_2 \dots \a_{2s})} \big ) \Big )~, \\
	\bm \nabla^{\2\b} \d \bar {G}_{\b\a(2s-1)}| &=& \frac{\ri}{2s} \Big ( 2s(2s+1)|\m|\nabla_{(\a_1}\z_{\a_2...\a_{2s-1})}  + (s+1)\nabla_{\b(\a_1}\nabla^\b \z_{\a_2 ... \a_{2s-1})} ~  \\
	&+& (4s^2-1)|\m| \big ( \t_{\a(2s-1)} + \ri  \tilde{\t}_{\a(2s-1)} \big ) + \frac{\ri}{2}(2s+1)\nabla^2\big ( \t_{\a(2s-1)} + \ri  \tilde{\t}_{\a(2s-1)} \big ) ~ \non \\ 
	&+& (2s-1) \nabla_{(\a_1}{}^\b \big ( \t_{\a_2 ... \a_{2s-1})\b} + \ri \tilde{\t}_{\a_2 ... \a_{2s-1})\b} \big ) - (s-1)\nabla_{(\a_1}{}^\b \nabla_{\a_2} \z_{\a_3 ... \a_{2s-1})\b} \Big ) ~ , \non 
	\eea
\end{subequations}
where we have defined
\be \label{RIGcomp}
\bm \nabla^\2{}_{(\a_1} \g _{\a_2 ... \a_{2s-1})} | = \t_{\a{(2s-1)}} + \ri \tilde { \t}_{\a{(2s-1)}}~.
\ee

It is useful to decompose the complex $\cN = 1$ superfields \eqref{4.13} into real and imaginary parts
\begin{subequations}
	\bea
	\bar {G}_{\a(2s)}| &=& -\frac{1}{2} \big ( H_{\a(2s)} - 2\ri \tilde{H}_{\a(2s)} \big  )~,\\
	\bm \nabla^{\2\b}\bar{ G}_{\b\a(2s-1)}|  &=& \F_{\a(2s-1)} -2 \ri \O_{\a(2s-1)}  ~.
	\eea
\end{subequations}
The gauge transformations of these four real $\cN=1$ superfields immediately follow from gauge transformations \eqref{4.15}
\begin{subequations}  \label{LongGaugeTrans}
	\bea
	\d H_{\a(2s)} &=& \nabla_{(\a_1} \r_{\a_2 ... \a_{2s})}~ , \\
	\d \tilde{H}_{\a(2s)} &=&  \nabla_{(\a_1} \t_{\a_2 ... \a_{2s})}~, \\
	\d \F_{\a(2s-1)}&=& - \frac{1}{8s} \Big ( 2(2s-1)\nabla_{(\a_1}{}^\b \r_{\a_2 ... \a_{2s-1})\b} + 2(4s^2-1)|\m| \r_{\a(2s-1)} \\
	&&+ (2s+1)\ri \nabla^2 \r_{\a(2s-1)} - 8\ri(4s^2-1)|\m|\nabla_{(\a_1}\z_{\a_2...\a_{2s-1})} \Big ) \non~, \\
	\d \O_{\a(2s-1)} &=& - \frac{1}{8s} \Big ( 2(2s-1) \nabla_{(\a_1}{}^\b \t_{\a_2 ... \a_{2s-1})\b} +2 (4s^2-1)|\m|\t_{\a(2s-1)} ~ \\
	&&+ \ri(2s+1)\nabla^2 \t_{\a(2s-1)} \Big )~, \non 
	\eea
\end{subequations}
where we have introduced the gauge parameter redefinition
\be
\r_{\a(2s-1)} := 2 \tilde{\t}_{\a(2s-1)} + \ri \nabla_{(\a_1}\z_{\a_2 ... \a_{2s-1})}~.
\ee

We are now left with the remaining unconstrained real $\cN = 1$ superfields
\begin{subequations} \label{4.9}
	\bea
	\cU_{\b;\a(2s-2)} := \ri \bm \nabla^{\2}_\b U_{\a(2s-2)}|~,\\
	V_{\a(2s-2)} := -\frac{\ri}{8s}(\bm \nabla^\2)^2U_{\a(2s-2)}|~.
	\eea
\end{subequations}

Utilising the gauge transformations \eqref{4.3a}, in conjunction with the residual gauge symmetry \eqref{4.10} and defintion \eqref{RIGcomp}, we find that the superfields \eqref{4.9} possess the following gauge freedom
\begin{subequations}\label{4.11}
	\bea \label{ULong}
	\d \cU_{\b;\a(2s-2)}  &=& -\r_{\b\a(2s-2)} + \ri \nabla_\b \z_{\a(2s-2)}~,\\
	\d V_{\a(2s-2)} &=& \frac{1}{2s}\nabla^\b\t_{\b\a(2s-2)}~.
	\eea
\end{subequations}

It then becomes apparent from gauge transformations \eqref{LongGaugeTrans} and \eqref{4.11} that we are dealing with two different gauge theories. The first of these gauge theories is formulated in terms of the dynamical variables
\be \label{4.17}
\cV^\perp_{(s)} = \big \{H_{\a(2s)}, \cU_{\b;\a(2s-2)} , \F_{\a(2s-1)} \big \}~,
\ee
which are defined modulo gauge transformations
\begin{subequations} \label{4.18}
	\bea\label{5.18a}
	\d H_{\a(2s)} &=& \nabla_{(\a_1} \r_{\a_2 ... \a_{2s})}  ~,\\
	\d \cU_{\b;\a(2s-2)}  &=& -\r_{\b\a(2s-2)} + \ri \nabla_\b \z_{\a(2s-2)}~,\label{5.18b}	\\
	\d \F_{\a(2s-1)}&=& -\frac{1}{8s} \Big ( 2(2s-1)\nabla_{(\a_1}{}^\b \r_{\a_2 ... \a_{2s-1})\b} + 2(4s^2-1)|\m| \r_{\a(2s-1)} \\
	&&+ (2s+1)\ri \nabla^2 \r_{\a(2s-1)} - 8\ri(4s^2-1)|\m|\nabla_{(\a_1}\z_{\a_2...\a_{2s-1})} \Big ) ~. \non
	\eea
\end{subequations}
The other gauge model is described by the superfields
\be \label{4.19}
\cV^\parallel_{(s)} = \big \{\tilde{H}_{\a(2s)}, V_{\a(2s-2)}, \O_{\a(2s-1)} \big \}~,
\ee
which possesses the following gauge freedom
\begin{subequations} \label{4.20}
	\bea 
	\d \tilde{H}_{\a(2s)} &=& \nabla_{(\a_1}\t_{\a_2...\a_{2s})}\label{4.20b} ~,\\
	\d V_{\a(2s-2)} &=& \frac{1}{2s}\nabla^\b\t_{\b\a(2s-2)}~, \label{4.20a}\\
	\d \O_{\a(2s-1)} &=& - \frac{1}{8s} \Big ( 2(2s-1) \nabla_{(\a_1}{}^\b \t_{\a_2 ... \a_{2s-1})\b} + 2(4s^2-1)|\m|\t_{\a(2s-1)} ~ \\
	&&+ \ri(2s+1)\nabla^2 \t_{\a(2s-1)} \Big )~. \non
	\eea
\end{subequations}
Applying the reduction prescription to the action \eqref{4.4}, we obtain two decoupled $\cN =1$ supersymmetric actions, which are formulated in terms of superfields \eqref{4.17} and \eqref{4.19} respectively.
\bea
\mathbb{S}^{\parallel}_{(s)} [U_{\a(2s-2)},G_{\a(2s)}, \bar{G}_{\a(2s)}] &=& S^\perp_{(s)}[H_{\a(2s)}, \cU_{\b;\a(2s-2)} ,  \F_{\a(2s-1)}] \non ~\\ 
&+& S^\parallel_{(s)}[\tilde{H}_{\a(2s)}, V_{\a(2s-2)}, \O_{\a(2s-1)}]~. \label{723}
\eea
The explicit form of the $\cN =1$ decoupled actions are provided in the following subsection.

Let us point out that there also appeared a new off-shell formulation for the massless integer superspin multiplet in (1,1) AdS superspace \cite{HKO}. This formulation proves to be a generalised version of the longitudinal action \eqref{long-action-int}, for the gauge-invariant action involves not only $U_{\a(2s-2)}, \J_{\a(2s-1)}$ and $\bar \J_{\a(2s-1)}$, but also new compensating superfields. Furthermore, the prepotential $\J_{\a(2s-1)}$ (associated to the longitudinal linear field strength $G_{\a(2s)}$), enjoys a larger gauge symmetry, which is that of the superconformal complex superspin-$s$ multiplet \cite{HKO}. Upon reduction to ${\cN}=1$ AdS superspace, we found that this new formulation decoupled into two ${\cN}=1$ supersymmetric higher-spin models which coincide with the right-hand side of eq.~\eqref{723}. 


\subsection{Massless higher-spin $\cN=1$ supermultiplets}
The first $\cN=1$ supersymmetric action which is described by the dynamical variables \eqref{4.17} takes the form
\bea 
&&S^\perp_{(s)}[H_{\a(2s)}, \cU_{\b;\a(2s-2)} ,  \F_{\a(2s-1)}] = \bigg ( -\frac{1}{2} \bigg )^s \frac{1}{4}  \int \rd^{3|2}z\,E\, \bigg \{-\ri \,\cU^{\b; \, \a(2s-2)} \Box \cU_{\b; \, \a(2s-2)} \non\\
&&-\hf \cU^{\b; \, \a(2s-2)} \de^2 \de_{\b}{}^{\g}\cU_{\g; \, \a(2s-2)} + \hf  (2s-3)|\m| \cU^{\b; \, \a(2s-2)} \de^2 \cU_{\b; \, \a(2s-2)}  \non\\
&&+ 2  (s-1)|\m| \cU_{\b;}{}^{\b; \, \a(2s-3)} \de^2 \cU^{\g;}{}_{\g \a(2s-3)} + \ri  (4s^2-1) |\m|^2\cU^{\b; \, \a(2s-2)}\cU_{\b; \, \a(2s-2)}  \non\\
&&- 8 \ri  (s-1)|\m|^2 \cU_{\b;}{}^{\b; \, \a(2s-3)} \cU^{\g;}{}_{\g \a(2s-3)} \non \\
&&- \frac{2s}{2s+1} \cU^{\b; \, \a(2s-2)} \Big( -2 \de^{\g \d} \de_{\g}H_{\d \b \a(2s-2)} + \de^{\g \d} \de_{\b} H_{\g \d \a(2s-2)}\non\\
&&+ \de^2 \F_{\b \a(2s-2)} + 2 \ri \, \de_{\b}{}^{\g} \F_{\a(2s-2)\g} \non\\
&&-4  s |\m| \de^{\g}H_{\g \b \a(2s-2)} + 2\ri (2s+1) |\m|  \F_{\b \a(2s-2)} \Big)\non\\
&&-\frac{s}{(2s+1)^2 (2s-1)}\Big( -\ri(3s+1) H^{\a(2s)} \de^2 H_{\a(2s)} \non\\
&&+ 2s(4s+1) H^{\a(2s-1)\b} \de_{\b}{}^{\g} H_{\a(2s-1)\g}+ 16s^2 H^{\a(2s)} \de_{(\a_1}\F_{\a_2 \dots \a_{2s})} \non\\
&&- 8\ri s \F^{\a(2s-1)} \F_{\a(2s-1)}-4s (4s^2+3s+1)|\m| H^{\a(2s)}H_{\a(2s)}\Big )
\bigg\}~.
\eea
It is apparent that the superfield $\F_{\a(2s-1)}$ is auxiliary, thus, upon elimination via its equation of motion,
\bea
\F_{\a(2s-1)} &=& -\frac{\ri}{8s}\Big ( 2\ri (2s+1)(4s^2-1)|\m| \cU_{\a(2s-1)} + (4s^2-1)\nabla^2 \cU_{\a(2s-1)}~ \\
&-&2\ri(4s^2-1) \nabla_{\b(\a_1} \cU^{\b;}{}_{\a_2 ...\a_{2s-1})} + 8s^2\nabla^\b H_{\b\a(2s-1)} \Big ) ~,\non
\eea
we arrive at the resulting action
\bea \label{5.24}
&&S^\perp_{(s)}[H_{\a(2s)}, \cU_{\b;\a(2s-2)}]= \bigg ( -\frac{1}{2} \bigg )^s \frac{1}{4}  \int \rd^{3|2}z\,E\, \bigg \{-\frac{\ri s(s-1)}{2s-1}H^{\a(2s)} \de^2 H_{\a(2s)} \non\\
&&- \frac{2s^2}{2s-1}H^{\a(2s-1) \b} \de_{\b}{}^{\g} H_{\a(2s-1)\g} -  \frac{4s^2(s-1)}{2s-1}|\m|H^{\a(2s)}H_{\a(2s)} \non\\
&&-2s \,\cU^{\d\a(2s-2)} \de^{\b \g} \Big( \de_{(\d} H_{\a_1 \dots \a_{2s-2} ) \b \g} - 2 \de_{\b} H_{\g \d \a(2s-2)}\Big)\non\\
&&+ \frac{4s(s-1)}{2s-1} \cU_{\l;}{}^{\l \a(2s-3)} \de^{\b \g} \de^{\d} H_{\b \g \d \a(2s-3)} - 8s^2|\m| \, \cU^{\b;\, \a(2s-2)} \de^{\g} H_{\g \b \a(2s-2)}\non\\
&&+ \cU^{\a(2s-1)} \bigg( -2\ri s \Box \cU_{\a(2s-1)} -s \de^2 \de^{\b}{}_{(\a_1} \cU_{\a_2 \dots \a_{2s-1})\b}\non\\
&&-\ri(s-1) \de_{(\a_1 \a_2}\de^{\b \g}\cU_{\a_3 \dots \a_{2s-1}) \b \g} + 2\ri \frac{(s-1)(2s-3)}{2s-1}  \de_{(\a_1 \a_2} \de_{\a_3}{}^{\b} \cU^{\g;}{}_{\a_4 \dots \a_{2s-1}) \b \g} \non\\
&&+ \frac{(s-1)(2s+1)}{2s-1} \de^2 \de_{(\a_1 \a_2} \cU^{\b;}{}_{\a_3 \dots \a_{2s-1}) \b} -s(2s+1)|\m|  \de^2 \cU_{\a(2s-1)} \non\\
&&- 4 \ri\,  s(2s-1) |\m|\de^{\b}{}_{(\a_1} \cU_{\a_2 \dots \a_{2s-1}) \b} + 2\ri  (s-1)(2s+1) |\m|\de_{(\a_1 \a_2} \cU^{\b;}{}_{\a_3 \dots \a_{2s-1}) \b} \non\\
&&- 2 \ri  s(2s-1)(2s+1)|\m|^2 \cU_{\a(2s-1)}\bigg) \non\\
&&+ \cU_{\l;}{}^{\l \a(2s-3)} \bigg(2 \ri \frac{(s-1)(6s-5)}{(2s-1)^2} \Box \cU^{\b;}{}_{\b \a(2s-3)} \non\\
&&- \frac{(s-1)(2s-3)}{(2s-1)^2} \Big(2 \ri (s-2) \de^{\b \g} \de_{(\a_1 \a_2} \cU^{\d;}{}_{\a_3 \dots \a_{2s-3}) \b \g \d} \non\\
&&+ \de^2 \de^{\b}{}_{(\a_1} \cU^{\g;}{}_{\a_2 \dots \a_{2s-3}) \b \g}\Big) + \frac{(s-1)(2s+1)}{(2s-1)}|\m| \de^2 \cU^{\b;}{}_{\b \a(2s-3)} \non\\
&&- 2 \ri  \frac{s-1}{2s-1} (4s^2 +16s-17)|\m|^2 \, \cU^{\b;}{}_{\b \a(2s-3)} \bigg)
\bigg\}~. 
\eea
Here we have made use of the notation $\cU_{\a(2s-1)} = \cU_{(\a_1;\a_2 ... \a_{2s-1})}$. It is a tedious but straightforward exercise to verify that the above ${\cN}=1$ action coincides with the transverse formulation for massless superspin-$s$ multiplet given by \eqref{TIAction}. To prove this, one first needs to introduce the ${\cN}=1$ field strength $\cW_{\b;\, \a(2s-2)}$ associated with the prepotential $\cU_{\b; \a(2s-2)}$,
\bea
\cW_{\b;\, \a(2s-2)} = -\ri( \de^{\g} \de_{\b} - 4\ri |\m| \d_{\b}{}^{\g})\, \cU_{\g; \, \a(2s-2)}~, \qquad \nabla^\b  \cW_{\b;\, \a(2s-2)}=0~.
\eea
The action \eqref{5.24} can then be written more compactly as
\bea
&&S^\perp_{(s)}[H_{\a(2s)}, \cU_{\b;\a(2s-2)}]
= \Big(-\hf \Big)^{s} \bigg (\frac{2s}{2s-1} \bigg )
\int \rd^{3|2}z\,
E\, \bigg\{\frac{1}{2} H^{\a(2s)} (\ri \nabla^2 +8 s |\m|) H_{\a(2s)}
\non \\
&& - \ri s \nabla_{\b} H^{\b \a(2s-1)} \nabla^{\g}H_{\g \a(2s-1)} 
-(2s-1) \cW^{\b ;\,\a(2s-2)} \nabla^{\g} H_{\g \b \a(2s-2)}
\non \\
&& -\frac{\ri}{2} (2s-1)\Big(\cW^{\b ;\, \a(2s-2)} \cW_{\b ;\, \a(2s-2)}+\frac{s-1}{s} \cW_{\b;}\,^{\b \a(2s-3)} \cW^{\g ;}\,_{\g \a(2s-3)} \Big) 
\non\\
&&
-2 \ri (2s-1) |\m| \, \cU^{\b ;\, \a(2s-2)} \cW_{\b ; \, \a(2s-2)}
\bigg\}~,
\eea
which is exactly \eqref{TIAction} modulo an overall normalisation factor. 

The other decoupled $\cN=1$ action which is formulated in terms of the dynamical variables \eqref{4.19} assumes the form
\bea \label{long_int_aux_action_11}
&&S^\parallel_{(s)}[\tilde{H}_{\a(2s)}, V_{\a(2s-2)}, \O_{\a(2s-1)}] = \bigg ( -\frac{1}{2} \bigg )^s \int \rd^{3|2}z\,E\, \bigg \{2s^2V^{\a(2s-2)} \big ( \ri \nabla^2 + 4|\m| \big ) V_{\a(2s-2)}~ \non\\
&&-\frac{4s^2}{2s+1}V^{\a(2s-2)} \big( {\nabla}^{\b \g}\tilde{H}_{\b \g \a(2s-2)} + 2{\nabla}^{\b} \O_{\b \a(2s-2)} \big )\non\\
&&+\frac{s}{(2s+1)^2 (2s-1)} 
\Big((2s+1)^2 \tilde{H}^{\a(2s)} (\ri \de^2 + 4 |\m|) \tilde{H}_{\a(2s)} \non \\
&&- 4\ri s(s+1) \de_{\b} \tilde{H}^{\b \a(2s-1)}\de^{\g}\tilde{H}_{\g \a(2s-1)} - 8s \tilde{H}^{\a(2s)} \de_{(\a_1}\O_{\a_2 \dots \a_{2s})}\non\\
&&+16 \ri s^2 \O^{\a(2s-1)}\O_{\a(2s-1)}\Big)
\bigg \}~. 
\eea
It is clear from the action \eqref{long_int_aux_action_11} that the superfield $\O_{\a(2s-1)}$ is auxiliary. Integrating it out by using its equation of motion,
\be
\O_{\a(2s-1)} = \frac{\ri}{4s} \Big ( (4s^2-1) \nabla_{(\a_1}V_{\a_2 ...\a_{2s-1})} - \nabla^\b \tilde{H}_{\b\a(2s-1)} \Big )~,
\ee
we find that the resulting action, modulo a normalisation factor, coincides with \eqref{action-t3}.


\subsection{Second supersymmetry transformations}
We begin by computing the second supersymmetry transformations \eqref{SST} of the superfields \eqref{4.17} and \eqref{4.19} 
\begin{subequations}
\bea
\d_\e U_{\a(2s-2)}| &=& -\ri \e^\b \cU_{\b;\a(2s-2)}~, \label{SSTLI1}\\
	\d_\e \cU_{\b;\a(2s-2)} &=& 4s \e_\b V_{\a(2s-2)} ~, \\
	\d_\e V_{\a(2s-2)} &=& \frac{\ri}{4s} \Big ( (2s-1)|\m|\e^\b \cU_{\b;\a(2s-2)} - \e^\b \nabla_\b{}^\g \cU_{\g;\a(2s-2)} ~ \\
	&+& 4(s-1)|\m|\e_{(\a_1}\cU^{\b;}{}_{\a_2...\a_{2s-2})\b} \Big ) ~, \non \\
	\d_\e H_{\a(2s)} &=& -2 \e^\b \nabla_{(\b}\tilde{H}_{\a(2s))} - \frac{8\ri s}{2s+1}\e_{(\a_1} \O_{\a_2 ... \a_{2s})}~,  \\
\d_\e \tilde{H}_{\a(2s)} &=& \hf \e^\b \nabla_{(\b}H_{\a(2s))}+\frac{2\ri s}{2s+1}\e_{(\a_1} \F_{\a_2 ... \a_{2s})} ~, \\ 
\d_\e \F_{\a(2s-1)} &=& \frac{1}{2(2s+1)} \e^\b \Big ( 2 (4s^2+3s+1)|\m|  \tilde{H}_{\b\a(2s-1)}  + \ri \nabla^2 \tilde{H}_{\b\a(2s-1)}~\\
&-& 2(4s+3) \nabla_\b{}^\g \tilde{H}_{\g\a(2s-1)}-2(2s-1) \nabla_{(\a_1}{}^\g \tilde{H}_{\a_2...\a_{2s-1})\b\g} ~ \non \\ 
&+&8s \nabla_{(\b}\O_{\a(2s-1))}  \Big ) ~, \non \\ 
\d_\e \O_{\a(2s-1)} &=& -\frac{1}{8(2s+1)}\e^\b \Big (4(4s^2+3s+1)|\m| H_{\b\a(2s-1)} + \ri \nabla^2 H_{\b\a(2s-1)} ~ \\
&-&2 (4s+3) \nabla_\b{}^\g H_{\g\a(2s-1)} - 2(2s-1) \nabla_{(\a_1}{}^\g H_{\a_2...\a_{2s-1})\b\g}  ~ \non \\
&+&8s \nabla_{(\b} \F_{(\a(2s-1))} \Big ) ~. \non
\eea
\end{subequations}

It is clear from the variation \eqref{SSTLI1} that the second supersymmetry transformation \eqref{SST} breaks the WZ gauge \eqref{4.8}. To restore this gauge condition, it is necessary to supplement the second supersymmetry transformation with the $\e$-dependent gauge transformations:
\bsubeq
\bea
\d_{\g (\e)} U_{\a(2s-2)} &=& \g_{\a(2s-2)}(\e) + \bar{\g}_{\a(2s-2)}(\e) ~,\\
\d_{\g (\e)} G_{\a(2s)} 
	&=& \bar{\cD}_{(\a_{1}} \cD_{\a_{2}}\bar{\g}_{\a_3 \dots \a_{2s})} (\e) \non\\
	&=& \ri \Big (\bm \nabla^\1{}_{(\a_1}\bm \nabla^\2{}_{\a_2} + \bm \nabla_{(\a_1\a_2}\Big)\g_{\a_3...{\a_{2s})}} (\e)~,
\eea
\esubeq
with
\be
\g_{\a(2s-2)}(\e)| = \frac{\ri}{2}\e^\b \cU_{\b;\a(2s-2)}~.
\ee
The modified second supersymmetry transformations have the form
\bsubeq \label{MSSTLI}
\bea 
\hat{\d}_\e U_{\a(2s-2)} &=& \d_\e U_{\a(2s-2)} + \d_{\g (\e)} U_{\a(2s-2)}~,\\
\hat{\d}_\e G_{\a(2s)} &=& \d_\e G_{\a(2s)} + \d_{\g (\e)} G_{\a(2s)}~.
\eea
\esubeq
They act on the ${\cN}=1$ gauge superfields \eqref{4.17} and \eqref{4.19} by the rule
\begin{subequations}
\bea
\hat{\d}_\e \cU_{\b;\a(2s-2)} &=& 4s \e_\b V_{\a(2s-2)} ~, \\
\hat{\d}_\e V_{\a(2s-2)} &=& - \frac{\ri}{8s} \Big ( 2(2s+1)|\m|\e^\b \cU_{\b;\a(2s-2)} +2 \e^\b \nabla_\b{}^\g \cU_{\g;\a(2s-2)} \\
&-&8(s-1)|\m|\e_{(\a_1}\cU^{\b;}{}_{\a_2 ... \a_{2s-2})\b} + \ri \e^\b \nabla^2 \cU_{\b;\a(2s-2)} \Big )~, \non \\
\hat{\d}_\e H_{\a(2s)} &=& -2 \e^\b \nabla_{(\b}\tilde{H}_{\a(2s))} - \frac{8\ri s}{2s+1}\e_{(\a_1} \O_{\a_2 ... \a_{2s})} ~, \non \\
\hat{\d}_\e \tilde{H}_{\a(2s)} &=& \hf \e^\b \nabla_{(\b}H_{\a(2s))}+\frac{2\ri s}{2s+1}\e_{(\a_1} \F_{\a_2 ... \a_{2s})} 
+\ri |\m| \e_{(\a_1}\cU_{\a_2 ... \a_{2s})} \\ &-& \frac{\ri}{2}\e_\b \nabla_{(\a_1 \a_2}\cU^{\b;}{}_{\a_3...\a_{2s})} ~ , \non \\
\hat{\d}_\e \F_{\a(2s-1)} &=& \frac{1}{2(2s+1)} \e^\b \Big ( 2 (4s^2+3s+1)|\m|  \tilde{H}_{\b\a(2s-1)}  + \ri \nabla^2 \tilde{H}_{\b\a(2s-1)}~\\
&-& 2(4s+3) \nabla_\b{}^\g \tilde{H}_{\g\a(2s-1)}-2(2s-1) \nabla_{(\a_1}{}^\g \tilde{H}_{\a_2...\a_{2s-1})\b\g} ~ \non \\ 
&+&8s \nabla_{(\b}\O_{\a(2s-1))}  \Big ) ~, \non \\ 
\hat{\d}_\e \O_{\a(2s-1)} &=& -\frac{1}{8(2s+1)}\e^\b \Big (4(4s^2+3s+1)|\m| H_{\b\a(2s-1)} + \ri \nabla^2 H_{\b\a(2s-1)} ~ \\
&-&2 (4s+3) \nabla_\b{}^\g H_{\g\a(2s-1)} - 2(2s-1) \nabla_{(\a_1}{}^\g H_{\a_2...\a_{2s-1})\b\g}  ~ \non \\
&+&8s \nabla_{(\b} \F_{(\a(2s-1))} \Big ) ~ \non \\
&+& \frac{1}{4s}\Big ( (2s+1) |\m| \e^\b \nabla_\b \cU_{\a(2s-1)} +(s+1)\e_\b \nabla_{\g(\a_1}\nabla^\g\cU^{\b;}{}_{\a_2...\a_{2s-1})} \non ~ \\
&+&2(s-1)|\m| \e_{(\a_1} \nabla_{\a_2} \cU^{\b;}{}_{\a_3 ...\a_{2s-1})\b} - (s-1) \e_\b \nabla_{(\a_1}{}^\g \nabla_{\a_2}\cU^{\b;}{}_{\a_3 ...\a_{2s-1})\g} \non \\
&+&(6s^2+3s+2)|\m|\e_\b \nabla_{(\a_1}\cU^{\b;}{}_{\a_2...\a_{2s-1})} \Big )~. \non
\eea
\end{subequations}

\section{Massless integer superspin: Transverse formulation} \label{Section8}
We now apply the same reduction procedure to the off-shell transverse formulation for the massless multiplets of integer superspin.
\subsection{Transverse formulation}
For any integer $s\geq 1$, the transverse formulation for the massless superspin-$s$ multiplet is realised in terms of the following dynamical variables
\be
\cV^\perp_{(s)} = \big \{U_{\a(2s-2)}, \G_{\a(2s)},\bar{\G}_{\a(2s)} \big  \}~.
\ee
The real superfield $U_{\a(2s-2)}$ is the same as in \eqref{4.1} and the complex superfield $\G_{\a(2s)}$ is transverse linear. The transverse formulation for the massless superspin-$s$ multiplet was derived by performing a superfield duality transformation on the longitudinal model reviewed in section \ref{Section7}. Thus, it can be shown that the transverse action takes the form
\bea \label {5.3}
&&\mathbb{S}^\perp_{(s)}[U_{\a(2s-2)}, \G_{\a(2s)}, \bar{\G}_{\a(2s)} ] = \Big (-\frac{1}{2} \Big )^s \int \text{d}^{3|4}z~\bm E~\bigg \{ 2s(s+1)\m\bar{\m}U^{\a(2s-2)}U_{\a(2s-2)} \non \\
&&+\frac{1}{8} U^{\a(2s-2)}\cD^\g(\bar{\cD}^2-6\m)\cD_\g U_{\a(2s-2)} - \ri U^{\a_1 ... \a_{2s-2}}\cD^{\a_{2s-1}\a_{2s}}(\G_{\a(2s)}-\bar{\G}_{\a(2s)}) ~\non \\
&&-\frac{2s-1}{16(2s+1)}\Big ( 8s\cD^{\a_1\a_2}U^{\a_3 ...\a_{2s}}\cD_{(\a_1\a_2}U_{\a_3...\a_{2s})} ~ \non \\
&&+ [\cD^{\a_1},\bar{\cD}^{\a_2}]U^{\a_3 ...\a_{2s}}[\cD_{(\a_1},\bar{\cD}_{\a_2}]U_{\a_3 ...\a_{2s})}\Big )~~ \non \\
&&- \frac{2}{2s-1}\bar{\G}^{\a(2s)}\G_{\a(2s)} +\frac{1}{2s+1}\big (\G^{\a(2s)}\G_{\a(2s)} + \bar{\G}^{\a(2s)}\bar{\G}_{\a(2s)}\big )\bigg \}~,
\eea
which is invariant under the gauge transformations
\begin{subequations}
	\bea \label{7.3a}
	\d_L U_{\a(2s-2)} &&= \cD^\b L_{\b\a(2s-2)} - \bar{\cD}^\b \bar{L}_{\b\a(2s-2)} \equiv \bar{\g}_{\a(2s-2)} +\g_{\a(2s-2)}~, \\
	\d_L \G_{\a(2s)} &&= -\frac{1}{4}(\bar{\cD}^2 + 4s\m)\cD_{(\a_1}\bar{L}_{\a_2 ... \a_{2s})} + \frac{\ri}{2}(2s+1)\bar{\cD}^\g \cD_{(\g\a_1}\bar{L}_{\a_2 ... \a_{2s})} \non \\ 
	&&\equiv \frac{1}{2}\cD_{(\a_1}\bar{\cD}_{\a_2}\g_{\a_3 ...\a_{2s})} - \frac{\ri}{2}(2s-1)\cD_{(\a_1\a_2}\g_{\a_3 ... \a_{2s})}~, \label{5.4b}
	\eea
\end{subequations}
where the gauge parameter $L_{\a(2s-1)}$ is complex unconstrained and the superfield $\g_{\a(2s-2)}$ is transverse linear, as in \eqref{4.3a}.

\subsection{Reduction of gauge prepotentials to AdS$^{3|2}$}
We begin by reducing the real superfield $U_{\a(2s-2)}$. Following the prescription endorsed in section \ref{Section7}, we have the freedom to impose the gauge condition
\be \label{GcU}
U_{\a(2s-2)}| = 0~.
\ee
The residual gauge symmetry which preserves this gauge is described by the real unconstrained $\cN=1$ superfield
\be
\g_{\a(2s-2)} | = -\frac{\ri}{2}\z_{\a(2s-2)}~, \qquad \z_{\a(2s-2)} = \bar{\z}_{\a(2s-2)}~.
\ee
As a result of choosing the gauge condition \eqref{GcU}, we are left with the remaining independent $\cN=1$ superfields with respect to $U_{\a(2s-2)}$
\begin{subequations} \label{TVar}
\bea 
\cU_{\b;\a(2s-2)} : &=& \ri \bm \nabla^\2_\b U_{\a(2s-2)}| ~, \\
V_{\a(2s-2)} :&=& - \frac{\ri}{8s}(\bm \nabla^\2)^2 U_{\a(2s-2)}|~.
\eea
\end{subequations}
Using \eqref{7.3a} enables the computation of the corresponding gauge transformations
\begin{subequations}
\bea \label{OldUTrans}
\d \cU_{\b;\a(2s-2)} &=& -2s \r_{\b\a(2s-2)} + \ri \nabla_\b \z_{\a(2s-2)} + \ri (2s-1)\nabla_{(\b}\z_{\a(2s-2))}~, \\
\d V_{\a(2s-2)} &=& \frac{1}{2s}\nabla^\b \t_{\b\a(2s-2)}~, \label{Vtrans}
\eea 
\end{subequations}
where we have introduced the following definition
\bea
\bm \nabla^\2{}_{(\a_1}\g_{\a_2 ... \a_{2s-1})}| = \t_{\a(2s-1)}
+ \ri s \tilde{\t}_{\a(2s-1)} ~, 
\eea
and gauge parameter redefinition
\be
\r_{\a(2s-1)} = \tilde{\t}_{\a(2s-1)} + \ri \nabla_{(\a_1}\z_{\a_2 ... \a_{2s-1})}~.
\ee

It is immediately obvious that the gauge transformation of $\cU_{\b;\a(2s-2)}$ does not coincide with its analogue from the longitudinal formulation \eqref{ULong}. Addressing this, we introduce the following ${\cN}=1$ superfield redefinition
\be 
\tilde{\cU}_{\b;\a(2s-2)} = \cU_{\b;\a(2s-2)} - \frac{2s-1}{2s}\cU_{\b\a(2s-2)}~.
\ee
By using the gauge transformation \eqref{OldUTrans}, it can be shown that $\tilde{\cU}_{\b;\a(2s-2)}$ possesses the desired gauge freedom
\be\label{NewUGT}
\d \tilde{\cU}_{\b;\a(2s-2)}  = - \r_{\b\a(2s-2)} + \ri \nabla_\b \z_{\a(2s-2)}~.
\ee

Next, we wish to go about reducing $\G_{\a(2s)}$ to $\cN=1$ AdS superspace. We begin by converting the transverse linear constraint $\G_{\a(2s)}$\eqref{TL} to the real representation \eqref{realrep}
\be \label{5.5}
\bm \nabla^{\2\b}\G_{\b\a(2s-1)} = \ri \bm \nabla^{\1\b} \G_{\b\a(2s-1)}~.
\ee
Taking a Taylor expansion of $\G_{\a(2s)}(\q^I)$ about $\q^\2$, and using constraint \eqref{5.5}, we find that the independent $\q^\2$~-components of $\G_{\a(2s)}$ are
\be \label{5.6}
\G_{\a(2s)}|~, \qquad \bm \nabla^\2{}_{(\a_1}\G_{\a_2 ... \a_{2s+1})}|~.
\ee
Using the gauge transformations \eqref{5.4b} and the real representation \eqref{realrep}, we find\begin{subequations} \label{gaugetransgamma}
	\bea 
	\d \G_{\a(2s)} &=& -\frac{\ri}{2} \bm \nabla^\1{}_{(\a_1}\bm \nabla^\2{}_{\a_2}\g_{\a_3 ... \a_{2s})} - \ri s \bm \nabla_{(\a_1 \a_2}\g_{\a_3 ... \a_{2s})}~, \\
	\bm  \nabla^\2{}_{(\a_1} \d \G_{\a_2 ... \a_{2s+1})} &=& - \frac{1}{2}\bm \nabla_{(\a_1\a_2}\bm \nabla^\1{}_{\a_3}\g_{\a_4 ...\a_{2s+1})} - \ri s\bm \nabla_{(\a_1 \a_2}\bm \nabla^\2{}_{\a_3} \g_{\a_4 ... \a_{2s+1})}~.
	\eea
\end{subequations}
Making use of \eqref{gaugetransgamma}  and the gauge freedom \eqref{4.10} allows us to directly compute the gauge transformations of the $\cN=1$ complex superfields \eqref{5.6}
\begin{subequations} \label{gtgamma}
	\bea
	\d \G_{\a(2s)}| &=& - \frac{\ri}{2} \nabla_{(\a_1}\t_{\a_2 ...\a_{2s})} + \frac{s}{2} \nabla_{(\a_1} \tilde{\t}_{\a_2 ... \a_{2s})}- \frac{s}{2} \nabla_{(\a_1 \a_2}\z_{\a_3 ... \a_{2s})} ~, \\
	\bm \nabla^\2{}_{(\a_1}\d  \G_{\a_2 ... \a_{2s+1})}| &=& \frac{\ri}{4} \nabla_{(\a_1 \a_2} \nabla_{\a_3}\z_{\a_4 ... \a_{2s+1})} - \ri s \nabla_{(\a_1 \a_2}\t_{\a_3 ... \a_{2s+1})} \\
	&+& s^2 \nabla_{(\a_1 \a_2}\tilde \t_{\a_3 ... \a_{2s+1})}~. \non 
	\eea
\end{subequations}
Let us split the complex $\cN=1$ superfields \eqref{5.6} into real and imaginary parts\begin{subequations} \label{5.8}
	\bea
	\G_{\a(2s)}| = \frac{1}{2} \Big ( sH_{\a(2s)} - \ri \tilde{H}_{\a(2s)} \Big )~, \\
	\bm \nabla^\2{}_{(\a_1}\G_{\a_2 ... \a_{2s+1})}| = \F_{\a(2s+1)} - \ri s \O_{\a(2s+1)}~.
	\eea
\end{subequations}
The gauge transformations of these four real $\cN=1$ superfields can be read off from the gauge transformations \eqref{gaugetransgamma}
\begin{subequations} \label{ReducedTransGC}
	\bea 
	\d H_{\a(2s)} &=& \nabla_{(\a_1}\r_{\a_2 ... \a_{2s})}~, \\
	\d \tilde{H}_{\a(2s)} &=& \nabla_{(\a_1}\t_{\a_2 ... \a_{2s})}~, \\
	\d \F_{\a(2s+1)} &=& s^2 \nabla_{(\a_1\a_2}\r_{\a_3 ... \a_{2s+1})} -\frac{\ri}{4}(4s^2-1)\nabla_{(\a_1\a_2}\nabla_{\a_3}\z_{\a_4...\a_{2s+1})}~,\\
	\d \O_{\a(2s+1)} &=& \nabla_{(\a_1\a_2}\t_{\a_3...\a_{2s+1})}~.
	\eea
\end{subequations}

It then follows from gauge transformations \eqref{Vtrans}, \eqref{NewUGT} and \eqref{ReducedTransGC} that we are indeed dealing with two different gauge theories. The first of the models is formulated in terms of the dynamical variables
\be \label{5.10}
\cV^\perp_{(s)} = \big \{  H_{\a(2s)}, \tilde{\cU}_{\b;\a(2s-2)} , \F_{\a(2s+1)} \big \}~,
\ee
which are defined modulo gauge transformations
\begin{subequations} \label{7.21}
	\bea
	\d H_{\a(2s)} &=&  \nabla_{(\a_1}\r_{\a_2...\a_{2s})}~, \label{5.11b} \\
	\d \tilde{\cU}_{\b;\a(2s-2)}  &=& - \r_{\b\a(2s-2)} + \ri \nabla_\b \z_{\a(2s-2)}~, \label{5.11a} \\
	\d \F_{\a(2s+1)} &=& s^2\nabla_{(\a_1\a_2}\r_{\a_3 ... \a_{2s})} - \frac{\ri}{4}(4s^2-1)\nabla_{(\a_1\a_2}\nabla_{\a_3}\z_{\a_4 ...\a_{2s-1})}~.
	\eea
\end{subequations}
The other gauge theory is described by the superfields
\be \label{5.12}
\cV^\parallel_{(s)} = \big \{ \tilde H_{\a(2s)}, V_{\a(2s-2)}, \O_{\a(2s+1)} \big \}~,
\ee
which has the corresponding the gauge freedom
\begin{subequations} \label{7.22}
	\bea
	\d \tilde{H}_{\a(2s)} &=& \nabla_{(\a_1}\t_{\a_2 ... \a_{2s})}~, \label{5.13a} \\
	\d V_{\a(2s-2)} &=& \frac{1}{2s}\nabla^\b \t_{\b\a(2s-2)}~, \label{5.13b} \\
	\d \O_{\a(2s+1)} &=& \nabla_{(\a_1\a_2}\t_{\a_3 ... \a_{2s+1})}~.
	\eea
	From the perspective of $\cN=1$ AdS superspace, the reduction of the action \eqref{5.3} gives rise to two decoupled supersymmetric actions, which are formulated in terms of superfields \eqref{5.10} and \eqref{5.12} respectively
\end{subequations}
\bea \label{t-int}
\mathbb{S}^\perp_{(s)}[U_{\a(2s-2)}, \G_{\a(2s)},\bar{\G}_{\a(2s)} ] &=& S^\perp_{(s)}[ H_{\a(2s)}, \tilde{\cU}_{\b;\a(2s-2)} , \F_{\a(2s+1)} ] ~ \non \\
&+& S^\parallel_{(s)}[\tilde{H}_{\a(2s)}, V_{\a(2s-2)}, \O_{\a(2s+1)}]~.
\eea
We provide the explicit realisations of these two $\cN=1$ decoupled actions in the following subsection.


\subsection{Massless higher-spin $\cN=1$ supermultiplets}
The first of the decoupled $\cN=1$ supersymmetric actions which is formulated in terms of the dynamical variables \eqref{5.10} takes the form 
\bea \label{5.15}
&&S^\perp_{(s)}[ H_{\a(2s)}, \tilde{\cU}_{\b;\a(2s-2)} , \F_{\a(2s+1)} ] = \bigg ( - \frac{1}{2} \bigg )^s \frac{1}{4}  \int \rd^{3|2}z~E~ \bigg \{ \tilde{\cU}^{\a(2s-1)} \Big ( -2\ri s (4s-1) \Box \tilde{\cU}_{\a(2s-1)}  ~ \non \\ 
&&-s \de^2 \de^{\b}{}_{(\a_1} \tilde{\cU}_{\a_2 \dots \a_{2s-1})\b} +4\ri s (s-1)^2 \de_{(\a_1 \a_2}\de^{\b \g}\tilde{\cU}_{\a_3 \dots \a_{2s-1}) \b \g} -s(2s+1)|\m|\nabla^2\tilde{\cU}_{\a(2s-1)} ~ \non\\
&&+  \frac{2\ri (s-1)(2s-3)}{2s-1}  \de_{(\a_1 \a_2} \de_{\a_3}{}^{\b} \tilde{\cU}^{\g;}{}_{\a_4 \dots \a_{2s-1}) \b \g} + \frac{(s-1)(2s+1)}{2s-1} \de^2 \de_{(\a_1 \a_2} \tilde{\cU}^{\b;}{}_{\a_3 \dots \a_{2s-1}) \b} ~  \non \\
&&- 4 \ri  s(2s-1) |\m|\de_{(\a_1}{}^{\b} \, \tilde{\cU}_{\a_2 \dots \a_{2s-1}) \b} + 2\ri  (s-1)(2s+1) |\m|\de_{(\a_1 \a_2} \tilde{\cU}^{\b;}{}_{\a_3 \dots \a_{2s-1}) \b} \non \\
&&+ 2\ri s(2s-1)(2s+1)(4s-3)|\m|^2 \, \tilde{\cU}_{\a(2s-1)}\Big ) ~ \non \\
&&+ \tilde{\cU}_{\b;}{}^{\b \a(2s-3)} \bigg(2 \ri \frac{(s-1)(6s-5)}{(2s-1)^2} \Box \tilde{\cU}^{\g;}{}_{\g \a(2s-3)}+ \frac{(s-1)(2s+1)}{(2s-1)}|\m| \de^2 \tilde{\cU}^{\g;}{}_{\g \a(2s-3)} ~ \non \\
&&- \frac{(s-1)(2s-3)}{(2s-1)^2} \Big(2 \ri (s-2) \de^{\g \d} \de_{(\a_1 \a_2} \tilde{\cU}^{\l;}{}_{\a_3 \dots \a_{2s-3}) \g \d \l} + \de^2 \de^{\g}{}_{(\a_1} \tilde{\cU}^{\d;}{}_{\a_2 \dots \a_{2s-3}) \g\d}\Big) \non\\
&&- \frac{2\ri(s-1)}{2s-1} (4s^2 +16s-17)|\m|^2 \, \tilde{\cU}^{\g;}{}_{\g \a(2s-3)} \bigg) \non \\
&&-\frac{8s}{(2s+1)^2(2s-1)}  \Big ( s(2s^3-2s-1)|\m|H^{\a(2s)}H_{\a(2s)} +s^2(s+1)H^{\b\a(2s-1)}\nabla_\b{}^\g H_{\g\a(2s-1)}\non \\
&&+ \frac{\ri s}{4}(2s^2-1)H^{\a(2s)}\nabla^2H_{\a(2s)}  + (2s+1)H^{\a(2s)}\nabla^\b \F_{\b\a (2s)} + 2\ri  (2s+1)\F^{\a(2s+1)}\F_{\a(2s+1)} \Big ) \non \\
&&+\frac{8s}{2s+1}\tilde{\cU}^{\a(2s-1)} \Big ( s\nabla^{\b\g}\nabla_\b H_{\g\a(2s-1)}-s\nabla^{\b\g}\nabla_{(\a_1}H_{\a_2 ... \a_{2s-1})\b\g}+s(2s+1)|\m|\nabla^\b H_{\b\a(2s-1)} \non \\
&&+ \ri (2s+1) \nabla^{\b\g}\F_{\b\g\a(2s-1)} \Big ) + \frac{s(s-1)}{2s-1}\tilde{\cU}_{\b;}{}^{\b\a(2s-3)}\nabla^{\b\g}\nabla^\d H_{\b\g\d\a(2s-3)}  \bigg \} ~, 
\eea
which is invariant under the gauge transformations \eqref{7.21}. It is apparent that the superfield $\F_{\a(2s+1)}$ is auxiliary, so upon elimination via its equation of motion,
\be
\F_{\a(2s+1)} = - \frac{1}{4}\Big ( (4s^2-1)\nabla_{(\a_1\a_2}\tilde{\cU}_{\a_3...\a_{2s+1})} + \ri \nabla_{(\a_1}H_{\a_2 ... \a_{2s+1})} \Big ) ~,
\ee
one arrives at the action, which, up to an overall normalisation factor, coincides with \eqref{5.24}.

The second $\cN=1$ gauge theory formulated in terms of the dynamical variables \eqref{5.12} assumes the form
\bea \label{5.16}
&&S^\parallel_{(s)}[\tilde{H}_{\a(2s)}, V_{\a(2s-2)}, \O_{\a(2s+1)}] = \Big ( - \frac{1}{2} \Big )^s \int \rd^{3|2}z~ E ~ \Big \{ 4s^2(s+1)|\m|V^{\a(2s-2)}V_{\a(2s-2)}  \\
&&+2s(s-1)V^{\a(2s-3)\b}\nabla_\b{}^\g V_{\g\a(2s-3)} -2sV^{\a(2s-2)}\nabla^{\b\g}\tilde{H}_{\b\g\a(2s-2)} + \ri s^2 V^{\a(2s-2)}\nabla^2 V_{\a(2s-2)} \non \\
&&-\frac{1}{2(2s+1)^2(2s-1)} \Big ( 4\ri s^2(2s+1)\O^{\a(2s+1)}\O_{\a(2s+1)} +2s(4s+1)\tilde{H}^{\a(2s-1)\b}\nabla_\b{}^\g \tilde{H}_{\g\a(2s-1)} \non ~\\
&&- \ri s \tilde{H}^{\a(2s)}\nabla^2 \tilde{H}_{\a(2s)}-4s(4s^2+3s+1)|\m|\tilde{H}^{\a(2s)}\tilde{H}_{\a(2s)} + 8s^2(2s+1)\tilde{H}^{\a(2s)}\nabla^\b\O_{\b\a(2s)} \Big ) \Big \}~. \non
\eea
The superfield $\O_{\a(2s+1)}$ is auxiliary, so integrating it out using its equation of motion, 
\be
\O_{\a(2s+1)} = -\ri \nabla_{(\a_1}\tilde{H}_{\a_2...\a_{2s+1})}~,
\ee
we obtain action \eqref{action-t3} up to an overall normalisation factor.


\subsection{Second supersymmetry transformations}
We begin by computing the second supersymmetry transformations of the real $\cN=1$ fields \eqref{5.10} and \eqref{5.12}
\begin{subequations}
	\bea
	\d_\e U_{\a(2s-2)} |&=&-2\ri s \e^\b \tilde{\cU}_{\b\a(2s-2)} + \frac{2\ri(s-1)}{2s-1}\e_{(\a_1}\tilde{\cU}^{\b;}{}_{\a_2...\a_{2s-2})\b} ~, \label{TLBG}\\
	\d_\e  \tilde{\cU}_{\b;\a(2s-2)} &=& 2(2s-1)\e_\b V_{\a(2s-2)} -4(s-1)\e_{(\a_1}V_{\a_2...\a_{2s-2})\b}~, \\
	\d_\e V_{\a(2s-2)} &=& \frac{\ri}{2s} \Big ( s(2s-1)|\m|\e^\b \tilde{\cU}_{\b\a(2s-2)}+(s-1)|\m|\e_{(\a_1} \tilde{\cU}^{\b ;}{}_{\a_2 ... \a_{2s-2})\b}  ~ \\
	&-& s\e^\b \nabla_\b{}^\g \tilde{\cU}_{\g\a(2s-2)} +\frac{s-1}{2s-1}\e^\b \nabla_{\b(\a_1}\tilde{\cU}^{\g ;}{}_{\a_2 ... \a_{2s-2})\g} \Big )~, \non \\
	\d_\e H_{\a(2s)} &=& -2\ri \e^\b \O_{\b\a(2s)} - \frac{2 }{2s+1}\e_{(\a_1}\nabla^\b \tilde{H}_{\a_2 ...\a_{2s})\b}   ~, \\
	\d_\e \tilde{H}_{\a(2s)} &=& 2 \ri \e^\b\F_{\b\a(2s)} + \frac{2 s^2}{2s+1}\e_{(\a_1}\nabla^\b H_{\a_2 ... \a_{2s})\b} ~ , ~\\
	\d_\e \F_{\a(2s+1)} &=& \hf \e^\b \nabla_{\b(\a_1} \tilde{H}_{\a_2 ... \a_{2s+1})} -s \e_{(\a_1}\nabla^\b \O_{\a_2 ... \a_{2s+1})\b} ~\\
	&+&\frac{1}{4(2s+1)} \Big ( \ri \e_{(\a_1}\nabla^2 \tilde{H}_{\a_2 ... \a_{2s+1})} -4s \e_{(\a_1} \nabla_{\a_2}{}^\b \tilde{H}_{\a_3 ... \a_{2s+1})\b} \non ~\\
	&+& 4(4s^2 + 5s +2)|\m| \e_{(\a_1}\tilde{H}_{\a_2 ... \a_{2s+1})} \Big )~, \non \\
	\d_\e \O_{\a(2s+1)} &=& - \hf \e^\b \nabla_{\b(\a_1}H_{\a_2 ... \a_{2s+1})} + \frac{1}{s}\e_{(\a_1}\nabla^\b \F_{\a_2 ... \a_{2s+1})\b} ~ \\
	&-& \frac{1}{4(2s+1)} \Big ( \ri \e_{(\a_1} \nabla^2 H_{\a_2 ... \a_{2s+1})}
	 -4s\e_{(\a_1}\nabla_{\a_2}{}^\b H_{\a_3 ...\a_{2s+1})\b} ~ \non  \\
	 &+&4(4s^2+5s+2)|\m|\e_{(\a_1}H_{\a_2 ... \a_{2s+1})}\Big )~. \non 
	\eea
\end{subequations}

It is clear from the variation \eqref{TLBG} that the second supersymmetry transformation \eqref{SST} breaks the gauge condition \eqref{GcU}. We should then supplement the second supersymmetry transformation with the $\e$-dependent gauge transformations:
\bsubeq
\bea
\d_{\g (\e)} U_{\a(2s-2)} &=& \g_{\a(2s-2)}(\e) + \bar{\g}_{\a(2s-2)}(\e) ~,\\
\d_{\g (\e)} \G_{\a(2s)} 
	&=& \hf {\cD}_{(\a_{1}} \bar \cD_{\a_{2}}{\g}_{\a_3 \dots \a_{2s})} (\e)-\frac{\ri}{2} (2s-1) \cD_{(\a_1 \a_2} \g_{\a_3 \dots \a_{2s})} (\e) ~,
\eea
\esubeq
with
\be
\g_{\a(2s-2)}(\e)| = \ri s \e^\b \tilde{\cU}_{\b\a(2s-2)} - \frac{\ri(s-1)}{2s-1}\e_{(\a_1}\tilde{\cU}^{\b;}{}_{\a_2...\a_{2s-2})\b}~.
\ee
The modified second supersymmetry transformations have the form
\bsubeq 
\bea 
\hat{\d}_\e U_{\a(2s-2)} &=& \d_\e U_{\a(2s-2)} + \d_{\g (\e)} U_{\a(2s-2)}~,\\
\hat{\d}_\e \G_{\a(2s)} &=& \d_\e \G_{\a(2s)} + \d_{\g (\e)} \G_{\a(2s)}~.
\eea
\esubeq
They act on the ${\cN}=1$ gauge superfields \eqref{5.10} and \eqref{5.12} by the rule
\begin{subequations}
\bea
\hat{\d}_\e  \tilde{\cU}_{\b;\a(2s-2)} &=&  2(2s-1)\e_\b V_{\a(2s-2)} -4(s-1)\e_{(\a_1}V_{\a_2...\a_{2s-2})\b}~, \\
\hat{\d}_\e V_{\a(2s-2)} &=& -\frac{\ri}{4} \Big ( 2(2s+1)|\m| \e^\b \tilde{\cU}_{\b\a(2s-2)} + 2 \e^\b \nabla_\b{}^\g \tilde{\cU}_{\g\a(2s-2)} ~ \\
&+& \ri\e^\b \nabla^2 \tilde{\cU}_{\b\a(2s-2)} - \frac{(s-1)}{s(2s-1)} \big ( 2(6s-1)|\m| \e_{(\a_1}\tilde{\cU}^{\b;}{}_{\a_2 ... \a_{2s-2})\b} \non ~ \\
&+&2 \e^\b \nabla_{\b (\a_1}\tilde{\cU}^{\g;}{}_{\a_2...\a_{2s-2})\g} +\ri \e_{(\a_1}\nabla^2\tilde{\cU}^{\b;}{}_{\a_2...\a_{2s-2})\b} \big ) \Big )~, \non \\
\hat{\d}_\e H_{\a(2s)} &=&-2\ri \e^\b \O_{\b\a(2s)} - \frac{2 }{2s+1}\e_{(\a_1}\nabla^\b \tilde{H}_{\a_2 ...\a_{2s})\b}  ~, \\
\hat{\d}_\e \tilde{H}_{\a(2s)} &=&  2 \ri \e^\b\F_{\b\a(2s)} + \frac{2 s^2}{2s+1}\e_{(\a_1}\nabla^\b H_{\a_2 ... \a_{2s})\b} +4\ri s^2 |\m| \e_{(\a_1}\tilde{\cU}_{\a_2 ... \a_{2s})} ~ \\
&+& 2\ri s^2 \e^\b \nabla_{(\a_1 \a_2} \tilde{\cU}_{\a_3 ... \a_{2s})\b} - \frac{2\ri s}{2s-1}(s-1)\e_{(\a_1}\nabla_{\a_2 \a_3} \tilde{\cU}^{\b;}{}_{\a_4 ... \a_{2s})\b}~,  \non \\
\hat{\d}_\e \F_{\a(2s+1)} &=& \hf \e^\b \nabla_{\b(\a_1} \tilde{H}_{\a_2 ... \a_{2s+1})} -s \e_{(\a_1}\nabla^\b \O_{\a_2 ... \a_{2s+1})\b} ~\\
&+&\frac{1}{4(2s+1)} \Big ( \ri \e_{(\a_1}\nabla^2 \tilde{H}_{\a_2 ... \a_{2s+1})} -4s \e_{(\a_1} \nabla_{\a_2}{}^\b \tilde{H}_{\a_3 ... \a_{2s+1})\b} \non ~\\
&+& 4(4s^2 + 5s +2)|\m| \e_{(\a_1}\tilde{H}_{\a_2 ... \a_{2s+1})} \Big )~, \non \\
\hat{\d}_\e \O_{\a(2s+1)} &=& - \hf \e^\b \nabla_{\b(\a_1}H_{\a_2 ... \a_{2s+1})} + \frac{1}{s}\e_{(\a_1}\nabla^\b \F_{\a_2 ... \a_{2s+1})\b} ~ \\
&-& \frac{1}{4(2s+1)} \Big ( \ri \e_{(\a_1} \nabla^2 H_{\a_2 ... \a_{2s+1})}
-4s\e_{(\a_1}\nabla_{\a_2}{}^\b H_{\a_3 ...\a_{2s+1})\b} ~  \non \\
&+&4(4s^2+5s+2)|\m|\e_{(\a_1}H_{\a_2 ... \a_{2s+1})}\Big ) -\hf \e^\b \nabla_{(\a_1 \a_2}\nabla_{\a_3}\tilde{\cU}_{\a_4 ... \a_{2s+1})\b} \non \\
&-& |\m|\e_{(\a_1}\nabla_{\a_2}\tilde{\cU}_{\a_3 ... \a_{2s+1})}  + \frac{1}{2s(2s-1)}(s-1)\e_{(\a_1}\nabla_{\a_2 \a_3}\nabla_{\a_4} \tilde{\cU}^{\b;}{}_{\a_5 ... \a_{2s+1})\b}~. \non
\eea
\end{subequations}


\section{Discussion} \label{Section9}

This section provides a summary of the results obtained in sections \ref{Section5}$-$\ref{Section8}, where we carried out the (1,1) $\rightarrow$ (1,0) AdS superspace reduction of the four massless higher-spin multiplets with (1,1) AdS supersymmetry. 

It is useful to recall from section \ref{Section3} that there exist four off-shell series of massless higher-spin ${\cN}=1$ multiplets in AdS$_{3}$ (two series for each half-integer $(\hat{s}=s+\frac{1}{2})$ and integer $(\hat{s}=s)$ 
superspin $\hat s \geq1$):
\bsubeq \label{N1duals}
\bea
S^{\parallel}_{(s+\hf)} [H_{\a(2s+1)}, L_{\a(2s-2)}] ~, &&\, S^{\perp}_{(s+\hf)}[{H}_{\a(2s+1)} ,\U_{\b; \,\a(2s-2)}]~; \\
S^{\parallel}_{(s)} [H_{\a(2s)}, V_{\a(2s-2)}] ~, \quad &&S^{\perp}_{(s)}[{H}_{\a(2s)} ,\J_{\b; \,\a(2s-2)} ]~.
\eea
\esubeq
As demonstrated in \cite{HK19}, the models $S^{\parallel}_{(s+\hf)}[ H_{\a(2s+1)}, L_{\a(2s-2)}]$ and $S^{\perp}_{(s+\hf)}[ H_{\a(2s+1)}, \U_{\b; \a(2s-2)}]$ are dually equivalent in the flat superspace limit. However, this duality no longer holds in AdS$^{3|2}$. The same feature characterises the two formulations for the massless integer superspin multiplet: $S^\perp_{(s)}[H_{\a(2s)}, \J_{\b; \a(2s-2)}]$ is dual to $S^\parallel_{(s)}[H_{\a(2s)}, V_{\a(2s-2)}]$ \textit{only} in flat superspace. In appendix \ref{AppB}, we study in detail the component structure of these four massless ${\cN}=1$ supersymmetric  higher-spin gauge models in AdS$_3$. Upon elimination of the corresponding auxiliary fields, the results are as follows:
\bsubeq\label{compresults}
\bea
S^{\perp}_{(s+\hf)}[{H}_{\a(2s+1)} ,\U_{\b; \,\a(2s-2)} ] &=& S_{\rm bos} [h_{\a(2s+2)},\F_{\b\g;\a(2s-2)}]  \label{96a} \\
&&+ S_{\text{FF}}^{(2s+1,+)}[                h_{\a(2s+1)}, y_{\a(2s-1)}, y_{\a(2s-3)}] ~. \non\\
S^{\parallel}_{(s+\hf)}[H_{\a(2s+1)} ,L_{\a(2s-2)} ] &=& S^{(2s+2)}_{\text{F}}[h_{\a(2s+2)}, y_{\a(2s-2)}]  \label{96b} \\
&&+ S_{\text{FF}}^{(2s+1,-)}[h_{\a(2s+1)}, y_{\a(2s-1)}, y_{\a(2s-3)}]~. \non\\
S^{\perp}_{(s)}[H_{\a(2s)} ,{\Psi}_{\b; \,\a(2s-2)} ] &=& \frac{2s-1}{s-1}S^{(2s)}_{\text{F}}[h_{\a(2s)},y_{\a(2s-4)}] \\
&&+ \, S^{(2s+1,-)}_{\text{FF}}[h_{\a(2s+1)},y_{\a(2s-1)},y_{\a(2s-3)}]~.  \label{96c}
\non\\
S^{\parallel}_{(s)}[H_{\a(2s)} ,V_{\a(2s-2)} ] &=& \Big ( -\frac{1}{2} \Big )^s 
\int \rd^3x\,e\, 
\Big \{ h^{\a(2s)} h_{\a(2s)} + 2s(2s-1)y^{\a(2s-2)}y_{\a(2s-2)} \Big \} \non\\
&&+ S_{\text{FF}}^{(2s+1,+)}\,[h_{\a(2s+1)},y_{\a(2s-1)},y_{\a(2s-3)}]~.  \label{96d}
\eea
\esubeq
Let us consider the bosonic sector. The action $S_{\rm bos} [h_{\a(2s+2)},\F_{\b\g;\a(2s-2)}]$ is dual to the Fronsdal action $S^{(2s+2)}_{\text{F}}[h_{\a(2s+2)}, y_{\a(2s-2)}]$ in the flat-space limit, see eq.~\eqref{eqb43}. The actions $S^{(2s+2)}_{\text{F}}[h_{\a(2s+2)}, y_{\a(2s-2)}]$ and $S^{(2s)}_{\text{F}}[h_{\a(2s)},y_{\a(2s-4)}]$ correspond to the Fronsdal actions in AdS$_3$ describing massless spin-$(s+1)$ and spin-$s$ fields, respectively.  Although the bosonic fields $h_{\a(2s)}$ and $y_{\a(2s-2)}$ in eq.~\eqref{96d} are pure auxiliary, the action is still consistent with the fact that the (Fang-)Fronsdal action in three dimensions has no propagating degrees of freedom. In the fermionic sector, our results agree with the fact that there are two AdS$_3$ extensions of the (Fang-)Fronsdal action, $S_{\text{FF}}^{(2s+1,+)}$ and  $S_{\text{FF}}^{(2s+1,-)}$. They are related to each other by the replacement $|\m| \to -|\m|$. 

Sections \ref{Section5} and \ref{Section6} detail the (1,1) $\rightarrow$ (1,0) AdS superspace reduction of the transverse and longitudinal formulations for the massless \textit{half-integer} superspin multiplets, respectively. In (1,1) AdS superspace, these off-shell formulations are dually equivalent.  
When reduced to $\cN=1$ AdS superspace, we found that the actual difference between the two models lies in the structure of auxiliary superfields. Let us refer to eq.~\eqref{ReductionTransHI} which shows that the transverse formulation gives rise to two decoupled $\cN=1$ supersymmetric theories:
\bea 
\mathbb{S}^{\perp}_{(s+\hf)} [\mathfrak{H}_{\a(2s)},\G_{\a(2s-2)}, \bar{\G}_{\a(2s-2)}] &=& S^\parallel_{(s+\frac{1}{2})}[H_{\a(2s+1)},L_{\a(2s-2)},  \F_{\a(2s-1)}] \non ~\\ 
&&+ S^\parallel_{(s)}[H_{\a(2s)},V_{\a(2s-2)}, \O_{\a(2s-1)}]~.
\eea
In the longitudinal case, from eq.~\eqref{RATHI} we have that
\bea 
\mathbb{S}^{\parallel}_{(s+\hf)} [\mathfrak{H}_{\a(2s)},G_{\a(2s-2)}, \bar{G}_{\a(2s-2)}] =&& S^\parallel_{(s+\frac{1}{2})} [H_{\a(2s+1)},L_{\a(2s-2)},  \F_{\a(2s-3)}] \non ~\\ 
&&+ S^\parallel_{(s)}[H_{\a(2s)},V_{\a(2s-2)}, \O_{\a(2s-3)}]~.
\eea
We further showed that the superfields $\F_{\a(2s-1)}, \O_{\a(2s-1)}, \F_{\a(2s-3)}, \O_{\a(2s-3)}$ are all auxiliary. Once they are eliminated, each formulation then leads to the same ${\cN}=1$ supersymmetric higher-spin actions:
\bsubeq
\begin{equation}
\begin{tikzpicture}[level distance=40mm, baseline=(current  bounding  box.center)]
\tikzstyle{level 2}=[sibling distance=10mm, ->,black]
\node (a) at (0,0) {Transverse:~~~~~~~~} [parent anchor=east]
[child anchor=west,grow=east]
child {node (c) at (0,0) {$\mathbb{S}^{\perp}_{(s+\hf)}[\mathfrak{H}_{\a(2s)}, \G_{\a(2s-2)}, \bar \G_{\a(2s-2)}]$}  edge from parent[draw=none]
                [child anchor=west,grow=east]
                child [level distance=61mm] {node {$S^{\parallel}_{(s)}[H_{\a(2s)}, V_{\a(2s-2)}]~,$}}
                child [level distance=64.5mm] {node {$S^{\parallel}_{(s+\hf)}[H_{\a(2s+1)}, L_{\a(2s-2)}]$}}
}; \label{92a}
\end{tikzpicture}
\end{equation}
\begin{equation}
\begin{tikzpicture}[level distance=43mm, baseline=(current  bounding  box.center)]
\tikzstyle{level 2}=[sibling distance=10mm, ->,black]
\node (b) at (0,0) {Longitudinal: } [parent anchor=east]
[child anchor=west,grow=east]
child {node {$\mathbb{S}^{\parallel}_{(s+\hf)}[\mathfrak{H}_{\a(2s)}, G_{\a(2s-2)}, \bar G_{\a(2s-2)}]$} edge from parent[draw=none]
                child [level distance=61mm] {node {$S^{\parallel}_{(s)}[H_{\a(2s)}, V_{\a(2s-2)}]~.$}}
                child [level distance=64.5mm] {node {$S^{\parallel}_{(s+\hf)}[H_{\a(2s+1)}, L_{\a(2s-2)}]$}}
};
\end{tikzpicture}
\end{equation}
\esubeq

Sections \ref{Section7} and \ref{Section8} concern the (1,1) $\rightarrow$ (1,0) AdS superspace reduction of the longitudinal and transverse formulations for the massless \textit{integer} superspin multiplets, respectively. In (1,1) AdS superspace, they are dual to each other. As in the half-integer case, we demonstrated that upon reduction to $\cN$=1 AdS superspace, these formulations differ by the structure of the auxiliary superfields. 
As shown in eq.~\eqref{723}, the longitudinal model is equivalent to
\bea
\mathbb{S}^{\parallel}_{(s)} [U_{\a(2s-2)},G_{\a(2s)}, \bar{G}_{\a(2s)}] &=& S^\perp_{(s)}[H_{\a(2s)}, \cU_{\b;\a(2s-2)} ,  \F_{\a(2s-1)}] \non ~\\ 
&&+ S^\parallel_{(s)}[\tilde{H}_{\a(2s)}, V_{\a(2s-2)}, \O_{\a(2s-1)}]~.
\eea
On the other hand, eq.~\eqref{t-int} shows that the transverse model yields
\bea
\mathbb{S}^\perp_{(s)}[U_{\a(2s-2)}, \G_{\a(2s)},\bar{\G}_{\a(2s)} ] &=& S^\perp_{(s)}[ H_{\a(2s)}, \tilde{\cU}_{\b;\a(2s-2)} , \F_{\a(2s+1)} ] ~ \non \\
&&+ S^\parallel_{(s)}[\tilde{H}_{\a(2s)}, V_{\a(2s-2)}, \O_{\a(2s+1)}]~.
\eea
The superfields $\F_{\a(2s-1)}, \O_{\a(2s-1)}, \F_{\a(2s+1)}, \O_{\a(2s+1)}$ are all auxiliary and thus can be integrated out from their corresponding actions. We showed that the resulting ${\cN}=1$ actions are the same in both formulations:
\bsubeq \label{TLRed}
\begin{equation}
\begin{tikzpicture}[level distance=33.5mm, baseline=(current  bounding  box.center)]
\tikzstyle{level 2}=[sibling distance=10mm, ->,black]
\node (b) at (0,0) {Transverse:~~~~~~~~~~~~~~~~~~} [parent anchor=east]
[child anchor=west,grow=east]
child {node {$\mathbb{S}^{\perp}_{(s)}[U_{\a(2s-2)}, \G_{\a(2s)}, \bar \G_{\a(2s)}]$} edge from parent[draw=none]
	child [level distance=57.6mm] {node {$S^{\parallel}_{(s)}[H_{\a(2s)}, V_{\a(2s-2)}]~,$}}
	child [level distance=59mm] {node {$S^{\perp}_{(s)}[H_{\a(2s)}, \J_{\b;\,\a(2s-2)}]$}}
};
\end{tikzpicture}
\end{equation}
\begin{equation}
\begin{tikzpicture}[level distance=44mm, baseline=(current  bounding  box.center)]
\tikzstyle{level 2}=[sibling distance=10mm, ->,black]
\node (a) at (0,0) {Longitudinal: } [parent anchor=east]
[child anchor=west,grow=east]
child {node (c) at (0,0) {$\mathbb{S}^{\parallel}_{(s)}[U_{\a(2s-2)}, G_{\a(2s)}, \bar G_{\a(2s)}]$}  edge from parent[draw=none]
                [child anchor=west,grow=east]
                child [level distance=57.6mm] {node {$S^{\parallel}_{(s)}[H_{\a(2s)}, V_{\a(2s-2)}].$}}
                child [level distance=59mm] {node {$S^{\perp}_{(s)}[H_{\a(2s)}, \J_{\b;\,\a(2s-2)}]$}}
};
\end{tikzpicture}
\end{equation}
\esubeq

We are now in a position to compare the above (1,1) $\rightarrow$ (1,0) AdS reduction results with the ${\cN}=1$ supersymmetric higher-spin gauge theories obtained via (2,0) $\to$ (1,0) AdS reduction \cite{HK19}. There exist two off-shell formulations for massless \textit{half-integer} superspin multiplets with (2,0) AdS supersymmetry, which are not dual to each other.\footnote{Off-shell construction for the massless integer superspin (2,0) multiplets remains an open problem.} They are known as type II and type III series \cite{HK18}. Upon reduction to ${\cN}=1$ superspace, type II series yields
\begin{equation} \label{type2red}
\begin{tikzpicture}[level distance=35mm, baseline=(current  bounding  box.center)]
\tikzstyle{level 2}=[sibling distance=10mm, ->,black]
\node (a) at (0,0) {Type \RomanNumeralCaps{2}:~~~~~~~} [parent anchor=east]
[child anchor=west,grow=east]
child{node (c) at (0,0) {$\mathbb{S}^{\text{\RomanNumeralCaps{2}}}_{(s+\hf)}[\mathfrak{H}_{\a(2s)}, \mathfrak{L}_{\a(2s-2)}]$}  edge from parent[draw=none]
                [child anchor=west,grow=east]
                child [level distance=56.8mm] {node {$S^{\perp}_{(s)}[H_{\a(2s)}, \J_{\b ;\a(2s-2)}]~.$}}
                child [level distance=59.3mm] {node {$S^{\parallel}_{(s+\hf)}[H_{\a(2s+1)}, L_{\a(2s-2)}]$}}
};
\end{tikzpicture}
\end{equation}
On the other hand, type III series leads to
\begin{equation} \label{type3red}
\begin{tikzpicture}[level distance=40mm, baseline=(current  bounding  box.center)]
\tikzstyle{level 2}=[sibling distance=10mm, ->,black]
\node (b) at (0,0) {~~~~Type \RomanNumeralCaps{3}:} [parent anchor=east]
[child anchor=west,grow=east]
child {node {$\mathbb{S}^{\text{\RomanNumeralCaps{3}}}_{(s+\hf)}[\mathfrak{H}_{\a(2s)}, \mathfrak{V}_{\a(2s-2)}]$} edge from parent[draw=none]
                child [level distance=55mm] {node {$S^{\parallel}_{(s)}[H_{\a(2s)}, V_{\a(2s-2)}]~.$}
                }
                child [level distance=61mm] {
               node {$S^{\perp}_{(s+\hf)}[H_{\a(2s+1)}, \U_{\b ;\a(2s-2)}]$}
                }};
\end{tikzpicture}
\end{equation}
We point out that unlike the situation for the (1,1) AdS multiplets, the $\cN=1$ models obtained from the $(2,0) \to (1,0)$ AdS reduction do not involve any auxiliary superfields.
Additionally, reductions of the (1,1) AdS multiplets only produce three of the four series of off-shell $\cN=1$ multiplets, with the exception of the transverse half-integer model, $S^{\perp}_{(s+\hf)}[ H_{\a(2s+1)}, \U_{\b;\, \a(2s-2)}]$. The latter can only be derived via (2,0) $\to$ (1,0) AdS reduction of the type III series. This formulation corresponds to the linearised supergravity model (for $s=1$), which does not admit a non-linear extension \cite{KT-M11}. In accordance with \eqref{96a} (and the component analysis in \ref{apB3}), the bosonic sector of $S^{\perp}_{(s+\hf)}[ H_{\a(2s+1)}, \U_{\b;\, \a(2s-2)}]$ does not correspond to the standard Fronsdal action, but its dual version. However, this duality is only present in the flat-space limit.

The dualities relating the four ${\cN}=1$ supersymmetric models in the flat limit, $|\mu| \to 0$, deserve further comment. In flat superspace, one may take any of the above ${\cN}=2$ superfield formulations and construct several dual models. This can be done by replacing its ${\cN}=1$ sector(s) by its dual versions. For concreteness, let us consider the transverse formulation for the massless half-integer superspin, eq.~\eqref{92a}, in flat superspace. We denote the flat superspace action by $\mathbb{S}^{\perp}_{(s+\hf)}[\mathfrak{H}_{\a(2s)}, \G_{\a(2s-2)}, \bar \G_{\a(2s-2)}]_{\rm FS}$.
We can dualise one of its ${\cN}=1$ sectors, or both. This procedure gives us four different gauge theories:
\begin{enumerate}[label=(\roman*)]
\item $S^{\parallel}_{(s+\hf)} [H_{\a(2s+1)}, L_{\a(2s-2)}]_{\rm FS} +
S^{\parallel}_{(s)} [H_{\a(2s)}, V_{\a(2s-2)}]_{\rm FS}~,$
\item $S^{\parallel}_{(s+\hf)} [H_{\a(2s+1)}, L_{\a(2s-2)}]_{\rm FS} +
S^{\perp}_{(s)} [H_{\a(2s)}, \J_{\b;\,\a(2s-2)}]_{\rm FS}~,$
\item $S^{\perp}_{(s+\hf)} [H_{\a(2s+1)}, \U_{\b;\,\a(2s-2)}]_{\rm FS} +
S^{\parallel}_{(s)} [H_{\a(2s)}, V_{\a(2s-2)}]_{\rm FS}~,$
\item $S^{\perp}_{(s+\hf)} [H_{\a(2s+1)}, \U_{\b;\,\a(2s-2)}]_{\rm FS} +
S^{\perp}_{(s)} [H_{\a(2s)}, \J_{\b;\,\a(2s-2)}]_{\rm FS}~.$
\end{enumerate}
Case (i) describes $\mathbb{S}^{\perp}_{(s+\hf)}[\mathfrak{H}_{\a(2s)}, \G_{\a(2s-2)}, \bar \G_{\a(2s-2)}]_{\rm FS}$. Cases (ii) and (iii) correspond to $\mathbb{S}^{\rm II}_{(s+\hf)}[\mathfrak{H}_{\a(2s)}, \mathfrak{L}_{\a(2s-2)}]_{\rm FS}$ and $\mathbb{S}^{\rm III}_{(s+\hf)}[\mathfrak{H}_{\a(2s)}, \mathfrak{V}_{\a(2s-2)}]_{\rm FS}$, respectively.
At this stage, it is unknown if there exists a ${\cN}=2$ superfield formulation which leads to the two transverse models described by (iv). This will be an interesting open problem to investigate further.

The massless higher-spin ${\cN}=1$ models \eqref{N1duals} do not have any propagating degrees of freedom. However, they can be deformed in order to generate off-shell topologically massive higher-spin supersymmetric theories. There has been recent progress made in the construction of these off-shell theories with various supersymmetries. It is based on the approaches developed in \cite{KT} for ${\cN}=1$ and in \cite{KO} for ${\cN}=2$ Poincar\'e supersymmetries. They were later extended to $\rm AdS_3$ for the following cases: $\cN=1$ \cite{KP,HK19}, $\cN=(1,1)$ \cite{HKO} and $\cN=(2,0)$ \cite{HK18}. The gauge-invariant actions for such massive multiplets are obtained by adding a superconformal and massless higher-spin action together, following the philosophy of topologically massive theories \cite{WS,JS,DJT1,DJT2}.

For a positive integer $n$, the conformal superspin-$\frac{n}{2}$ action in AdS$^{3|2}$ \cite{KP,KT,SK} is
\be \label{1SCS}
S^{(n/2)}_{\text{SCS}}[H_{\a(n)}] = - \frac{\ri^n}{2^{\lfloor n/2 \rfloor+1}} 
\int \rd^{3|2} z \, E\,H^{\a(n)}W_{\a(n)}(H)~.
\ee
Here $H_{\a(n)}$ is the superconformal gauge prepotential and $W_{\a(n)}(H)$ 
is the higher-spin $\cN=1$ super-Cotton tensor. The latter possesses the defining properties of being (i) gauge-invariant, $\d_\z W_{\a(n)} = 0$; and (ii) transverse, $\nabla^\b  W_{\b \a_1 \dots \a_{n-1}}=0$.  The action \eqref{1SCS} is invariant under the gauge transformations \eqref{SCSGI}. The explicit expression for $W_{\a(n)}(H)$ was given for the first time
in $\cN=1$ Minkowski superspace in \cite{SK,KT}. A few years later, it was generalised to an arbitrary conformally flat superspace in \cite{KP19} by using the formalism of $\cN=1$ conformal superspace
\cite{BKNT-M1}. Recently, the closed form expression for $W_{\a(n)}(H)$ in AdS$^{3|2}$ has been derived in \cite{KP21}.

In accordance with \cite{HK19,KP}, there exist four off-shell formulations for topologically massive higher-spin ${\cN}=1$ multiplets in AdS${}_3$. For a positive integer $s$, there are two off-shell gauge-invariant models for a topologically 
massive superspin-$(s+\hf)$ multiplet in AdS${}_3$:
\begin{subequations} \label{8.5}
	\bea
	S^{\parallel}_{(s+\hf)}[H_{\a(2s+1)},L_{\a(2s-2)}|m] &=& \kappa
	S^{(s+\hf)}_{\text{SCS}}[H_{\a(2s+1)}] \\ 
	&&+m^{2s-1}S^{\parallel}_{(s+\hf)}[H_{\a(2s+1)},L_{\a(2s-2)}]~, \label{N2Massive1} \non
	\\
	S^{\perp}_{(s+\hf)}[H_{\a(2s+1)},\U_{\b;\a(2s-2)}|m] &=& \kappa S^{(s+\hf)}_{\text{SCS}}[H_{\a(2s+1)}]\\
	&&+m^{2s-1}S^{\perp}_{(s+\hf)}[H_{\a(2s+1)},\U_{\b;\a(2s-2)}]~. \non
	\eea
\end{subequations}
Here $\kappa$ is a dimensionless parameter and $m$ is a real massive parameter.
For a topologically massive superspin-$s$ multiplet in AdS${}_3$, we have  the following models:
\begin{subequations} \label{8.4}
\bea
S^{\parallel}_{(s)}[H_{\a(2s)} ,V_{\a(2s-2)} |m]
&=& {S}_{\rm{SCS}}^{(s)} [ H_{\a(2s)}] 
+m^{2s-1}S^{\parallel}_{(s)}[H_{\a(2s)} ,V_{\a(2s-2)} ]~, \label{8.4a} \\
S^{\perp}_{(s)}[H_{\a(2s)} ,{\Psi}_{\b; \,\a(2s-2)} |m]
&=& {S}_{\rm{SCS}}^{(s)} [ H_{\a(2s)}] 
+m^{2s-1}S^{\perp}_{(s)}[H_{\a(2s)} ,{\Psi}_{\b; \,\a(2s-2)}]~. \label{8.4b}
\eea
\end{subequations}

Another way of deriving the four massive ${\cN}=1$ models, \eqref{8.5}-\eqref{8.4}, is by performing either a (1,1) $\to$ (1,0) or (2,0) $\to$ (1,0) AdS reduction of the off-shell topologically massive higher-spin supersymmetric theories presented in \cite{HKO} and \cite{HK18}, respectively. In what follows, we will recall the structure of off-shell massive higher-spin supermultiplets in (1,1) AdS superspace \cite{HKO} and take a closer look at their reduction.

Given a positive integer $n$, the superconformal higher-spin action in AdS$^{(3|1,1)}$ \cite{KO,HKO,KP,BHHK} is given by 
\be \label{SCSAct}
\mathbb{S}_{\text{SCS}}[\mathfrak{H}_{\a(n)}] = -\frac{\ri^n}{2^{\lfloor n/2 \rfloor+1}} \int \rd^{3|4}z\, \bm E \,\mathfrak{H}^{\a(n)}\mathfrak{W}_{\a(n)}(\mathfrak{H})~,
\ee
where $\mathfrak{H}_{\a(n)}$ is the superconformal gauge multiplet, 
and $\mathfrak{W}_{\a(n)}(\mathfrak{H})$
is the higher-spin $\cN=2$ super-Cotton tensor.
The action \eqref{SCSAct} is invariant under the gauge transformations
\be \label{N2GT}
 \d_\l \mathfrak{H}_{\a(n)}= \cDB_{(\a_1}\l_{\a_2...\a_n)}-(-1)^n \cD_{(\a_1}\bar{\l}_{\a_2...\a_n)}~,
 \ee
where $\l_{\a(n-1)}$ is unconstrained complex. To check the gauge invariance, one needs to make use of the properties that $\mathfrak{W}_{\a(n)}(\mathfrak{H})$ is (i) transverse linear, $\bar \cD^\b {\mathfrak W}_{\b \a_1 \dots \a_{n-1}}=
\cD^\b {\mathfrak W}_{\b \a_1 \dots \a_{n-1}}=0$; and (ii) gauge invariant, $\d_\l \mathfrak{W}_{\a(n)} = 0$. In $\cN=2$ Minkowski superspace, the closed form expression for
$\mathfrak{W}_{\a(n)}(\mathfrak{H})$ was given for the first time in \cite{KO}, 
and a much simpler expression was derived in \cite{BHHK}.
It was generalised to an arbitrary conformally flat superspace in \cite{KP19} by using the formalism of $\cN=2$ conformal superspace
\cite{BKNT-M1}.

In (1,1) AdS superspace, let us consider the two dually equivalent off-shell formulations for a massive superspin-$(s+\hf)$ multiplet \cite{HKO}:
\begin{subequations} \label{11Massive}
	\bea
	\mathbb{S}^{\perp}_{(s+\hf)}[\mathfrak{H}_{\a(2s)},\G_{\a(2s-2)},\bar{\G}_{\a(2s-2)}|m] &=& \k \mathbb{S}_{\text{SCS}}[\mathfrak{H}_{\a(2s)}] ~ \label{11MassiveTrans} \\ 
	&&+m^{2s-1}\mathbb{S}^{\perp}_{(s+\hf)}[\mathfrak{H}_{\a(2s)},\G_{\a(2s-2)},\bar{\G}_{\a(2s-2)}]~, \non \\
	\mathbb{S}^{\parallel}_{(s+\hf)}[\mathfrak{H}_{\a(2s)},G_{\a(2s-2)},\bar{G}_{\a(2s-2)}|m]&=&\k \mathbb{S}_{\text{SCS}}[\mathfrak{H}_{\a(2s)}]~  \label{11MassiveLong} \\
 &&+m^{2s-1}\mathbb{S}^{\parallel}_{(s+\hf)}[\mathfrak{H}_{\a(2s)},G_{\a(2s-2)},\bar{G}_{\a(2s-2)}]~. \non
	\eea
\end{subequations}
The superconformal action $\mathbb{S}_{\text{SCS}}[\mathfrak{H}_{\a(2s)}]$ is given by \eqref{SCSAct}, for the case $n=2s$. The transverse $\mathbb{S}^{\perp}_{(s+\hf)}[\mathfrak{H}_{\a(2s)},\G_{\a(2s-2)},\bar{\G}_{\a(2s-2)}]$ and longitudinal $\mathbb{S}^{\parallel}_{(s+\hf)}[\mathfrak{H}_{\a(2s)},G_{\a(2s-2)},\bar{G}_{\a(2s-2)}]$ actions are given by \eqref{2.5} and \eqref{3.3} respectively.

As pointed out in \cite{HKO}, it is expected that the topologically massive actions \eqref{11MassiveTrans} and \eqref{11MassiveLong} describe the on-shell massive supermultiplets in (1,1) AdS superspace \cite{KNG}.
Given a positive integer $n>0$, 
a massive  on-shell multiplet of superspin
$(n+1)/2$ is realised in terms of a real symmetric rank-$n$ spinor  
$T_{\a (n)}$ constrained by 
\begin{subequations}  \label{5.10ab}
\bea
\cD^\b T_{ \a_1 \cdots \a_{n-1} \b} 
= \bar \cD^\b T_{ \a_1 \cdots \a_{n-1} \b}&=&0~, \\ 
\Big( \frac{\ri}{2} \cD^\g\bar \cD_\g + m \Big) T_{\a_1 \cdots \a_n} &=&0~.
\eea
\end{subequations}
It may be shown that 
\bea
\Big(\frac{\ri}{2}\cD^\g\cDB_\g\Big)^2 T_{\a_1\cdots\a_n}
&=&
\Big( \cD^{a} \cD_{a} 
+2(n+2) |\mu|^2 \Big) T_{\a_1\cdots\a_n}
~.~~~~~~~~~~~~
\eea

One can also construct a massive model which leads directly to the equations
\eqref{5.10ab}, as an extension of the flat-space bosonic constructions of \cite{BHT,BKRTY}. It is given by 
\bea
{\mathbb{S}}_{\rm{massive}}
[ {\mathfrak H}_{\a(n)}] 
= - \frac{\ri^n}{2^{\left \lfloor{n/2}\right \rfloor +1}} \frac{\k}{m} 
   \int \rd^{3|4}z\, \bm E\,{\mathfrak W}^{\a(n) }( {\mathfrak H}) 
  \Big\{ m
+\frac{\ri}{2} \cD^\g \bar \cD_\g  \Big\} {\mathfrak H}_{\a(n)}~.
\label{5.13}
\eea
The action \eqref{5.13} is invariant under gauge transformations \eqref{N2GT}. To check gauge invariance, one makes use of the properties that 
${\mathfrak W}_{\a(n) }( {\mathfrak H}) $ is (i) gauge-invariant; and 
(ii) transverse linear, 
$\bar \cD^\b {\mathfrak W}_{\b \a_1 \dots \a_{n-1}}=
 \cD^\b {\mathfrak W}_{\b \a_1 \dots \a_{n-1}}=0$.
The action  \eqref{5.13} becomes superconformal in the $m \to\infty$ limit. 

We now turn to describing the (1,1) $\to$ (1,0) AdS reduction of the massive models \eqref{11Massive}. The reduction of the massless transverse \eqref{2.5} and longitudinal \eqref{3.3} formulations was achieved in sections \ref{Section5} and \ref{Section6} respectively. Thus, we need only carry out the reduction of the superconformal action.
The superconformal gauge multiplet $\mathfrak{H}_{\a(2s)}$ was reduced in section 5.2. Upon reduction, the superfield $\mathfrak{H}_{\a(2s)}$ is equivalent to two unconstrained real $\cN=1$ superfields \eqref{2.9}, defined modulo gauge transformations of the form \eqref{2.11}. Since $\mathfrak{W}_{\a(2s)}$ is a real transverse linear superfield, it follows that $\mathfrak{W}_{\a(2s)}$ is equivalent to two real $\cN=1$ superfields, defined as:
\be
W_{\a(2s)}(H) = -\mathfrak{W}_{\a(2s)}(\mathfrak{H})|~, \qquad W_{\a(2s+1)}(H) = \frac{\ri}{2}\nabla_{(\a_1}\mathfrak{W}_{\a_2 ... \a_{2s+1})}(\mathfrak{H})|~.
\ee
Here each of the ${\cN}=1$ superfields are transverse linear and gauge invariant. Applying the reduction procedure to the superconformal action \eqref{SCSAct} yields two decoupled $\cN=1$ supersymmetric theories
\be \label{SCSRed}
\mathbb{S}_{\text{SCS}}[\mathfrak{H}_{\a(2s)}] = - \frac{(-1)^s}{2^{s+1}}
\int \rd^{3|2}z\,E\, \big \{ H^{\a(2s)}W_{\a(2s)} 
+ \ri H^{\a(2s+1)}W_{\a(2s+1)} \big \}~.
\ee
These two gauge-invariant theories can be identified as the $\cN=1$ superconformal actions in AdS$_3$ \eqref{1SCS}, for the cases $n=2s$ and $n=2s+1$ respectively. Recalling the reduction results \eqref{92a}, it is apparent that both \eqref{11MassiveTrans} and \eqref{11MassiveLong} decouple into two off-shell topologically massive $\cN=1$ theories: $S^{\parallel}_{(s+\hf)}[H_{\a(2s+1)} ,L_{\a(2s-2)} |m]$ and $S^{\parallel}_{(s)}[H_{\a(2s)} ,V_{\a(2s-2)} |m]$. Thus, the (1,1) $\rightarrow$ (1,0)  AdS reduction of \eqref{11Massive} is only able to generate two of the gauge-invariant formulations for topologically massive $\cN=1$ supermultiplets, given by eqs.\,\eqref{8.5}-\eqref{8.4}.

In (2,0) AdS superspace, the following off-shell gauge-invariant massive models were proposed \cite{HK18}:
\begin{subequations} \label{20mass}
\bea
\mathbb{S}_{(s+\hf)}^{\rm II}[{\mathfrak H}_{\a(2s)} ,{\mathfrak L}_{\a(2s-2)}|m]&=& 
\k {\mathbb{S}}_{\rm SCS} [ {\mathfrak H}_{\a(2s)}] 
+ m^{2s-1} \mathbb{S}^{\rm II}_{(s+\hf)} [{\mathfrak H}_{\a(2s)} ,{\mathfrak L}_{\a(2s-2)} ]~, 
\label{5.7a} \\
\mathbb{S}_{(s+\hf)}^{\rm III}[{\mathfrak H}_{\a(2s)} ,{\mathfrak V}_{\a(2s-2)}|m]&=& 
\k {\mathbb{S}}_{\rm SCS} [ {\mathfrak H}_{\a(2s)}] 
+ m^{2s-1} \mathbb{S}^{\rm III}_{(s+\hf)} [{\mathfrak H}_{\a(2s)} ,{\mathfrak V}_{\a(2s-2)} ]~.
\label{5.7b}
\eea
\end{subequations} 
Here 
\bea
{\mathbb{S}}_{\rm{SCS}}
[ {\mathfrak H}_{\a(2s)}] 
= - \frac{(-1)^s}{2^{s+1}}
   \int 
  \rd^{3|4}z 
   \, \bm E\,
 {\mathfrak H}^{\a(2s)} 
{\mathfrak W}_{\a(2s) }( {\mathfrak H})
\label{2.34}
\eea
is the superconformal higher-spin action \cite{HKO}, with 
${\mathfrak W}_{\a(2s) }( {\mathfrak H}) = \bar {\mathfrak W}_{\a(2s) }( {\mathfrak H}) $ 
being the higher-spin super-Cotton tensor in (2,0) AdS superspace. 

The (2,0) AdS analogue of the model \eqref{5.13}, which was described in \cite{HK18}, takes the form
\bea
{\mathbb{S}}_{\rm{massive}}
[ {\mathfrak H}_{\a(n)}] 
= - \frac{\ri^n}{2^{\left \lfloor{n/2}\right \rfloor +1}} \frac{\k}{m} 
   \int \rd^{3|4}z\, \bm E\,{\mathfrak W}^{\a(n) }( {\mathfrak H}) 
  \Big\{ m
+\frac{\ri}{2} \bf{D}^\g \bar {\bf{D}}_\g  \Big\} {\mathfrak H}_{\a(n)}~,
\label{mass20}
\eea
where $\bf{D}_{\a}, \bar{\bf{D}}_{\a}$ denote the spinor covariant derivatives of (2,0) AdS superspace. One can then obtain the following set of constraints describing an on-shell massive superspin-$(n+1)/2$ multiplet in (2,0) AdS superspace:
\begin{subequations} 
\bea
{\bf{D}}^\b T_{ \a_1 \cdots \a_{n-1} \b} 
= \bar{\bf{D}}^\b T_{ \a_1 \cdots \a_{n-1} \b}&=&0~, \\ 
\Big( \frac{\ri}{2} {\bf{D}}^\g\bar {\bf{D}}_\g + m \Big) T_{\a_1 \cdots \a_n} &=&0~.
\eea
\end{subequations}
We note that the models \eqref{5.13} and \eqref{mass20} are not dual to each other since they have different types of supersymmetry, namely (1,1) and (2,0) AdS supersymmetry, respectively. 

Recalling the reduction result \eqref{type2red}, it is clear that \eqref{5.7a} decouples into two off-shell massive ${\cN}=1$ models: $S^{\parallel}_{(s+\hf)}[H_{\a(2s+1)} ,L_{\a(2s-2)} |m]$ and $S^{\perp}_{(s)}[H_{\a(2s)} ,{\Psi}_{\b; \,\a(2s-2)} |m]$.
On the other hand, it follows directly from \eqref{type3red} that the reduction of \eqref{5.7b} leads to $S^{\perp}_{(s+\hf)}[H_{\a(2s+1)} ,\U_{\b;\, \a(2s-2)}| m]$ and $S^{\parallel}_{(s)}[H_{\a(2s)} ,V_{\a(2s-2)} |m]$.
Thus, the (2,0) $\to$ (1,0) AdS reduction of \eqref{20mass} allows us to recover all four off-shell gauge-invariant formulations for topologically massive ${\cN}=1$ supermultiplets, which are described by \eqref{8.5} and \eqref{8.4}.

It is worth pointing out that there also exists an on-shell construction of gauge-invariant Lagrangian formulations for massive higher-spin ${\cN}=1$ supermultiplets in $\mathbb{R}^{2,1}$ and AdS$_{3}$ \cite{BSZ3, BSZ4, BSZ5}, extending previous works on the non-supersymmetric cases \cite{BSZ1, BSZ2}. These frame-like formulations are based on the gauge-invariant approach (also known as the Stueckelberg approach) to the dynamics of massive higher-spin fields, which were proposed by Zinoviev \cite{Zinoviev1, Zinoviev2} and Metsaev \cite{Metsaev}. As demonstrated in \cite{BSZ3, BSZ4, BSZ5}, in three dimensions it is possible to rewrite the corresponding Lagrangians in terms of a set of gauge-invariant curvatures, resulting in a more elegant formulation. An interesting open problem is to understand if there exists an off-shell uplift of these models. 
\\

\noindent
{\bf Acknowledgements:}\\
We are grateful to the referee of this work for pointing out important references as well as for the suggestion to work out the component results described in appendix \ref{AppB}.
The work of JH and SMK is supported in part by the Australian 
Research Council, project No. DP200101944.
The work of DH is supported by the Jean Rogerson Postgraduate Scholarship and an Australian Government Research Training Program Scholarship at The University of Western Australia.


\appendix
\section{3D notation and AdS identities} \label{AppendixA}

Our 3D notation and conventions correspond to
\cite{KLT-M11}. The 3D Minkowski metric is
$\eta_{ab}=\mbox{diag}(-1,1,1)$ and the Levi-Civita tensor is normalised as
$\varepsilon^{012}=-\varepsilon_{012}=1$.

The $\rm SL(2,\mathbb{R})$ invariant tensors
\bea
\ve_{\a\b}=\left(\begin{array}{cc}0~&-1\\1~&0\end{array}\right)~,\qquad
\ve^{\a\b}=\left(\begin{array}{cc}0~&1\\-1~&0\end{array}\right)~,\qquad
\ve^{\a\g}\ve_{\g\b}=\d^\a_\b
\eea
are used to raise and lower the spinor indices:
\bea
\psi^{\a}=\ve^{\a\b}\psi_\b~, \qquad \psi_{\a}=\ve_{\a\b}\psi^\b~.
\label{sp}
\eea

We make use of real Dirac $\g$-matrices,  $\g_a := \big( (\g_a)_\a{}^\b \big)$, which are defined by
\bea
(\g_a)_\a{}^\b := \ve^{\b \g} (\g_a)_{\a \g} = (-\ri \s_2, \s_3, \s_1)~.
\eea
They have the following properties:
\bsubeq
\bea
\gamma_a \gamma_b &=& \eta_{ab}{\mathbbm{1}} + \varepsilon_{abc}
\gamma^c~,\\
(\gamma^a)_{\alpha\beta}(\gamma_a)^{\rho\sigma}
&=&-(\delta_\alpha^\rho\delta_\beta^\sigma
+\delta_\alpha^\sigma\delta_\beta^\rho)~, \\
\ve_{abc}(\g^b)_{\a\b}(\g^c)_{\g\d}&=&
\ve_{\g(\a}(\g_a)_{\b)\d}
+\ve_{\d(\a}(\g_a)_{\b)\g}
~,
\\
\tr[\g_a\g_b\g_{c}\g_d]&=&
2\eta_{ab}\eta_{cd}
-2\eta_{ac}\eta_{db}
+2\eta_{ad}\eta_{bc}
~.
\eea
\esubeq

A three-vector $x_a$ can be equivalently realised as a symmetric second-rank spinor $x_{\a\b}$
defined as
\bea
x_{\a\b}:=(\g^a)_{\a\b}x_a=x_{\b\a}~,\qquad
x_a=-\hf(\g_a)^{\a\b}x_{\a\b}~.
\eea

The relationships between Lorentz generators with two vector indices ($M_{ab} =-M_{ba}$), one vector index ($M_a$)
and two spinor indices ($M_{\a\b} =M_{\b\a}$) are:
\bea
M_{ab} = -\ve_{abc}M^c~, \,\,\, M_a=\hf \ve_{abc}M^{bc}~, \,\,\, M_{\a\b}=(\g^a)_{\a\b}M_a~, \,\,\, M_{a}= -\hf (\g_a)^{\a \b} M_{\a \b}~.~~~~~
\eea
These generators 
act on a vector $V_c$ 
and a spinor $\J_\g$ by the rules
\bea
M_{ab}V_c=2\eta_{c[a}V_{b]}~, ~~~~~~
M_{\a\b}\J_{\g}
=\ve_{\g(\a}\J_{\b)}~.
\label{generators}
\eea

The (brackets) parentheses correspond to (anti-)symmetrisation of tensor or spinor indices, which encode a normalisation factor, for example
\bea
V_{[a_1 a_2 \dots a_n]} := \frac{1}{n!}\sum_{\pi \in S_n} \mbox{sgn}(\pi) V_{a_{\pi(1)} \dots a_{\pi(n)}}~, \quad V_{(\a_1 \dots \a_n )} := \frac{1}{n!} \sum_{\pi \in S_n} V_{\a_{\pi(1)} \dots \a_{\pi(n)}}~,~~~~~~
\eea
with $S_n$ being the symmetric group of $n$ elements.

We provide a summary of essential identities for (1,1) AdS covariant derivatives, which we denote by $\bm \nabla^I_A = (\bm \nabla_a, \bm \nabla^I_\a)$, where $I=\1,\2$. Making use of the algebra in the real basis \eqref{1.3}, we can derive the following results:
\begin{subequations}  
	\bea \label{A.9a}
	\bm \nabla^\1_\a \bm \nabla^\1_\b &=& \ri \bm \nabla_{\a\b} - 2\ri|\m|M_{\a\b}+\frac{1}{2}\ve_{\a\b}\bm (\bm \nabla^\1)^2~,\\
	\bm \nabla^{\1\b} \bm \nabla^\1_\a \bm \nabla^\1_\b &=& 4\ri |\m|\bm \nabla^\1_\a~, \qquad \ [ \bm \nabla^\1_\a,\bm \nabla_{\b\g}] = 2|\m|\ve_{\a(\b}\bm \nabla^\1{}_{\g)}~,\\
	\{ \bm (\bm \nabla^\1)^2, \bm \nabla^\1_\a  \} &=& 4\ri |\m|\bm \nabla^\1_\a \qquad	\ [\bm \nabla^\1_\a ,  \Box] = 2|\m|\bm \nabla_{\a\b}\bm \nabla^{\1\b} + 3|\m|^2\bm \nabla^\1_\a~,  \\
	\bm (\bm \nabla^\1)^2 \bm \nabla^\1_\a &=& 2\ri |\m|\bm \nabla^\1_\a + 2\ri \bm \nabla_{\a\b}\bm \nabla^{\1\b} - 4\ri |\m|\bm \nabla^{\1\b} M_{\a\b}~, \label{A.9d}\\
	\label{A.9e}
	-\frac{1}{4}\bm (\bm \nabla^\1)^2 \bm (\bm \nabla^\1)^2 &=&  \Box - 2\ri|\m|\bm (\bm \nabla^\1)^2+2|\m|\bm \nabla^{\a\b}M_{\a\b} -2|\m|^2M^{\a\b}M_{\a\b}~,\\
	\bm \nabla^\2_\a\bm \nabla^\2_\b &=&  \ri \bm \nabla_{\a\b} + 2\ri|\m|M_{\a\b}+\frac{1}{2}\ve_{\a\b}(\bm \nabla^\2)^2~,\\
	\bm \nabla^{\2\b}\bm \nabla^\2_\a \bm \nabla^\2_\b &=& -4\ri|\m|\bm \nabla^\2_\a,~  \qquad \ [ \bm \nabla^\2_\a,\bm \nabla_{\b\g}] = -2|\m|\ve_{\a(\b}\bm \nabla^\2{}_{\g)}~, \\
	\{ (\bm \nabla^\2)^2 , \bm \nabla^\2_\a \} &=& -4\ri|\m|\bm \nabla^\2_\a ~, \qquad \ [\bm \nabla^\2_\a ,  \Box] = - 2|\m|\bm \nabla_{\a\b}\bm \nabla^{\2\b} + 3|\m|^2\bm \nabla^\2_\a~, \\
	(\bm \nabla^\2)^2\bm \nabla^\2_\a&=& -2\ri|\m|\bm \nabla^\2_\a + 2\ri \bm \nabla_{\a\b} \bm \nabla^{\2\b}+4\ri |\m|\bm \nabla^{\2\b}M_{\a\b}~, \\
	-\frac{1}{4}(\bm \nabla^\2)^2 (\bm \nabla^\2)^2 &=&  \Box + 2\ri|\m|(\bm \nabla^\2)^2- 2|\m|\bm \nabla^{\a\b}M_{\a\b} -2|\m|^2M^{\a\b}M_{\a\b}~,	
	\eea
\end{subequations}


\section{Component structure of $\cN=1$ supersymmetric higher-spin actions in AdS$_3$} \label{AppB}

In this appendix, we study the component structure of the longitudinal and transverse formulations for both the half-integer and integer superspin multiplets which are reviewed in section \ref{Section3}. In accordance with \eqref{1.8}, any $\cN=1$ 
supersymmetric action in $\rm AdS_3$ can be reduced to components by the rule  
\bea
S = \frac{1}{4} \int \rd^3x \, e \, (\ri \nabla^2 + 8|\m|) L\,\big|_{\,\q = 0}~.
\eea
In what follows, we will denote the torsion-free covariant derivative on  AdS$_3$ by
${\mathfrak D}_a$. It is related to the vector covariant derivative $\nabla_a$ in \eqref{N10}
by the simple rule ${\mathfrak D}_{a} := \nabla_{a}|_{\q = 0}$, provided an appropriate Wess-Zumino gauge chosen.

\subsection{Fronsdal-type actions in AdS$_3$}

We begin by reviewing the AdS$_3$ counterparts of the (Fang-)Fronsdal actions in AdS$_4$ \cite{F,FF}, mostly following the presentation in \cite{KP}.

The Fang-Fronsdal action in AdS$_4$ \cite{FF} is first-order in derivatives. It can be generalised to AdS$_3$ using two different gauge-invariant actions, 
$S^{(n,+)}_{\text{FF}}$ and $S^{(n,-)}_{\text{FF}}$. Given an integer $n\geq 4$, such a model in AdS$_3$
is realised in terms of three real fields
\be
\cV^{(n, \pm)}_{\text{FF}} = \big \{ h_{\a(n)}, y_{\a(n-2)}, y_{\a(n-4)}  \big \}~,
 \ee
which are defined modulo gauge transformations of the form
\begin{subequations}\label{B.3}
	\bea
	\d h_{\a(n)}&=&\mathfrak{D}_{(\a_1 \a_2}\l_{\a_3 ... \a_n)}~,\\
	\d y_{\a(n-2)}&=&\frac{1}{n}\mathfrak{D}_{(\a_1}{}^\b \l_{\a_2 ... \a_{n-2})\b}\pm|\m|\l_{\a(n-2)}~, \\
	\d y_{\a(n-4)}&=&\mathfrak{D}^{\b\g} \l_{\b\g\a(n-4)}~.
	\eea
\end{subequations}
Here, the gauge parameter $\l_{\a(n-2)}$ is real and unconstrained. Modulo an overall constant, the corresponding gauge-invariant  action is
\bea \label{VTIrred}
S^{(n,\pm)}_{\text{FF}}&=&\frac{\ri^n}{2^{\lfloor n/2 \rfloor +1}}\int \rd^3x\,e\,\bigg \{ h^{\b\a(n-1)}\mathfrak{D}_{\b}{}^{\g}h_{\g\a(n-1)}+2(n-2)y^{\a(n-2)}\mathfrak{D}^{\b\g}h_{\b\g\a(n-2)} ~ \non \\
&&+4(n-2)y^{\b\a(n-3)}\mathfrak{D}_\b{}^\g y_{\g\a(n-3)}+\frac{2n(n-3)}{n-1}y^{\a(n-4)}\mathfrak{D}^{\b\g}y_{\b\g\a(n-4)}~ \non \\
&&- \frac{(n-3)(n-4)}{(n-1)(n-2)}y^{\b\a(n-5)}\mathfrak{D}_\b{}^\g y_{\g\a(n-5)}
\pm |\m| \Big ( (n-2)  h^{\a(n)}h_{\a(n)}~ \non \\
&&- 4n(n-2) y^{\a(n-2)}y_{\a(n-2)}- \frac{n(n-3)}{n-1} y^{\a(n-4)}y_{\a(n-4)} \Big ) \bigg \} ~.
\eea
Only the action $S^{(n,+)}_{\text{FF}}$ was considered in \cite{KP}.
Both actions $S^{(n,+)}_{\text{FF}}$ and $S^{(n,-)}_{\text{FF}}$ are  the $d=3$ counterparts of the Fang-Fronsdal action provided $n$ is odd, 
$n = 2s +1$, with $s\geq 2$ an integer. 

It must be noted that the action \eqref{VTIrred} can be expressed in a more compact way in terms of a reducible frame field,
\be
\bm h_{\b\g;\a{(n-2)}} :=(\g^b)_{\b\g} \bm h_{b;\a{(n-2)}}~,
\ee
which is defined modulo gauge transformations of the form
\bea
\d \bm{h}_{\b\g;\a(n-2)}=\mathfrak{D}_{\b\g}\l_{\a(n-2)}\pm(n-2)|\m|\big ( \varepsilon_{\b(\a_1}\l_{\a_2 ... \a_{n-2})\g}+\varepsilon_{\g(\a_1}\l_{\a_2 ... \a_{n-2})\b} \big ) ~.
\label{frameB.6}
\eea
The corresponding gauge-invariant action takes the form
\bea \label{VT}
S^{(n,\pm)}_{\text{FF}}&=&\frac{\ri^n}{2^{\lfloor n/2 \rfloor +1}}\int \rd^3x\,e\,\bigg \{ \bm{h}^{\b\g;\a(n-2)} \mathfrak{D}_\b{}^\d \bm{h}_{\g\d;\a(n-2)}\pm(n-2)|\m|\big ( \bm{h}^{\b\g;\a(n-2)}\bm{h}_{\b\g;\a(n-2)} ~\non \\
&&+2\bm{h}^{\b\g;\a(n-3)}{}_\b \bm{h}_{\g\d;}{}_{\a(n-3)}{}^\d \big ) \bigg \}~.
\eea
In the flat-space limit, the action \eqref{VT} reduces to the model considered
by Tyutin and Vasiliev 
\cite{VT} (see also \cite{KP} for a review).
Ref. \cite{BSZ2} also made use of the action \eqref{VT} in the frame-like formulation, but only a particular choice of sign was considered. 
In order to relate the action \eqref{VT} to \eqref{VTIrred}, it is necessary to decompose the dynamical field $\bm{h}_{\b\g;\a(n-2)}$ into its irreducible components. 
The irreducible fields contained in $\bm{h}_{\b\g;\a(n-2)}$ can be defined as follows:
\begin{subequations} \label{Irred}
	\bea
	h_{\a(n)}&:=&\bm{h}_{(\a_1 \a_2;\a_3 ... \a_n)}~,\\
	y_{\a(n-2)}&:=&\frac{1}{n}\bm{h}^{\b}{}_{(\a_1 ; \a_2 ... \a_{n-2})\b}~, \\
	y_{\a(n-4)}&:=&\bm{h}^{\b\g;}{}_{\b\g\a(n-4)}~.
	\eea
\end{subequations}
The gauge transformation laws for these fields, which follow from \eqref{frameB.6},  are given by those in  \eqref{B.3}.   
Upon expressing the action \eqref{VT} in terms of the  fields \eqref{Irred}, it can be shown that the resulting action coincides with \eqref{VTIrred}.

The Fronsdal action in AdS$_4$ \cite{F} is second-order in derivatives.
 It can be generalised to AdS$_3$ as follows. Given an integer $n\geq 4$, 
 we introduce two real dynamical variables 
\be
\cV_{\text{F}} = \big \{ h_{\a(n)},  y_{\a(n-4)} \big \}~,
\ee
which are defined modulo gauge transformations of the form
\begin{subequations} \label{fg00}
	\bea
	\d h_{\a(n)}&=&\mathfrak{D}_{(\a_1 \a_2}\l_{\a_3 ... \a_n)}~ ,\\
	\d y_{\a(n-4)}&=&\frac{n-2}{n-1}\mathfrak{D}^{\b\g}\l_{\b \g \a(n-4)}~,
	\eea
\end{subequations}
with the gauge parameter $\l_{\a(n-2)}$ being real unconstrained. 
 Modulo an overall constant, the gauge-invariant  action is given by
\bea \label{Fronsdal}
S^{(n)}_{\text{F}}&=& \frac{\ri^n}{2^{\lfloor n/2 \rfloor +1}}\int \rd^3x\,e\,\bigg \{ h^{\a(n)} \cQ h_{\a(n)}-\frac{n}{4}\mathfrak{D}_{\b \g}h^{\b\g\a(n-2)}\mathfrak{D}^{\d\l}h_{\d\l\a(n-2)}~ \\
&&-\frac{n-3}{2}y^{\a(n-4)}\mathfrak{D}^{\b\g}\mathfrak{D}^{\d\l}h_{\b\g\d\l\a(n-4)}-2n(n-2)|\m|^2~   h^{\a(n)} h_{\a(n)} \non ~ \\
&&-\frac{n-3}{n} \bigg ( 2 y^{\a(n-4)} \cQ y_{\a(n-4)}-4(n^2+2)|\m|^2y^{\a(n-4)}y_{\a(n-4)} ~ \non \\
&&+\frac{(n-4)(n-5)}{4(n-2)}\mathfrak{D}_{\b\g}y^{\b\g\a(n-6)}\mathfrak{D}^{\d\l}y_{\d\l\a(n-6)} \bigg ) \bigg \}~, \non
\eea
where $\mathcal{Q}$ is the following Casimir operator of the AdS group $\sSO(2, 2)$:
\be
\mathcal{Q}:=\mathfrak{D}^a \mathfrak{D}_a -2|\m|^2M^{\b\g}M_{\b\g}~,\quad [\mathcal{Q},\mathfrak{D}_a]=0~.
\ee
The model \eqref{Fronsdal} is the $d=3$ counterpart of the Fronsdal action provided $n$ is even, $n = 2s $, with $s\geq 2$ an integer.


\subsection{Dual formulation of the Fronsdal-type action in Minkowski space} \label{apB2}

Here we introduce a one-parameter family of dual formulations for the flat-space counterpart
of the model \eqref{Fronsdal}. These results will be important for our analysis
in the next subsection. 

Let us consider the flat-space limit of the model \eqref{Fronsdal}
\bea 
S_{\rm F}^{(n)} 
&=& \frac{\ri^n}{2^{\lfloor n/2 \rfloor +1}}\int \rd^3x\,\bigg \{ 
\hf \pa^{\b\g} h^{\a(n)} \pa_{\b\g} h_{\a(n)}-\frac{n}{4}\pa_{\b \g}h^{\b\g\a(n-2)}\pa^{\d\l}h_{\d\l\a(n-2)}~  \non\\
&&+\frac{n-3}{n} \Big( \frac{n}{2} \pa^{\b\g} y^{\a(n-4)}\pa^{\d\l}h_{\b\g\d\l\a(n-4)}
- \pa^{\b\g}  y^{\a(n-4)} \pa_{\b\g} y_{\a(n-4)}
~ \non \\
&&\qquad -\frac{(n-4)(n-5)}{4(n-2)}\pa_{\b\g}y^{\b\g\a(n-6)}\pa^{\d\l}y_{\d\l\a(n-6)} \Big ) \bigg \}~,
\label{flat-n}
\eea
where $\partial_{\a\b}$ are the partial derivatives of three-dimensional Minkowski space. This action is invariant under the gauge transformations
\begin{subequations} \label{fg00-flat}
	\bea
	\d_\l h_{\a(n)}&=&\pa_{(\a_1 \a_2}\l_{\a_3 ... \a_n)}~ , \label{fg00-flat-a} \\
	\d_\l y_{\a(n-4)}&=&\frac{n-2}{n-1}\pa^{\b\g}\l_{\b \g \a(n-4)}~. \label{fg00-flat-b}
	\eea
\end{subequations}
Both fields $h_{\a(n)}$ and $y_{\a(n-4)}$ appear in \eqref{flat-n} with derivatives, 
\bea
S_{\rm F}^{(n)} =\frac{\ri^n}{2^{\lfloor n/2 \rfloor +1}}  \int \rd^{3} x \, \cL \big( \pa_{\b\g} h_{\a(n)}, \pa_{\b \g} y_{\a(n-4)} \big)~. 
\label{b00}
\eea 
and therefore a duality transformation may be performed upon each of them. Here we will only dualise $y_{\a(n-4)}$ and keep $h_{\a(n)}$ intact.\footnote{The gauge transformation law \eqref{fg00-flat-a} allows us to interpret $h_{\a(n)}$ as a conformal 
spin-$\frac{n}{2}$ gauge field, while \eqref{fg00-flat-b} is compatible with the interpretation of 
$y_{\a(n-4) }$ as a conformal compensator (following the modern supergravity terminology). Performing a duality transformation on 
$y_{ \a(n-4) }$ is equivalent to the introduction of  an alternative conformal compensator.   }

We may think of 
$\pa_{\b \g} y_{\a(n-4)}$ as a longitudinal tensor field, by analogy with a longitudinal vector field. 
Our dual formulation for \eqref{b00} is obtained by introducing the following first-order action 
\bea \label{b01}
S_{\text{first-order}} = \frac{\ri^n}{2^{\lfloor n/2 \rfloor +1}} \int \rd^{3}x \,  \Big\{ \cL \big( \pa_{\b\g} h_{\a(n)}, \cH_{\b \g; \, \a(n-4)} \big)
 \non\\
 -(n-3) \cH^{\b \g; \, \a(n-4)} F_{\b \g;\, \a(n-4)} \Big\}~, 
\eea
where we have introduced the Lagrangian
\bea
&& \cL \big( \pa_{\b\g} h_{\a(n)}, \cH_{\b \g; \, \a(n-4)} \big) = \hf \pa^{\b\g} h^{\a(n)} \pa_{\b\g} h_{\a(n)}-\frac{n}{4}\pa_{\b \g}h^{\b\g\a(n-2)}\pa^{\d\l}h_{\d\l\a(n-2)}~  \non\\
&&+\frac{n-3}{n} \Big( \frac{n}{2} \cH^{\b\g;\, \a(n-4)} \pa^{\d\l}h_{\b\g\d\l\a(n-4)}
- \cH^{\b\g;\, \a(n-4)}  \cH_{\b\g;\, \a(n-4)}
~ \non \\
&&\qquad + A \, \cH^{\b\g;\, \d \l \a(n-6)} \cH_{\d\l;\, \b \g \a(n-6)} + B\, \cH_{\b \g;}\,^{\b \g \a(n-6)} \cH^{\d \l;}\,_{\d \l\a(n-6)}\Big )
\eea
which involves the real coefficients $A$ and $B$ constrained by
\bea
A+ B = -\frac{(n-4)(n-5)}{4(n-2)}~.
\label{B.17}
\eea
The field $\cH_{\b \g;\, \a(n-4)}$ in \eqref{b01}  is unconstrained and the Lagrange multiplier $ F_{\b \g; \, \a(n-4)} $ is given by 
\bea
 F_{\b \g; \, \a(n-4)} = \pa_{(\b}{}^{\d} \Phi_{\g) \d; \, \a(n-4)} \quad \implies \quad 
 \pa^{\b \g}F_{\b \g; \, \a(n-4)}=0~,
\label{b02}
 \eea
for some unconstrained prepotential $\Phi_{\g \d; \, \a(n-4)}$. 
Varying the first-order action \eqref{b01} with respect to $\Phi^{\g \d;\, \a(n-4)}$ gives
\bea
\pa_{(\d}{}^{\b} \cH_{\g) \b; \, \a(n-4)} = 0
\quad \implies \quad \cH_{\b \g; \,\a(n-4)} =   \pa_{\b \g} y_{\a(n-4)}~,
\eea
and hence, $S_{\text{first-order}} $ reduces to the original  action \eqref{b00}, 
in accordance with \eqref{flat-n} and \eqref{B.17}.
On the other hand, we can integrate out
$\cH_{\b \g; \,\a(n-4)}$ using its equation motion, 
\bea
\frac{\pa}{\pa \cH^{\b \g; \,\a(n-4)}} \cL \big( \pa_{\d\l} h_{\a(n)}, \cH_{\d \l; \, \a(n-4)} \big) + F_{\b \g;\, \a(n-4)} = 0~,
\eea
which is equivalent to 
\bea
&&\cH_{\b \g; \, \a(n-4)}
-\frac{B}{2} \Big( \ve_{\b (\a_1} \ve_{|\g| \a_2} \cH^{\d \l;}\,_{\a_3 \dots \a_{n-4})\d \l}+ \ve_{\g (\a_1} \ve_{|\b| \a_2} \cH^{\d \l;}\,_{\a_3 \dots \a_{n-4})\d \l} \Big)\non\\
&&\qquad \quad - A \cH_{(\a_1 \a_2; \, \a_3 \dots \a_{n-4}) \b \g}
= \frac{n}{4} \pa^{\d \l} h_{\d \l \b \g \a(n-4)} - \frac{n}{2} F_{\b \g; \, \a(n-4)}~.
\eea
This equation allows us to express $\cH_{\b \g;\, \a(n-4)}$ in terms of $h_{\a(n)}$ and $F_{\b \g;\, \a(n-4)}$:
\bea
&&\cH_{\b \g; \a(n-4)} = \frac{n}{4(1-A)}\Big\{ \pa^{\d \l} h_{\d \l \b \g \a(n-4)} -2 F_{\b \g; \a(n-4)} \non\\
&&+ c_1 (A) \Big( \ve_{\b (\a_1} F^{\d}{}_{\a_2; \,\a_3 \dots \a_{n-4})\g\d} + \ve_{\g (\a_1} F^{\d}{}_{\a_2; \,\a_3 \dots \a_{n-4})\b\d} \Big) \non\\
&&+ c_2 (A) \Big( \ve_{\b (\a_1} \ve_{|\g| \a_2} F^{\d \l;}\,_{\a_3 \dots \a_{n-4})\d \l}+ \ve_{\g (\a_1} \ve_{|\b| \a_2} F^{\d \l;}\,_{\a_3 \dots \a_{n-4})\d \l} \Big) \Big\} ~,~~~~~~
\label{eomcH}
\eea
where we have defined
\begin{subequations}
\bea
c_1 (A) &=& \frac{2A(n-4)}{(n+2(A-2))}~,\\
c_2 (A) &=& -\frac{(n-4) \Big( 2A \,\big(2A+B(n-3)\big)+ (n-4)(B-A)\big(n+2(A-2)\big) \Big)}{(n+2(A-2)) \Big((n-4)\big(n-5-B(n-3)\big)-2A \Big) }~.~~
\eea
\end{subequations}
Substituting \eqref{eomcH} into $S_{\text{first-order}}$ results in a dual action
\bea
S_{\rm A}^{(n)}[ h_{\a(n)}, \Phi_{\g \d; \, \a(n-4)} ]= \frac{\ri^n}{2^{\lfloor n/2 \rfloor +1}}  \int \rd^{3}x \, 
\cL_{\rm dual} \Big(  \pa_{\b \g} h_{\a(n)}, F_{\b \g;\, \a(n-4)}  \Big)~.
\label{b03}
\eea
We have obtained an one-parameter family of dual actions. 

As described above, the one-parameter first-order action \eqref{b01} is equivalent to the original model \eqref{b00}. The action \eqref{b01} is invariant under the gauge $\l$-transformations \eqref{fg00-flat-a}, which act on $\cH_{\b \g;\, \a(n-4)}$ and $\F_{\b \g;\, \a(n-4)}$ as follows:
\begin{subequations}
\bea
\d_{\l} \mathcal{H}_{\b \g; \, \a(n-4)}&=&\frac{n-2}{n-1}\partial_{\b\g}\partial^{\d\rho}\l_{\d \rho \a(n-4)}~, \\
\d_{\l} \F_{\b\g;\a(n-4)} &=& -\frac{2}{n(n-1)}  \Big( \big( 2n-5-2B(n-2)\big) \pa_{(\b}{}^{\d} \l_{\g) \a(n-4)\d} \non\\
&& -\big( n-4-2B(n-2)\big) \partial_{(\a_1}{}^{\d} \l_{\a_2 \dots \a_{n-4}) \b \g \d } \Big) ~.
\eea
\label{b04}
\end{subequations}
The dual action \eqref{b03} is also invariant under  the gauge $\rho$-transformation,
\begin{subequations}
\bea
\d_{\rho} {\Phi}_{\g \d; \,\a(n-4)} &=& \pa_{\g \d} \rho_{\a(n-4)} 
\quad \implies \quad \d_{\rho} {F}_{\b \g; \,\a(n-4)} = 0~,\\
\d_{\rho} h_{\a(n-4)} &=& 0~.
\eea
\end{subequations}
We emphasise that the parameters $A$ and $B $ are constrained by \eqref{B.17}.


\subsection{Transverse formulation of the superspin-$(s+\hf)$ multiplet}
 \label{apB3}

The transverse formulation of the massless superspin-$(s+\hf)$ multiplet is described by the action \eqref{TransHIAct}, which is invariant under the gauge transformations \eqref{GTTHI}. The gauge freedom \eqref{GTTHI} can be used to impose the following Wess-Zumino gauge
\be \label{CGT}
H_{\a(2s+1)}|=0~, \qquad \nabla^{\b}H_{\b\a(2s)}|=0~, \qquad \U_{\b;\a(2s-2)}|=0~, \qquad \nabla^\b \U_{\b;\a(2s-2)}|=0~.
\ee
The residual gauge freedom preserving the gauge conditions \eqref{CGT} is given by
\begin{subequations}
	\bea
	\nabla_{(\a_1}\z_{\a_2...\a_{2s+1})}|&=&0~,\\
	\nabla^2\z_{\a(2s)}|&=&-\frac{2\ri s}{s+1}\big (\nabla_{(\a_1}{}^\b \z_{\a_2 ... \a_{2s})\b}+2(s+1)|\m|\z_{\a(2s)} \big )|~, \\
	\nabla_{\b}\eta_{\a(2s-2)}|&=&\nabla_{(\b}\eta_{\a(2s-2))}|=-\frac{1}{2s+1}\nabla^\g \z_{\b\g\a(2s-2)}|~, \\
	\nabla^2 \eta_{\a(2s-2)}|&=&-\frac{\ri}{2s+1}\nabla^{\b\g}\z_{\b\g\a(2s-2)}|~.
	\eea
\end{subequations}
This implies that there are three real independent gauge parameters at the component level, which we choose as
\bea
\x_{\a(2s)}:=\z_{\a(2s)}|~, \quad \l_{\a(2s-1)}:=-\frac{\ri s}{2s+1}\nabla^\b \z_{\b\a(2s-1)}|~, \quad  \r_{\a(2s-2)}:= -\eta_{\a(2s-2)}|~.
\eea
The next task is to identify the remaining independent component fields of $H_{\a(2s+1)}$ and $\U_{\b;\a(2s-2)}$  in the Wess-Zumino gauge \eqref{CGT}. 

Let us consider the fermionic sector. We begin by decomposing the superfield $\U_{\b;\a(2s-2)}$ into irreducible components
\be \label{DI}
\U_{\b;\a(2s-2)}:= Y_{\b\a(2s-2)}+\sum_{k=1}^{2s-2}\varepsilon_{\b\a_k}Z_{\a_1 ... \hat{\a}_k ... \a_{2s-2}}~,
\ee
where we have introduced the irreducible superfields
\begin{subequations}
	\bea
	Y_{\a(2s-1)}&:=& \U_{(\a_1;\a_2 \dots \a_{2s-1})}~,\\
	Z_{\a(2s-3)}&:=& \frac{1}{2s-1}\U^{\b;}{}_{\b\a(2s-3)}~.
	\eea
\end{subequations}
We find that the remaining independent fermionic fields are given by
\begin{subequations}
	\bea
	h_{\a(2s+1)}&:=&\frac{\ri}{4}\nabla^2H_{\a(2s+1)}|~, \\
	y_{\a(2s-1)}&:=& \frac{\ri}{8}\nabla^2 Y_{\a(2s-1)}|~, \\
	y_{\a(2s-3)}&:=& \frac{\ri s}{2}(2s-1) \nabla^2 Z_{\a(2s-3)}|~,
	\eea
\end{subequations}
Their gauge transformations are
\begin{subequations}
	\bea
	\d h_{\a(2s+1)}&=& \mathfrak{D}_{(\a_1 \a_2}\l_{\a_3 ... \a_{2s+1})}~, \\
	\d y_{\a(2s-1)}&=&\frac{1}{2s+1}\mathfrak{D}_{(\a_1}{}^\b \l_{\a_2 ... \a_{2s-1})\b} + |\m|\l_{\a(2s-1)}~, \\
	\d y_{\a(2s-3)}&=& \mathfrak{D}^{\b\g}\l_{\b\g\a(2s-3)}~.
	\eea
\end{subequations}
Upon the component reduction of the action \eqref{TransHIAct}, we find that the fermionic sector coincides with the Fang-Fronsdal-type action, $S_{\text{FF}}^{(2s+1,+)}$ given by $\eqref{VTIrred}$.

We now turn to the bosonic sector. To start with, we point out that the fourth  condition in \eqref{CGT} is equivalent to
\be
\nabla_\b \U_{\g;\a(2s-2)} |=\nabla_\g \U_{\b;\a(2s-2)} |~.
\ee
We then choose the following independent bosonic fields
\begin{subequations}
	\bea
	h_{\a(2s+2)}&:=&-\nabla_{(\a_1}H_{\a_2 ... \a_{2s+2})}|~,\\
	\F_{\b\g;\a(2s-2)}&:=&\nabla_\b \U_{\g;\a(2s-2)}|=\nabla_{(\b} \U_{\g);\a(2s-2)}|~.
	\eea
\end{subequations}
Upon the component reduction of the action \eqref{TransHIAct}, we obtain the bosonic action
\bea \label{BosComp}
S_{\text{bos}} &=& \Big (-\frac{1}{2}\Big )^s \int \rd^3 x\,e\, \bigg \{-\frac{1}{4}h^{\a(2s+2)}\mathcal{Q} h_{\a(2s+2)}+\frac{3}{16}\mathfrak{D}_{\b\g}h^{\b\g\a(2s)}
\mathfrak{D}^{\d\r}h_{\d\r\a(2s)}~ \non \\
&&+\frac{1}{4}(2s-1)F^{(\b\g;\a(2s-2))}\mathfrak{D}^{\d\l}h_{\b\g\d\l\a(2s-2)} -\frac{1}{4}(2s-1)F^{\b\g;\a(2s-2)}F_{\b\g;\a(2s-2)}~ \non \\
&&+\hf(s-1)(2s-1)F^{\b\g ; \a(2s-3)}{}_\b F_{\d\g ; \a(2s-3)}{}^\d \non\\
&&-\hf (2s-1)|\m| \Big( h^{\a(2s+2)}\mathfrak{D}_{(\a_1}{}^\b h_{\a_2 \dots \a_{2s+2})\b} - (s-1) \F^{\b \g; \, \a(2s-2)} \mathfrak{D}^{\d \l} h_{\d \l \b \g \a(2s-2)} \non\\
&& -2 F^{\b \g;\, \a(2s-2)} \F_{\b \g; \, \a(2s-2)} + 8(s-1) F^{\b \g;\, \a(2s-3)}\,_{\b} \F_{\d \g;\, \a(2s-3)}{}^{\d} \non\\
&&-2 (s-1)(2s-3) F_{\b \g;}{}^{\b \g \a(2s-4)} \F^{\d \l;}{}_{\d \l \a(2s-4)} \Big) \non\\
&&+|\m|^2 (s-1) (2s-1)\Big( \frac{3s}{(s-1) (2s-1)}\, h^{\a(2s+2)} h_{\a(2s+2)} \non\\
&&+ (s+1) \F^{\b \g; \, \a(2s-2)} \F_{\b \g; \, \a(2s-2)} + 2(2s^2-s+3) \F^{\b \g; \, \a(2s-3)}\,_{\b} \F_{\g \d;\, \a(2s-3)}{}^{\d} \non\\
&&+(s-2)(2s-3) \F_{\b \g;}{}^{\b \g \a(2s-4)} \F^{\d \l;}{}_{\d \l \a(2s-4)} \Big)
 \bigg \}~,
\eea
where we have introduced the field strength $F_{\b\g ; \a(2s-2)}$, 
\be
F_{\b\g ; \a(2s-2)} := \mathfrak{D}_{(\b}{}^\d \F_{\g)\d;\a(2s-2)}~.
\ee
The bosonic action \eqref{BosComp} is invariant under the gauge transformations
\begin{subequations} \label{rt0}
	\bea
	\d h_{\a(2s+2)} &=& \mathfrak{D}_{(\a_1 \a_2}\x_{\a_3 ... \a_{2s+2})}~, \\
	\d \F_{\b\g;\a(2s-2)} &=& \mathfrak{D}_{\b\g}\r_{\a(2s-2)} + 2(s-1)|\m| \Big( \varepsilon_{\b  (\a_1}\r_{\a_2 ... \a_{2s-2})\g} + \varepsilon_{\g  (\a_1}\r_{\a_2 ... \a_{2s-2})\b} \Big) ~   \non\\
	&&-\frac{1}{(2s+1)(s+1)}\Big ( (s+2)\mathfrak{D}_{(\b}{}^\d \x_{\g)\d\a(2s-2)}+(s-1)\mathfrak{D}_{(\a_1}{}^\d \x_{\a_2 ... \a_{2s-2})\b\g\d} \non \\
	&&+2(s+1)(2s+1)|\m|\x_{\b\g\a(2s-2)}\Big ) ~. 
	\label{rt}
	\eea
\end{subequations}
From \eqref{rt}, it can be shown that the field strength transforms as
\bea
\d F_{\b\g;\a(2s-2)} &=&-2(s-1)|\m| \Big (\varepsilon_{(\a_1|(\b}\mathfrak{D}_{\g)|}{}^\d\r_{\a_2 ... \a_{2s-2})\d} + \mathfrak{D}_{(\b|(\a_1}\r_{\a_2 ... \a_{2s-2})|\g)} ~\\
                &&-4|\m|\varepsilon_{(\b|(\a_1}\r_{\a_2 ... \a_{2s-2})|\g)} \Big ) \non \\
                &&-\frac{1}{(2s+1)(s+1)}\Big ( \hf (s+2)\mathfrak{D}_{\b\g}\mathfrak{D}^{\d\r}\x_{\d\r \a(2s-2)}+(2s+1)\cQ\x_{\g\b \a(2s-2)} \non ~\\
                &&+(s-1)\mathfrak{D}_{(\b|(\a_1}\mathfrak{D}^{\d\r}\x_{\a_2 ... \a_{2s-2})|\g)\d\r} +2(s+1)(2s+1)|\m|\mathfrak{D}_{(\b}{}^\d \x_{\g)\a(2s-2)\d} ~ \non \\
                &&-4(s-1)(s+1)(2s+1)|\m|^2\x_{\b\g\a(2s-2)} \Big ) ~ \non~.\label{rt2}
\eea
Thus, the field strength $F_{\b\g; \a(2s-2)}$ is invariant under $\r$-gauge transformations \eqref{rt}, $\d_{\r} F_{\b \g; \a(2s-2)} = 0$, only in the case $|\m| \to 0$. 

The complete component action of our transverse half-integer model takes the form
\bea
S^{\perp}_{(s+\hf)}[{H}_{\a(2s+1)} ,\U_{\b; \,\a(2s-2)} ] &=& S_{\rm bos} [h_{\a(2s+2)},\F_{\b\g;\a(2s-2)}] \non\\
&+& S_{\text{FF}}^{(2s+1,+)}[h_{\a(2s+1)}, y_{\a(2s-1)}, y_{\a(2s-3)}] ~.
\label{b27}
\eea
By applying the duality transformation described in 
subsection \ref{apB2}, it can be shown that the flat-space counterpart of \eqref{BosComp} is 
the dual formulation \eqref{b03} of 
 the Fronsdal action $S_{\rm F}^{(2s+2)}$ \eqref{flat-n} provided we fix the coefficients $A$ and $B$ as
\bea
A= -\hf (s-1)~, \qquad B = \frac{3}{4s} (s-1)~.
\eea
That is, the dual action given in \eqref{b03} coincides with the bosonic action \eqref{BosComp} in Minkowski space
\be
S_{\rm bos} [h_{\a(2s+2)},\F_{\b\g;\a(2s-2)}] \Big|_{ |\m|=0}
=S_{\rm A}^{(2s+2)} [h_{\a(2s+2)},\F_{\b\g;\a(2s-2)}]~.
\label{eqb43}
\ee
However, it must be noted that this duality does not hold in the presence of a non-vanishing AdS curvature, since \eqref{BosComp} cannot be written solely in terms of the field strength $F_{\b \g; \, \a(2s-2)}$.


\subsection{Longitudinal formulation of the superspin-$(s+\hf)$ multiplet}
 \label{apB4}

The longitudinal formulation of the massless superspin-$(s+\hf)$ multiplet is described by the action \eqref{LongHalfIntAct}, which  is invariant under the gauge transformations \eqref{LFHIGT}.
 The component structure of this model was studied in\cite{KP} 
 only in the flat superspace limit. Here we extend these results to AdS$_3$.

The gauge freedom \eqref{GTTHI} can be used to impose the following Wess-Zumino gauge
\be \label{gc}
H_{\a(2s+1)}| = 0~, \qquad \nabla^\b H_{\b\a(2s)}|=0~.
\ee
The residual gauge freedom which preserves the conditions \eqref{gc} is described by 
\begin{subequations}
\bea
0 &=& \nabla_{(\a_1}\z_{\a_2 ... \a_{2s+1})}|~,\\
\nabla^2 \z_{\a(2s)}| &=&- \frac{2 \ri s}{s+1} \big ( \nabla_{(\a_1}{}^\b \z_{\a_2 ... \a_{2s})\b} +2(s+1)|\m|\z_{\a(2s)}\big )|~.
\eea
\end{subequations}
These conditions imply that there are only two independent gauge parameters, which we define as follows:
\be
\x_{\a(2s)} := \z_{\a(2s)}|~, \qquad \l_{\a(2s-1)}:=-\frac{\ri s}{2s+1}\nabla^\b \z_{\b\a(2s-1)}|~.
\ee
Upon imposing the gauge \eqref{gc}, we are left with the following  component fields:
\begin{subequations} \label{cf}
	\bea
	h_{\a(2s+1)}&:=&\frac{\ri}{4}\nabla^2 H_{\a(2s+1)}|~, \\
	h_{\a(2s+2)}&:=&-\nabla_{(\a_1}H_{\a_2 ... \a_{2s+2})}|~, \\
	y_{\a(2s-2)}&:=&-4L_{\a(2s-2)}|~ \\
	y_{\a(2s-1)} &:=& (2s+1)\ri \nabla_{(\a_1}L_{\a_2 ... \a_{2s-1})}|~, \\
	y_{\a(2s-3)}&:=&2\ri \nabla^\b L_{\b \a(2s-3)}|~, \\
	F_{\a(2s-2)}&:=&\frac{\ri}{4}\nabla^2 L_{\a(2s-2)}|~.
	\eea
\end{subequations} 

Upon component reduction of the higher-spin model \eqref{LongHalfIntAct}, 
we find that the theory decouples into bosonic and fermionic sectors. The bosonic action is given by
\bea 
&&S_{\text{bos}} = \Big ( -\frac{1}{2} \Big )^s \int \rd^3 x\,e\, \bigg \{ - \frac{1}{4}h^{\a(2s+2)} \mathcal{Q} h_{\a(2s+2)} +\frac{1}{8}(s+1)\mathfrak{D}_{\b\g}h^{\b\g\a(2s)} \mathfrak{D}^{\d\l}h_{\d\l\a(2s)} ~ \non \\
&&+\frac{1}{8}(2s-1)y^{\a(2s-2)}\mathfrak{D}^{\b\g}\mathfrak{D}^{\d\l}h_{\b\g\d\l\a(2s-2)}+\frac{1}{16s}(2s-1)(s+1)y^{\a(2s-2)}\mathcal{Q} y_{\a(2s-2)}~ \non \\
&&+ \frac{1}{s}(s-1)(2s-1)F^{\b\a(2s-3)}\mathfrak{D}_\b{}^\g y_{\g\a(2s-3)}+\frac{4}{s}(s+1)(2s-1)F^{\a(2s-2)}F_{\a(2s-2)}~ \non \\
&& +\frac{1}{4}(s-1)(2s-1)|\m|y^{\b\a(2s-3)}\mathfrak{D}_\b{}^\g y_{\g\a(2s-3)}+2(s-1)(2s-1)|\m|F^{\a(2s-2)}y_{\a(2s-2)} ~ \non \\
&&+2s(s+1)|\m|^2h^{\a(2s+2)} h_{\a(2s+2)}-\hf (2s-1)(s+1)|\m|^2 y^{\a(2s-2)}y_{\a(2s-2)}  \bigg \}~. \label{BA00}
\eea
The bosonic field $F_{\a(2s-2)}$ is auxiliary, so upon elimination via its equation of motion
\be 
F_{\a(2s-2)}=-\frac{1}{8(s+1)}(s-1)\mathfrak{D}_{(\a_1}{}^\b y_{\a_2 ... \a_{2s-2})\b} - \frac{s}{4}|\m|y_{\a(2s-2)} ~,
\ee
we find that the resulting action coincides with the massless spin-$(s+1)$ action, 
$S^{(2s+2)}_{\text{F}}$, which is obtained from \eqref{Fronsdal} by setting $n = 2s+2$.

It can be shown that the fermionic action emerging from the reduction procedure coincides with the Fang-Fronsdal type action $S_{\text{FF}}^{(2s+1,-)}$ \eqref{VTIrred}. 
Therefore, our component actions take the form
\bea
S^{\parallel}_{(s+\hf)}[H_{\a(2s+1)} ,L_{\a(2s-2)} ] &=& S^{(2s+2)}_{\text{F}}[h_{\a(2s+2)}, y_{\a(2s-2)}] \non\\
&&+ S_{\text{FF}}^{(2s+1,-)}[h_{\a(2s+1)}, y_{\a(2s-1)}, y_{\a(2s-3)}]~.
\label{b40}
\eea

Comparing \eqref{b40} with \eqref{b27}, it is seen that the fermionic sector resulting from the reduction of the longitudinal model is given by $S_{\text{FF}}^{(2s+1,+)}[ h_{\a(2s+1)}, y_{\a(2s-1)}, y_{\a(2s-3)}]$. 
In flat-space, the two bosonic actions, $S^{(2s+2)}_{\text{F}}[h_{\a(2s+2)}, y_{\a(2s-2)}]$ and $S_{\rm bos} [h_{\a(2s+2)},\F_{\b\g;\a(2s-2)}]$, are related to each other by a duality transformation as described in subsection B.2. The fermionic actions are now given by $S_{\text{FF}}^{(2s+1)}$, thus they coincide identically.


\subsection{Transverse formulation of the superspin-$s$ multiplet}

The transverse formulation of the massless superspin-$s$ multiplet is described by the action \eqref{TIAction} and is invariant under gauge transformations \eqref{TIGT}. The gauge freedom \eqref{TIGT} can be used to impose the following Wess-Zumino gauge conditions:
\be \label{TICGT}
H_{\a(2s)}|=0~, \qquad \nabla^{\b}H_{\b\a(2s-1)}|=0~, \qquad \J_{\b;\a(2s-2)}|=0~, \qquad \nabla^\b \J_{\b;\a(2s-2)}|=0~.
\ee
The residual gauge symmetry preserving \eqref{TICGT} is given by
\begin{subequations}
	\bea
	\nabla_{(\a_1}\z_{\a_2...\a_{2s})}|&=&0~,\\
	\nabla^2\z_{\a(2s-1)}|&=&-\frac{2\ri(2s-1)}{2s+1}\big (\nabla_{(\a_1}{}^\b \z_{\a_2 ... \a_{2s-1})\b}+(2s+1)|\m|\z_{\a(2s-1)} \big )|~, \\
	\nabla_{\b}\eta_{\a(2s-2)}|&=&\nabla_{(\b}\eta_{\a(2s-2))}|=-\ri  \z_{\b\a(2s-2)}|~, \\
	\nabla^2 \eta_{\a(2s-2)}|&=&-\ri \nabla^{\b}\z_{\b\a(2s-2)}|~.
	\eea
\end{subequations}
Thus, there are three real independent gauge parameters, which we define as
\begin{subequations}
\bea
&&\x_{\a(2s-1)}:=\z_{\a(2s-1)}|~, \quad \r_{\a(2s-2)}:=-\eta_{\a(2s-2)}|~,  \\  &&\l_{\a(2s-2)}:=-\frac{2s-1}{4s}\nabla^\b \z_{\b\a(2s-2)}|~, \qquad ~
\eea
\end{subequations}
We now wish to find the remaining independent component fields of $H_{\a(2s)}$ and $\J_{\b;\a(2s-2)}$  in the Wess-Zumino gauge \eqref{TICGT}. The last Wess-Zumino gauge condition \eqref{TICGT} yields the relation
\be
\nabla_\b \J_{\g;\a(2s-2)}=\nabla_\g \J_{\b;\a(2s-2)}~.
\ee
Thus, we choose to describe the bosonic sector in terms of the independent fields
\begin{subequations}
	\bea
	X_{\a(2s)}&:=&\frac{\ri}{4}\nabla^2 H_{\a(2s)}|~,\\
	\F_{\b\g;\a(2s-2)}&:=&\nabla_\b \J_{\g;\a(2s-2)}|=\nabla_{(\b} \J_{\g);\a(2s-2)}|~.
	\eea
\end{subequations}
Upon the component reduction of the action \eqref{TransHIAct}, we obtain the bosonic action
\bea \label{TIB}
&&S_{\text{bos}}= -2(2s-1) \Big (-\frac{1}{2}\Big )^s  \int \rd^3 x~e~ \bigg \{ \frac{s-1}{2s-1}X^{\a(2s)}X_{\a(2s)}+F^{\b\g;\a(2s-2)}X_{\b\g\a(2s-2)} \non ~ \\
&&+\hf F^{\b\g;\a(2s-2)}F_{\b\g;\a(2s-2)}+\frac{1}{2s}(s-1)F^{\b \g; \a(2s-3)}{}_\b F_{\g \d ; \a(2s-3)}{}^\d ~ \non \\
&&+2(s-1)|\m|\F^{\b\g;\a(2s-2)}X_{\b\g\a(2s-2)}+(2s-1)|\m|F^{\b\g;\a(2s-2)}\F_{\b\g;\a(2s-2)}~ \non \\
&&+2(s-1)|\m|F^{\b \g; \a(2s-3)}{}_\b \F_{\g \d; \a(2s-3)}{}^\d+\frac{2}{s}(s-1)^2|\m|F^{\b \g; \a(2s-3)}{}_{\b}\F_{\d (\g;\a(2s-3))}{}^\d ~ \non \\
&&+2s(s-1)|\m|^2\F^{\b\g;\a(2s-2)}\F_{\b\g;\a(2s-2)}+2s(s-1)|\m|^2\F^{\b \g; \a(2s-3)}{}_\b \F_{\g \d; \a(2s-3)}{}^\d ~\non \\
&&+\frac{2}{s}(s-1)^2(s-2)|\m|^2 \F^{\b \g; \a(2s-3)}{}_\b \F_{\d (\g;\a(2s-3))}{}^\d  \bigg \}~,
\eea
where we have introduced the field strength $F_{\b\g ; \a(2s-2)}$, 
\be
F_{\b\g ; \a(2s-2)} := \mathfrak{D}_{(\b}{}^\d \F_{\g)\d;\a(2s-2)}~.
\ee
It can be shown that the action \eqref{TIB} is invariant under the gauge transformations\begin{subequations}
	\bea
	\d X_{\a(2s)} &=& \mathfrak{D}_{(\a_1 \a_2}\l_{\a_3 ... \a_{2s})}~, \\
	\d \F_{\b\g;\a(2s-2)} &=& \mathfrak{D}_{\b\g}\r_{\a(2s-2)} +4 (s-1)|\m|\varepsilon_{(\b | (\a_1}\r_{\a_2 ... \a_{2s-2})|\g)} ~  \label{rgt1} \\
	&&+\frac{4}{2s-1}(s-1) \varepsilon_{(\b | (\a_1}\l_{\a_2 ... \a_{2s-2})|\g)} ~. \non
	\eea 
\end{subequations}
The field $X_{\a(2s)}$ is auxiliary, so upon elimination via its equation of motion
\be
X_{\a(2s)}=-\frac{2s-1}{2(s-1)}\big ( F_{\a(2s)}+2(s-1)|\m|\F_{(\a_1 \a_2;\, \a_3 \dots \a_{2s})} \big )~,
\ee
we obtain
\bea \label{BWA}
&&S_{\text{bos}}= \frac{2s-1}{s(2s-2)} \Big (-\frac{1}{2}\Big )^s  \int \rd^3 x~e~ \bigg \{ s \, F^{\b\g;\a(2s-2)}F_{\b\g;\a(2s-2)}  \non ~ \\
&&+ 2s(s-1)F^{\b \g; \a(2s-3)}{}_\b F_{\g \d ; \a(2s-3)}{}^\d +(2s-3)(s-1)F_{\b \g;}\,^{\b \g \a(2s-4)} F^{\d \l;}\,_{\d \l \a(2s-4)}  ~ \non \\ 
&&-4 (s-1)^2|\m|^2 \Big( s\, \F^{\b\g;\a(2s-2)}\F_{\b\g;\a(2s-2)} + 2 \,\F^{\b \g; \a(2s-3)}{}_\b \F_{\d \g; \a(2s-3)}{}^\d \non\\
&&- (2s-3)\F_{\b \g;}\,^{\b \g \a(2s-4)} \F^{\d \l;}\,_{\d \l \a(2s-4)} \Big) \bigg\}~.
\eea
We can express the field $\F_{\b\g;\a(2s-2)}$ in terms of its irreducible components
\begin{subequations}
\bea
h_{\a(2s)}&:=&\F_{(\a_1 \a_2;\a_3 ... \a_{2s})}~, \label{CF1} \\
y_{\a(2s-2)}&:=&\F^{\b}{}_{(\a_1 ; \a_2 ... \a_{2s-2})\b}~, \\
y_{\a(2s-4)}&:=&\frac{2(s-1)}{2s-1}\F^{\b\g;}{}_{\b\g\a(2s-4)}~. \label{CF2}
\eea
\end{subequations}
The gauge freedom \eqref{rgt1} can be used to gauge away $y_{\a(2s-2)}$. It follows that the remaining fields, \eqref{CF1} and \eqref{CF2}, have the corresponding gauge transformations\begin{subequations}
 \bea
 \d h_{\a(2s)} &=& \mathfrak{D}_{(\a_1 \a_2}\r_{\a_3 ... \a_{2s})}~, \\
 \d y_{\a(2s-4)}&=& \frac{2(s-1)}{2s-1}\mathfrak{D}^{\b\g}\r_{\b\g\a(2s-4)}~.
 \eea
\end{subequations}
Hence, by rewriting \eqref{BWA} in terms of the irreducible fields \eqref{CF1} and \eqref{CF2}, the resulting action proves to coincide (up to an overall factor) with the massless spin-$s$ action \eqref{Fronsdal}.

The study of the fermionic sector requires the decomposition of the reducible superfield $\J_{\b;\a(2s-2)}$ into irreducible parts. This procedure is completely analogous to that of the prepotential $\U_{\b;\a(2s-2)}$ \eqref{DI}. 
In this case, we find that the remaining independent fermionic fields are given by
\begin{subequations}
	\bea
	h_{\a(2s+1)}&:=&- \ri \nabla_{(\a_1}H_{\a_2 ... \a_{2s+1})}|~, \\
	y_{\a(2s-1)}&:=& \frac{\ri}{4}\nabla^2 Y_{\a(2s-1)}|~, \\
	z_{\a(2s-3)}&:=& \frac{\ri }{2}(2s-1) \nabla^2 Z_{\a(2s-3)}|~.
	\eea
\end{subequations}
Upon the component reduction of the action \eqref{TIAction}, it can be shown that the fermionic action obtained coincides with the Fang-Fronsdal-type action, $S^{(2s+1,-)}_{\text{FF}}$ \eqref{VTIrred}.
As a result, we find that the action corresponding to the transverse formulation of the massless superspin-$s$ multiplet in AdS$_3$ decomposes into
\bea \label{b65}
S^{\perp}_{(s)}[H_{\a(2s)} ,{\Psi}_{\b; \,\a(2s-2)} ] &=& \frac{2s-1}{s-1}S^{(2s)}_{\text{F}}[h_{\a(2s)},y_{\a(2s-4)}] \non\\
&&+ \, S^{(2s+1,-)}_{\text{FF}}[h_{\a(2s+1)},y_{\a(2s-1)},y_{\a(2s-3)}]~. 
\eea

\subsection{Longitudinal formulation of the superspin-$s$ multiplet}
The longitudinal formulation of the massless superspin-$s$ multiplet is described by the action \eqref{action-t3} and is invariant under gauge transformations \eqref{LongGTI}. This formulation corresponds to the massless first-order model, whose component reduction in the flat-superspace limit has been studied in \cite{KP}. 

The gauge freedom \eqref{LongGTI} can be used to impose the following Wess-Zumino gauge
\be  \label{GCLI}
H_{\a(2s)}| = 0~, \qquad \nabla^\b H_{\b\a(2s-1)}|=0~, \qquad V_{\a(2s-2)}| =0~.
\ee
The residual gauge symmetry which preserves the gauge transformations \eqref{GCLI} are
\begin{subequations} \label{RGLI}
\bea
0 &=& \nabla_{(\a_1}\z_{\a_2 ... \a_{2s})}|~,\\
0 &=& \nabla^\b \z_{\b\a(2s-2)}|~,\\
\nabla^2 \z_{\a(2s-1)}| &=&- \frac{2 \ri }{2s+1}(2s-1) \big ( \nabla_{(\a_1}{}^\b \z_{\b\a_2 ... \a_{2s-1})} +(2s+1)|\m|\z_{\a(2s-1)}\big )|~.
\eea
\end{subequations}
The relations \eqref{RGLI} indicate that there is only one independent gauge parameter, which we choose to define as:
\be
\x_{\a(2s-1)} := \z_{\a(2s-1)}|~.
\ee
Thus, we are left with the remaining independent component fields in the gauge \eqref{gc}:
\begin{subequations} 
	\bea
	h_{\a(2s+1)}&:=&- \ri \nabla_{(\a_1}H_{\a_2 ... \a_{2s+1})}|~, \\
	h_{\a(2s)}&:=&\frac{\ri}{4}\nabla^2 H_{\a(2s)}|~, \\
	y_{\a(2s-2)}&:=&\frac{\ri}{4}\nabla^2 V_{\a(2s-2)}|~ \\
	y_{\a(2s-1)} &:=& \frac{\ri}{2} \nabla_{(\a_1}V_{\a_2 ... \a_{2s-1})}|~, \\
	y_{\a(2s-3)}&:=&-2\ri s \nabla^\b V_{\b \a(2s-3)}|~. 
	\eea
\end{subequations}
It is easy to see that the gauge transformation laws for the bosonic fields are
\bea
\d h_{\a(2s)} = 0~, \qquad \d y_{\a(2s-2)} = 0~.
\eea
Upon reduction of the higher-spin model \eqref{LongHalfIntAct}, it can be shown that these bosonic fields appear without derivatives and hence, they are pure auxiliary fields. Indeed, we find that 
\bea
S^{\parallel}_{(s)}[H_{\a(2s)} ,V_{\a(2s-2)} ] &=& \Big ( -\frac{1}{2} \Big )^s \int \rd^3 x \, e\, \Big \{ h^{\a(2s)} h_{\a(2s)} + 2s(2s-1)y^{\a(2s-2)}y_{\a(2s-2)} \Big \} \non \\
&&+ S_{\text{FF}}^{(2s+1,+)}\,[h_{\a(2s+1)},y_{\a(2s-1)},y_{\a(2s-3)}]~.
\eea
Unlike the reduction of the transverse superspin-$s$ multiplet described by \eqref{b65}, here the fermionic action comes with a positive sign in the $|\m|$-dependent terms,  $S_{\text{FF}}^{(2s+1,+)}$. 

\begin{footnotesize}

\end{footnotesize}

\begin{thebibliography}{66}


\bibitem{HKO}
J.~Hutomo, S.~M.~Kuzenko, D.~Ogburn,
``${\cal N} = 2$ supersymmetric higher spin gauge theories and current multiplets in three dimensions,'' 
Phys.\ Rev. \ D {\bf 98} (2018) 125004 [arXiv:1807.09098 [hep-th]]. 


\bibitem{HK18}
J.~Hutomo and S.~M.~Kuzenko,
``Higher spin supermultiplets in three dimensions: (2,0) AdS supersymmetry,''
Phys. Lett. B \textbf{787} (2018) 175
[arXiv:1809.00802 [hep-th]].	
		
\bibitem{HK19}
J.~Hutomo and S.~M.~Kuzenko, ``Field theories with (2,0) AdS supersymmetry in ${\cal N}=1$ AdS superspace,''
Phys. Rev. D \textbf{100} (2019) 045010
[arXiv:1905.05050 [hep-th]].

\bibitem{Rosly}
 A.~A.~Rosly,
``Super Yang-Mills  constraints as integrability conditions,'' 
in Proceedings of the International 
Seminar {\it Group Theoretical 
Methods in Physics} (Zvenigorod, USSR, 1982),
M. A. Markov  (Ed.), 
Nauka, Moscow, 1983, Vol. 1, p. 263 (in Russian);
English translation: in {\it Group Theoretical 
Methods in Physics},'' M. A. Markov, V. I. Man'ko 
and A. E. Shabad  (Eds.), Harwood Academic Publishers, 
London, Vol. 3, 1987, p. 587.


		
\bibitem{GIKOS}
A.~Galperin, E.~Ivanov, S.~Kalitzin, V.~Ogievetsky and E.~Sokatchev,
``Unconstrained N = 2 matter, Yang-Mills and supergravity theories in harmonic
superspace,''
Class.\ Quant.\ Grav.\  {\bf 1} (1984) 469.
 

\bibitem{KLR}
A. Karlhede, U. Lindstr\"om and M. Ro\v cek,
``Self-interacting tensor multiplets in N = 2 superspace,''
Phys.\ Lett.\ B {\bf 147} (1984) 297. 

		


\bibitem{A-GF}
L.~Alvarez-Gaum\'e and D.~Z.~Freedman,
``Geometrical structure and ultraviolet finiteness in 
the supersymmetric sigma model,''
Commun.\ Math.\ Phys.\  {\bf 80} (1981) 443.


\bibitem{HKLR}
C.~M.~Hull, A.~Karlhede, U.~Lindstr\"om and M.~Ro\v{c}ek,
``Nonlinear sigma models and their gauging in and out of superspace,''
Nucl.\ Phys.\ B {\bf 266} (1986) 1.


\bibitem{BX}
J.~Bagger and C.~Xiong,  ``N = 2 nonlinear sigma models in N = 1 superspace:  Four 
and five  dimensions,''  arXiv:hep-th/0601165.


 
\bibitem{GIOS}
A.~S.~Galperin, E.~A.~Ivanov, V.~I.~Ogievetsky and E.~S.~Sokatchev,
{\it Harmonic Superspace}, Cambridge University Press,  2001.


\bibitem{LR-projective1}
U.~Lindstr\"om and M.~Ro\v{c}ek,
``New hyperk\"ahler  metrics  and new supermultiplets,''
 Commun.\ Math.\ Phys.\  {\bf 115}  (1988) 21.
 

\bibitem{LR-projective2}
U.~Lindstr\"om and M.~Ro\v{c}ek,
   ``N = 2 super Yang-Mills theory in projective superspace,''
Commun.\ Math.\ Phys.\  {\bf 128}  (1990) 191.

 \bibitem{GIOS2}
A.~Galperin, E.~Ivanov, V.~Ogievetsky and E.~Sokatchev,
``Hyperk\"ahler metrics and harmonic superspace,''
Commun. Math. Phys. \textbf{103} (1986) 515.

\bibitem{KT-M08}
S.~M.~Kuzenko and G.~Tartaglino-Mazzucchelli,
 ``Field theory in 4D N=2 conformally flat superspace,''  JHEP {\bf 0810} (2008) 001
  [arXiv:0807.3368 [hep-th]].


\bibitem{BKsigma1}
D.~Butter and S.~M.~Kuzenko,
 ``N=2 supersymmetric sigma-models in AdS,''
Phys.\ Lett.\  B {\bf 703} (2011) 620
  [arXiv:1105.3111 [hep-th]].


\bibitem{BKsigma2}
D.~Butter and S.~M.~Kuzenko,
  ``The structure of N=2 supersymmetric nonlinear $\s$-models in AdS${}_4$,''
JHEP {\bf 1111} (2011) 080
  [arXiv:1108.5290 [hep-th]].
  
 
\bibitem{BKLT-M} 
  D.~Butter, S.~M.~Kuzenko, U.~Lindstr\"om and G.~Tartaglino-Mazzucchelli,
  ``Extended supersymmetric sigma models in AdS${}_4$ from projective superspace,''
  JHEP {\bf 1205} (2012) 138
  [arXiv:1203.5001 [hep-th]].
 

\bibitem{AT}
A.~Ach\'ucarro and P.~K.~Townsend,
``A Chern-Simons action for three-dimensional anti-de Sitter supergravity
 theories,''  Phys.\ Lett.\  B {\bf 180} (1986) 89.
 

\bibitem{KLT-M12} 
S.~M.~Kuzenko, U.~Lindstr\"om and G.~Tartaglino-Mazzucchelli,
``Three-dimensional (p,q) AdS superspaces and matter couplings,''
JHEP {\bf 1208} (2012)  024 [arXiv:1205.4622 [hep-th]].


\bibitem{KLT-M11}
S.~M.~Kuzenko, U.~Lindstr\"om and G.~Tartaglino-Mazzucchelli,
``Off-shell supergravity-matter couplings in three dimensions,''
JHEP  {\bf 1103} (2011) 120 [arXiv:1101.4013 [hep-th]].
		
 \bibitem{KT-M11} 
 S.~M.~Kuzenko and G.~Tartaglino-Mazzucchelli,
``Three-dimensional N=2 (AdS) supergravity and associated supercurrents,''
JHEP {\bf 1112} (2011) 052  [arXiv:1109.0496 [hep-th]].

  \bibitem{KLRST-M} 
  S.~M.~Kuzenko, U.~Lindstr\"om, M.~Ro\v{c}ek, I.~Sachs 
  and G.~Tartaglino-Mazzucchelli,
  ``Three-dimensional N=2 supergravity theories: From superspace to components,''
  Phys.\ Rev.\ D {\bf 89} (2014) 085028
  [arXiv:1312.4267 [hep-th]].

\bibitem{BILS}
I.~A.~Bandos, E.~Ivanov, J.~Lukierski and D.~Sorokin,
``On the superconformal flatness of AdS superspaces,''
JHEP \textbf{06} (2002) 040
[arXiv:hep-th/0205104 [hep-th]].



  \bibitem{GGRS}
 S.~J.~Gates, Jr., M.~T.~Grisaru, M.~Ro\v{c}ek and W.~Siegel,
{\it Superspace, or One Thousand and One Lessons in Supersymmetry},
Front.\ Phys.\  {\bf 58}, 1 (1983) [arXiv:hep-th/0108200].


		
\bibitem{BKT-M}
D.~Butter, S.~M.~Kuzenko and G.~Tartaglino-Mazzucchelli,
``Nonlinear sigma models with AdS supersymmetry in three dimensions,''
JHEP {\bf 1302} (2013) 121  [arXiv:1210.5906 [hep-th]].

	
			
\bibitem{KO} 
S.~M.~Kuzenko and D.~X.~Ogburn,
``Off-shell higher spin N=2 supermultiplets in three dimensions,''
Phys.\ Rev.\ D {\bf 94}  (2016) 106010 
[arXiv:1603.04668 [hep-th]].


\bibitem{KSP}
S.~M.~Kuzenko,  V.~V.~Postnikov and A.~G.~Sibiryakov,
``Massless gauge superfields of higher half-integer superspins,''
JETP Lett.\  {\bf 57}     (1993) 534
[Pisma Zh.\ Eksp.\ Teor.\ Fiz.\  {\bf 57} (1993) 521].
  
\bibitem{KS93}
S.~M.~Kuzenko and A.~G.~Sibiryakov,
``Massless gauge superfields of higher integer superspins,''
JETP Lett.\  {\bf 57} (1993)    539 
[Pisma Zh.\ Eksp.\ Teor.\ Fiz.\  {\bf 57} (1993) 526].

\bibitem{KS94}
S.~M.~Kuzenko and A.~G.~Sibiryakov,
``Free massless higher-superspin superfields on the anti-de Sitter superspace"
Phys.\ Atom.\ Nucl.\  {\bf 57} (1994) 1257
   [Yad.\ Fiz.\  {\bf 57} (1994) 1326]
  [arXiv:1112.4612 [hep-th]].


\bibitem{BK} 
I.~L.~Buchbinder and S.~M.~Kuzenko,
{\it Ideas and Methods of Supersymmetry and
Supergravity or a Walk Through Superspace}, IOP, Bristol, 1995
(Revised Edition: 1998).




\bibitem{BV}
  I.~A.~Batalin and G.~A.~Vilkovisky,
  ``Quantization of gauge theories with linearly dependent generators,''
  Phys.\ Rev.\ D  {\bf 28} (1983) 2567.
  
 
\bibitem{BKS} 
  I.~L.~Buchbinder, S.~M.~Kuzenko and A.~G.~Sibiryakov,
  ``Quantization of higher spin superfields in the anti-de Sitter superspace,''
  Phys.\ Lett.\ B {\bf 352} (1995) 29 
  [hep-th/9502148].
 
  \bibitem{FP} 
  L.~D.~Faddeev and V.~N.~Popov,
  ``Feynman diagrams for the Yang-Mills field,''
  Phys.\ Lett.\ B {\bf 25} (1967) 29.

\bibitem{GGS} 
  M.~R.~Gaberdiel, R.~Gopakumar and A.~Saha,
  ``Quantum $W$-symmetry in AdS${}_3$,''
  JHEP {\bf 1102} (2011) 004
  [arXiv:1009.6087 [hep-th]].
  
 
\bibitem{McA2}
I.~N.~McArthur, 
``Super b(4) coefficients,''
  Phys.\ Lett.\ B {\bf 128} (1983) 194;
``Super b(4) coefficients in supergravity,''
Class.\ Quant.\ Grav.\  {\bf 1} (1984) 245.
 
  
\bibitem{BK86} 
  I.~L.~Buchbinder and S.~M.~Kuzenko,
  ``Matter superfields in external supergravity: Green functions, effective action and superconformal anomalies,''
  Nucl.\ Phys.\ B {\bf 274} (1986) 653. 

		
\bibitem{KP} 	S.~M.~Kuzenko and M.~Ponds, ``Topologically massive higher spin gauge theories,''	
JHEP {\bf 1810} (2018) 160 [arXiv:1806.06643 [hep-th]]. 

		

\bibitem{HitchinKLR}
N.~J.~Hitchin, A.~Karlhede, U.~Lindstr\"om and M.~Ro\v cek,
``Hyperk\"ahler metrics and supersymmetry,''
Commun.\ Math.\ Phys.\  {\bf 108} (1987) 535.

		
\bibitem{KT}
S.~M.~Kuzenko and M.~Tsulaia,
``Off-shell massive N=1 supermultiplets in three dimensions,''	
Nucl.\ Phys.\ B {\bf 914} (2017) 160 [arXiv:1609.06910 [hep-th]].
	

\bibitem{KNG}
S.~M.~Kuzenko, J.~Novak and G.~Tartaglino-Mazzucchelli,
``Higher derivative couplings and massive supergravity in three dimensions,''
JHEP \textbf{09} (2015) 081
[arXiv:1506.09063 [hep-th]].


\bibitem{SK}
S.~M.~Kuzenko,
``Higher spin super-Cotton tensors and generalisations of the linear\textendash{}chiral duality in three dimensions,'' Phys. Lett. B \textbf{763} (2016) 308
[arXiv:1606.08624 [hep-th]].


\bibitem{IS}
E.~A.~Ivanov and A.~S.~Sorin,
``Superfield formulation of OSp(1,4) supersymmetry,''
J.\ Phys.\ A  {\bf 13} (1980) 1159.

		
\bibitem{WS}
W.~Siegel,
``Unextended superfields in extended supersymmetry,"
Nucl. Phys. B {\bf 156} (1979) 135.

\bibitem{JS}
J.~F.~Schonfeld, 
``A mass term for three-dimensional gauge fields,"
Nucl.Phys. B {\bf 185} (1980) 157.

\bibitem{DJT1}
S.~Deser, R.~Jackiw and S.~Templeton,
``Three-dimensional massive gauge theories,''
Phys. Rev. Lett. \textbf{48}  (1982) 975.

\bibitem{DJT2}
S.~Deser, R.~Jackiw and S.~Templeton,
``Topologically massive gauge theories,''
Annals Phys. \textbf{140} (1982) 372
[Erratum-ibid. \textbf{185} (1988) 406].

\bibitem{KP19}
S.~M.~Kuzenko and M.~Ponds,
``Conformal geometry and (super)conformal higher-spin gauge theories,''
JHEP {\bf 1905} (2019)   113 
[arXiv:1902.08010 [hep-th]].


  \bibitem{BKNT-M1} 
  D.~Butter, S.~M.~Kuzenko, J.~Novak and G.~Tartaglino-Mazzucchelli,
  ``Conformal supergravity in three dimensions: New off-shell formulation,''
  JHEP {\bf 1309} (2013) 072
  [arXiv:1305.3132 [hep-th]].
  
\bibitem{KP21}
S.~M.~Kuzenko and M.~Ponds,
``Higher-spin Cotton tensors and massive gauge-invariant actions in AdS$_3$,''
[arXiv:2103.11673 [hep-th]].
  
  
  \bibitem{BHHK}
E.~I.~Buchbinder, D.~Hutchings, J.~Hutomo and S.~M.~Kuzenko,
``Linearised actions for $ \mathcal{N} $ -extended (higher-spin) superconformal gravity,''
JHEP \textbf{08} (2019) 077
[arXiv:1905.12476 [hep-th]].
  
  \bibitem{BHT}
E.~A.~Bergshoeff, O.~Hohm and P.~K.~Townsend,
``On higher derivatives in 3D gravity and higher spin gauge theories,''
Annals Phys. \textbf{325} (2010) 1118 [arXiv:0911.3061 [hep-th]].


\bibitem{BKRTY}
E.~A.~Bergshoeff, M.~Kovacevic, J.~Rosseel, P.~K.~Townsend and Y.~Yin,
``A spin-4 analog of 3D massive gravity,''
Class. Quant. Grav. \textbf{28} (2011) 245007
[arXiv:1109.0382 [hep-th]].


\bibitem{BSZ3} 
  I.~L.~Buchbinder, T.~V.~Snegirev and Y.~M.~Zinoviev,
  ``Lagrangian formulation of the massive higher spin supermultiplets in three dimensional space-time,''
  JHEP {\bf 1510} (2015) 148
  [arXiv:1508.02829 [hep-th]].

\bibitem{BSZ4} 
I.~L.~Buchbinder, T.~V.~Snegirev and Y.~M.~Zinoviev,
  ``Lagrangian description of massive higher spin supermultiplets in AdS$_{3}$ space,''
  JHEP {\bf 1708} (2017) 021
  [arXiv:1705.06163 [hep-th]].
  
\bibitem{BSZ5}
I.~L.~Buchbinder, T.~V.~Snegirev and Y.~M.~Zinoviev,
``Supersymmetric higher spin models in three dimensional spaces,''
Symmetry \textbf{10} (2017)  9
[arXiv:1711.11450 [hep-th]].


\bibitem{BSZ1} 
  I.~L.~Buchbinder, T.~V.~Snegirev and Y.~M.~Zinoviev,
  ``Gauge invariant Lagrangian formulation of massive higher spin fields in $(A)dS_3$ space,''
  Phys.\ Lett.\ B {\bf 716} (2012) 243
  [arXiv:1207.1215 [hep-th]].

\bibitem{BSZ2} 
  I.~L.~Buchbinder, T.~V.~Snegirev and Y.~M.~Zinoviev,
  ``Frame-like gauge invariant Lagrangian formulation of massive fermionic higher spin fields in $AdS_3$ space,''
  Phys.\ Lett.\ B {\bf 738} (2014) 258
  [arXiv:1407.3918 [hep-th]].


\bibitem{Zinoviev1} 
  Y.~M.~Zinoviev,  ``On massive high spin particles in AdS,''  hep-th/0108192.
  
\bibitem{Zinoviev2}
Y.~M.~Zinoviev,
``Frame-like gauge invariant formulation for massive high spin particles,''
Nucl. Phys. B \textbf{808} (2009) 185
[arXiv:0808.1778 [hep-th]].

\bibitem{Metsaev} 
  R.~R.~Metsaev,
  ``Gauge invariant formulation of massive totally symmetric fermionic fields in (A)dS space,''  Phys.\ Lett.\ B {\bf 643} (2006) 205
  [hep-th/0609029].
  
  \bibitem{F}
C.~Fronsdal,
``Singletons and massless, integral-spin fields on de Sitter Space,'' Phys. Rev. D \textbf{20} (1979), 848-856.



  
\bibitem{FF}
J.~Fang and C.~Fronsdal,
``Massless, half-integer spin fields in de Sitter space,''
Phys. Rev. D \textbf{22} (1980) 1361.


\bibitem {VT}
I.~V.~Tyutin and M.~A.~Vasiliev,
``Lagrangian formulation of irreducible massive fields of arbitrary spin in (2+1)-dimensions,''
Theor. Math. Phys. \textbf{113} (1997) 1244
[Teor. Mat. Fiz. \textbf{113N1} (1997) 45]
[arXiv:hep-th/9704132 [hep-th]].




\end{thebibliography}
\end{document}